\documentclass{cernrep}

 \usepackage{amsmath}        % if you are using this package,
\usepackage{amsthm}
\usepackage{amssymb}
\usepackage{slashed}
\usepackage{bm}
\usepackage{bbm}
\usepackage{axodraw}
\usepackage{graphicx}
\usepackage{wrapfig,rotating}

\def\a{\alpha}
\def\b{\beta}
\def\g{\gamma}\def\G{\Gamma}
\def\d{\delta}\def\D{\Delta}
\def\eps{\epsilon}
\def\ft#1#2{{\textstyle{\frac{#1}{#2}}}}
\def\im{{\rm i}} % from NORTH-HOLLAND

\def\SU#1{{\rm SU({#1})}}

\def\l{\lambda}

\def\m{\mu}

\def\n{\nu}

\def\o{\omega}\def\O{\Omega}
\def\P{\Psi}
\def\pa{\partial}

\def\rd{{\rm d}}
\def\s{\sigma}
\def\U1{{\rm U(1)}}
\def\slash#1{\rlap{\hbox{$\mskip 1 mu /$}}#1}      % good slash for lower case
\def\Slash#1{\rlap{\hbox{$\mskip 3 mu /$}}#1}      % " upper
\begin{document}
\title{QCD}
\author{E. Laenen}
\institute{Nikhef, Amsterdam, The Netherlands\\
    Institute for Theoretical Physics, University of Amsterdam, The Netherlands\\
   Institute of Theoretical Physics, Utrecht University, The Netherlands}
\maketitle

\begin{abstract}
In these lecture notes I describe the theory of
QCD and its application, through perturbation theory,  at particle colliders. 
\end{abstract}

\section{Introduction}
\label{sec:introduction}

In particle physics, we encounter QCD nearly everywhere. 
The main collider of our time, the LHC, collides protons,
which are made up of quarks, antiquarks and gluons, collectively called partons. Every proton 
collision involves partons, which readily produce a multitude of
further partons, all turning into hadrons of one type or another.
At present we are however mostly interested in \emph{rare} final
states, faint signals involving Higgs bosons, top quarks, vector bosons,
possibly new particles. Hence we must understand very well how to
separate the new from the known, to ``remove the foreground'', in 
cosmology-speak; particle physicists call it background. 

But it would do gross injustice to QCD and its dynamics to see it as
merely a background engine. It really is a beautiful theory by itself.
It is the only unbroken non-abelian gauge theory we know exists in
Nature. Its Lagrangian is compact, and elegant
\begin{equation}
\label{eq:22}
   {\cal L}_{\mathrm{QCD}} = -\frac{1}{4}\mathrm{Tr}(G_{\mu\nu}G^{\mu\nu})
 -\sum_{f=1}^{n_f} \overline{\psi_f}(\slashed{D} + m_f) \psi_f\,.
\end{equation}
We shall discuss the meaning of the various symbols in this expression
shortly, but one should not forget to be amazed at the complex outcomes that
this relatively simple expression generates \footnote{Of course, for
  that matter, the QED Lagrangian is even simpler, and yet it 
governs all of atomic physics, chemistry etc.}. 
For this reason, QCD dynamics is very interesting to study \emph{sui generis}.
In these lecture notes I shall visit a number \footnote{Some of the
  notes correspond to a forthcoming book: Field Theory in Particle
  Physics, by B. de Wit, E. Laenen and J. Smith.} of aspects of QCD,
as relevant in collider physics. The structure of these notes is as follows.
 In the next section the fundamental
degrees of freedom and symmetries of QCD are discussed. In section 3 we discuss 
aspects of perturbative QCD when going to higher fixed orders. Section 4 contains
an exposition of some modern methods of calculations, focussing in particular
on helicity methods. Section 5 discusses aspects of all-order resummation,
the underlying reasons and some applications. I conclude in Section 6. 
An appendix contains conventions and useful formulae.\footnote{\emph{Caveat emptor}: though I tried
to avoid them, there might be errors and inconsistencies
in the equations
below. In addition, I made no effort to be exhaustive in references. }

\section{Partons and hadrons}
\label{sec:partons-hadrons}

In this section we discuss both the spectroscopic evidence for the presence
of quarks and gluons in hadrons,  as well as the partonic picture
relevant for high-energy collisons.

\subsection{Spectroscopy and symmetries of QCD}
\label{sec:spectr-symm-qcd}

Six types (or flavours) of quarks are presently known to exist.
They are fermions and are denoted by 
$u$, $c$, $t$ and $d$, $s$, $b$, respectively, abbreviations of the 
 names `up', `charm', `top', and `down', `strange',
`bottom'. 
Three ($u,c,t$) have electric 
charge $\ft23$ and three ($d,s,b$) charge $-\ft13$ (measured in units 
of the elementary charge).
Because quarks are not detected as separate physical particles (they are
confined into hadrons), their masses are not
exactly known, but can be estimated
from hadron spectroscopy once the hadron composition in terms of
quarks is given.  The mass values thus obtained are called
"constituent masses".  One commonly introduces quantum numbers
such as isospin or strangeness to distinguish the quark flavour,  
which then explains the corresponding quantum numbers of the
hadronic bound states.  Of course, there are also corresponding
antiquarks $\bar{u}, \bar{c}, \bar{t}$ and 
$\bar{d}, \bar{s}, \bar{b}$, with opposite charges.  The lightest-mass
mesons and baryons are bound states of quarks and/or antiquarks with
zero angular momentum.

For the moment let us restrict our attention to a single quark 
flavour, whose interactions are given by a non-abelian gauge theory.
This choice is motivated by the fact that only such theories have
the property that the interactions become strong at low energies, and
can therefore explain confinement. We shall return to this further below.
In order to let a non-abelian gauge group act on the 
quark field, we are forced to extend the number of fields. According to
QCD, this gauge group is $\SU{3}$. We shall try to justify this
choice for the gauge group in a little while and first consider the
definition of the theory.
In order that $\SU{3}$ can act nontrivially on the quark field
$q(x)$, this field must have at least 
three components, so we write $q_\a(x) = (q_1(x), q_2(x),
q_3(x))$. Hence for a given quark flavour, we have three
different fields. These three varieties are called {\it colours}
and are commonly denoted by `red', `green' and `blue'.
Of course, at first sight, this assumption seems to make matters worse. We
started with one quark for each flavour, which cannot be observed
as a free particle; now we have three times as many unobservable
quarks. Actually, the problem is even more vexing. Because quarks rotate 
under an $\SU{3}$ symmetry group, one should expect a corresponding
degeneracy for the observed bound states. In other words, 
each hadronic state should in general be degenerate and carry 
colour, while all other properties such as mass, electric charge and 
the like are independent of colour. We clearly do not observe such 
an exact degeneracy in Nature. Nevertheless, let us for now 
ignore this  apparent proliferation of degrees of freedom and turn
to the other ingredients of the model. Because the group $\SU{3}$ is
eight-dimensional ($\SU{3}$ has eight generators), we must have 
eight gauge fields, denoted by $V_\m{}^a$. Under $\SU{3}$ the quark
fields transform in the fundamental, triplet representation, viz.
\begin{equation}
  \label{eq:16n.1}
q(x) \rightarrow q^\prime (x) = \exp \left({\ft12} \im \l_a \xi^a (x) \right)
q (x) \; , 
%\eqno(14.a1)
\end{equation}
where $\xi^a (x)$ are the eight transformation parameters of $\SU{3}$,  and
$q(x)$ represents the three-component column vector  $q_\a$
consisting of the three quark colours.  The conjugate quark
fields are represented by the row vector $\bar{q}_\a = (\bar{q}_1, \bar{q}_2,
\bar{q}_3)$ and transform according to
\begin{equation}
  \label{eq:16n.2}
\bar{q}(x) \rightarrow \bar{q}^\prime (x) = \bar{q}(x) \exp \left(-{\ft12}
\im \l_a \xi^a (x) \right) \, . 
\end{equation}
The invariant Lagrangian now takes the  form
\begin{equation}
  \label{eq:16n.3}
L = - \ft14 (G_{\m\n}{}^a)^2 - \bar{q}\, \slashed{D} q - 
 m\, \bar{q} q \,, 
\end{equation}
with
\begin{eqnarray}
  \label{eq:16n.4}
G_{\m\n}{}^a &=& \pa_\m V_\n{}^a - \pa_\n V_\m{}^a - gf_{bc}{}^a\,
V_\m{}^b V_\n{}^c \; , \nonumber \\ 
D_\m q &=& \pa_\m q - \ft12 \im g \,V_\m{}^a \l_a \,q \; .
\end{eqnarray}
The $\SU{3}$ generators $t_a=\ft12 i\l_a$ are expressed in terms
of a standard set of matrices $\l_a$, the Gell-Mann matrices,
which are generalizations of the Pauli matrices $\tau_a$.
The $\SU{3}$ structure constants $f_{bc}{}^a$ follow from the
commutators of these generators. Note that, 
we choose our generators to be anti-hermitian.

For other flavours, the QCD Lagrangian takes the same form as in
(\ref{eq:16n.3}), except that the actual value for the quark-mass
parameter is different. The full Lagrangian thus depends on the QCD
coupling constant $g$ and on the mass parameters $m$, one for each
flavour (quarks of different colour but of the same flavour should
have the same mass in order to conserve the $\SU{3}$ gauge
symmetry). Here we stress that the mass parameter in the Lagrangian
{\it cannot} be identified directly with the constituent mass, 
which should follow from solving the full QCD
field equations.  Obviously, the QCD interactions leave the flavour of
the quarks unchanged, and thus also strangeness and similar
quantum numbers. However,
with the exception of the electric charge, these quantum numbers are
not conserved by the weak interactions, and quarks can
change their flavour by emitting weak interaction bosons.
The gluons do not carry flavour,
but they do carry colour, since they transform under the $\SU{3}$
gauge group. We therefore see that the quark content of the hadrons
can be probed by weak and electromagnetic interactions
through deep-inelastic scattering experiments.

Of course,  quarks also carry spin
indices, as they are normal Dirac spinor fields, so they are
quite rich in indices. One index is the spinor index, which takes
four values. Then there is the colour index, denoted above by $\a,
\b,\ldots$, which takes three values. Finally we can assign a flavour
index, which takes six values corresponding to the different flavours.
As we shall discuss colour further below, let us here explore aspects
of quark flavour. By construction the QCD Lagrangian is invariant
under \emph{local} $\SU{3}$. However, depending on the values for the
mass parameters, there can also be a number of \emph{global} flavour
symmetries. The presence of these flavour symmetries has direct
consequences for the hadronic bound states.  The flavour symmetries
are most relevant for the light quarks.  As the mass parameters of the
$u$ and $d$ quarks are comparable in size, the QCD Lagrangian is
nearly invariant under global unitary rotations of the $u$ and $d$
quarks. These rotations form the group $\U1\otimes \SU{2}$.  The
invariance under $\U1$ is related to the conservation of baryon number
(quarks carry baryon number $\ft13$, antiquarks $-\ft13$). The
$\SU{2}$ transformations mix up and down quarks and are called isospin
transformations.  The breaking of isospin invariance is thus due to
the fact that the $u$ and $d$ mass parameters are not quite equal (an
additional but small breaking is caused by the electroweak
interactions, which we do not consider in this chapter). The $u$ and
$d$ mass parameters are not only nearly equal, they are also very
small, which implies that the Lagrangian has in fact even more
approximate flavour symmetries. To wit, for vanishing quark mass the
Lagrangian is also invariant under unitary transformations of the $u$
and $d$ fields that contain the matrix $\g_5$. Such transformations
are called {\it chiral} transformations. Because of the presence of
$\g_5$, these transformations of the quarks will depend on their
spin. We shall discuss these symmetries further 
below. These extra
transformations involving $\gamma_5$ actually quite subtle because the
chiral symmetry is realized in a so-called spontaneously broken
way. The fact that the pion mass is so small (as compared to the other
hadron masses) can then be explained by an approximate chiral symmetry
in Nature.
Obviously, we may follow the same strategy when including the $s$
quark and consider extensions of the flavour symmetry group.  Apart
from the phase transformations one then encounters an $\SU{3}$ flavour
group (not to be confused with the $\SU{3}$ colour group).  In view of
the fact that the $s$ quark has a much higher mass, flavour $\SU{3}$
is not as good a symmetry as isospin. Symmetry breaking effects are
usually of the order of $10\%$. Of course one may consider further
extensions by including $\g_5$ into the tranformation rules or by
including even heavier quarks. However, these extensions of the
flavour symmetries tend to be less and less useful as they are
affected by the large quark masses and thus no longer correspond to
usefully approximate symmetries of Nature.

In order to realize the
 $\SU{3}$ gauge tranformations on the quark fields, we 
 introduce three varieties of quarks, prosaically denoted by
 colours. However, it seems inevitable that the observed hadrons,
 bound states of quarks and antiquarks, will also exhibit the colour
 degeneracy. For instance, the pions are thought of as bound states of
 a $u$ or a $d$ quark with a $\bar u$ or a $\bar d$ antiquark. Since
 quarks and antiquarks come in three different colours, one has in
 principle {\it nine} types of pions of given electric charge, which
 must have equal mass. Altogether there should then be twenty-seven
 types of pions, rather then the three found in Nature!

 The reason why this colour degeneracy is not observed in Nature is a
 rather subtle one. To explain this phenomenon, let us start by
 considering quarks of a single flavour, say $u$ quarks, and construct
 the possible states consisting of three quarks, all at rest.
 Together they form a state with zero angular momentum. Depending on
 the properties of the forces acting between these quarks, the three
 quarks may or may not cluster into a hadronic bound state.  By
 comparing the properties of these three-quark states to those of the
 low-mass hadrons in Nature (in view of the centrifugal barrier one
 expects that states with nonzero angular momentum acquire higher
 masses) one may hope to unravel the systematics of quark spectroscopy
 and understand the nature of the forces that hold the hadrons
 together.

 Hence, when considering the possibility of the three quarks forming a
 bound state, one may expect the emergence of a spin-$\ft32$ bound
 state and/or one or two spin-$\ft12$ bound states. Of course,
 whether or not they are actually realized as bound states depends on
 the properties of the interquark forces.

 However, the above conclusions are invalidated as we are dealing with
 bound states of {\it identical} spin-$\ft12$ particles. Being
 fermions they satisfy Pauli's exclusion principle, according to which
 the resulting state should be {\it antisymmetric} under the exchange
 of any two such particles. It turns out that the spin-$\ft32$ bound
 state is, however, {\it symmetric} under the interchange of two
 fermions. This is easy to see for the states with $S_z=\pm\ft32$, as
 they correspond to the situation where all three quark spins are
 aligned in the same direction.  Hence a spin-$\ft32$ bound state
 cannot be realized because of Pauli's exclusion principle. However,
 the spin-$\ft12$ states cannot be realized either, as they are
 neither symmetric nor antisymmetric under the interchange of any two
 particles, but are of mixed symmetry (i.e., they can be
 (anti)symmetric under the exchange of two of the quarks, but not with
 respect to the third quark). Therefore, bound states of three
 identical spin-$\ft12$ particles with zero angular momentum cannot
 exist, it would seem.

 Surprisingly enough, when comparing the result of such quark model
 predictions to the low-mass baryons in Nature, one finds that there
 is in fact a bound state of three $u$ quarks with spin-$\ft32$,
 namely the $\D^{++}$ baryon with a mass of 1232~MeV$\,c^{-2}$, which
 is unstable and decays primarily into $p \,\pi^+$ with an average
 lifetime of $0.59 \times 10^{-23}$~s. On the other hand, no
 spin-$\ft12$ bound states of three $u$ quarks are found. At this
 point one could of course question the quark interpretation of the
 $\D^{++}$, were it not for the fact that this phenomenon is
 universal! When comparing the quark model to the data, it turns out
 that the baryons always correspond to bound states of quarks that are
 {\it symmetric} rather than antisymmetric under the interchange of
 two quarks. Therefore, one would conclude, the Pauli principle is
 violated in the simple quark model.

 Before resolving this puzzle, let us once more exhibit this
 phenomenon, but now for the slightly more general case of low-mass
 baryons consisting of $u$ and $d$ quarks. Each quark in the baryon
 now comes in four varieties: a $u$ quark with spin `up' or `down'
 (measured along some direction in space) or a $d$ quark with spin
 `up' or `down'.  Assuming again zero total angular momentum, there
 are thus $4^3=64$ possible spin states, twenty of which are symmetric
 under the interchange of two particles. These symmetric states
 decompose into sixteen states with both isospin and ordinary spin
 equal to $\ft32$, and four states with both isospin and ordinary spin
 equal to $\ft12$. The first sixteen states correspond to the baryons
 $\D^{++}(uuu)$, $\D^{+}(uud)$, $\D^0(udd)$ and $\D^-(ddd)$, which
 carry spin-$\ft32$ so that each one of them appears in four possible
 spin states (we listed the quark content in parentheses).  The latter
 four states correspond to the nucleons $p(uud)$ and $n(udd)$, which
 carry spin-$\ft12$ and thus appear in two varieties.\footnote{As
   explained above, the spin-$\ft12$ states are of mixed
   symmetry. However, the mixed symmetry in terms of the spin indices
   of the quarks can be combined with the mixed symmetry of the
   isospin indices in such a way that the resulting state becomes
   symmetric.}  No other states
 corresponding to bound states of three $u$ or $d$ quarks can be
 identified with baryons in Nature (for higher masses such bound
 states can be found, but those will have nonzero angular momentum).

Let us now stop exploring in detail the subtleties of the simple quark 
model, and turn to quantum chromodynamics. Because the 
quarks carry colour one can make the three-quark state 
antisymmetric by postulating total antisymmetry in the three 
colour indices. In this way the exclusion principle is 
preserved. This conjecture may seem rather ad hoc, and one may wonder 
whether there is an a priori reason for assuming antisymmetry in 
the colour indices. Indeed, it turns out that there is a 
principle behind this. When antisymmetrizing over the colour 
indices of a three-quark state, this state is a singlet under 
the $\SU{3}$ colour group. This follows from the tensor product of
three triplets
\begin{equation}
  \label{eq:16n.6}
{\bf 3} \otimes {\bf 3} \otimes {\bf 3} = {\bf 1} \oplus {\bf 8} 
\oplus{\bf 8} \oplus {\bf 10} \;,
%\eqno(14.b2)
\end{equation}
which yields a singlet state under colour $\SU{3}$ that is fully 
antisymmetric.\footnote{The interpretation of this product rule 
is as before. The $3^3=27$ states formed by all possible products 
of $\SU{3}$ triplet states decompose under the action of $\SU{3}$ 
in four different representations: the singlet representation, 
which is completely antisymmetric, the $\bf 10$ representation, 
which is completely symmetric, and two $\bf 8$ representations, 
which have mixed symmetry (the $\SU{3}$ representations are 
denoted by their dimension, unlike the representions of the 
rotation group, which are denoted by the value of the spin). To 
derive such product rules is more complicated for $\SU{3}$ than 
for the $\SU{2}$, the relevant group for spin and isospin.}  
Assuming that no hadrons carry colour (so that they are {\it  
invariant} under the  
colour gauge group) requires the three-quark states to be 
antisymmetric in the colour indices. By virtue of Pauli's 
exclusion principle, they must therefore be {\it symmetric} with 
respect to all other quantum numbers, such as spin and isospin. 

The principle that hadrons should be colourless can be put to a 
test when considering the low-mass mesons. As we mentioned at the 
beginning of this section, the mesons are bound states of a quark 
and an antiquark. Because of the three-fold degeneracy of the 
quarks associated with colour, each meson should appear in nine 
varieties, which differ in colour, but not in electric charge and 
mass. However, one particular combination of these states is again 
colourless. This follows from the tensor product rule
\begin{equation}
  \label{eq:16n.7}
{\bf 3} \otimes \bar {\bf{3}} = {\bf 1}\oplus {\bf 8}\; , 
%\eqno(14.b3)
\end{equation}
according to which the nine colour states decompose into a 
singlet state and eight states belonging to the octet 
representation. Only the singlet state is realized as a physical 
particle, so that the colour degeneracy is avoided. This
turns out to be a universal feature of all hadrons.  We simply 
never observe the colour degrees of freedom, but only bound 
states of quarks that are singlets of the colour symmetry group.  
In other words if we assign the primary colours to $\a=1,2,3$ 
then the observed hadrons must be ``white''.  Of course, this analogy 
is mostly picturesque and by no means necessary.

\subsection{Parton model}
\label{sec:parton-mode}

%%%%%%%%%%%%%%%%%%%%%%%%%%%%%%%%%%%%%%%%%%%
\begin{figure}
  \begin{center}
    \includegraphics[scale=0.7]{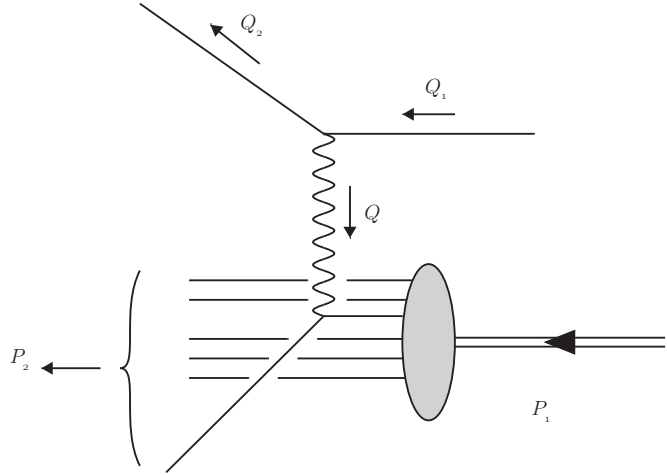}%
\caption{Parton picture of a deep-inelastic collision process. Note
that in the diagram time runs from right to left.}
\label{6.20}
  \end{center}
\end{figure}
%%%%%%%%%%%%%%%%%%%%%%%%%%%%%%%%%%%%%%%%%%%%
We mentioned above that deep-inelastic scattering reveals the
presence of weakly bound point-like parton constituents inside the
nucleon, which we will shortly identify as spin-$\ft12$ fractionally
charged quarks (gluons are neutral with respect to weak and
electromagnetic interactions, so they are not directly involved in
this process). To see this, we first examine a simple model in which
the fast-moving nucleon consists of a finite number of particles, each
carrying a certain fraction of its momentum.  These constituents are
so weakly bound that they may be regarded as free. For simplicity we
assume also that just one parton is subject to the interaction with
the photon that is exchanged in the inelastic process; the others are
neutral and play the role of spectators (see Fig.~\ref{6.20}).  The
charged constituent with momentum $p_\mu = \xi P_{1\mu}$ (we neglect
the transverse parton momenta) and mass $m = \xi M$ $(0 < \xi < 1$)
changes its momentum to $(\xi P_1 +Q)_\mu$ in the interaction with the
virtual boson; the mass-shell condition requires $(\xi P_1 +Q)^2 =
(\xi P_1)^2$ or $2\xi P_1\cdot Q +Q^2 = 0$.  We have then
\begin{equation}
  \label{eq:6n.69}
  \xi = - {{Q^2}\over{2P_1 \cdot Q}} \equiv x \,,
\end{equation}
where we have introduced the Bjorken scaling variable $x$, whose meaning
is clear from (\ref{eq:6n.69}).

\begin{figure}[htbp]
  \begin{center}
 \includegraphics[scale=1.0]{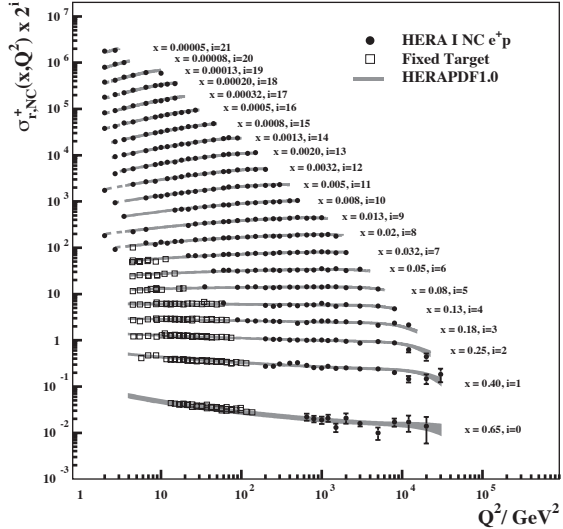}%
 \caption{The reduced cross section (corresponding to $F_2(x,Q^2)$ up to a
   small correction due to weak interaction effects). The data
   have been taken at the HERA collider \cite{Aaron:2009aa}.}
\label{fig:6.12}
  \end{center}
\end{figure}

Defining also the variable $y$ by the fractional energy loss of the
incoming lepton, i.e. in de target restframe by $(E'-E)/E$, one may
write the differential cross section for deep-inelastic scattering (DIS),
mediated by a photon, in terms of dimensionless structure functions as
\begin{eqnarray}
  \label{eq:6.52}
&& \Big( \frac{{\rm d}^2\sigma}{{\rm d} x\, {\rm d} y} \Big)^\gamma =
\frac{8\pi\alpha^2 ME }{(Q^2)^2} \bigg[
\frac{1 + (1-y)^2}{2} \,2 x F_1^\gamma(x,Q^2) \nonumber \\
&& \qquad + (1-y) \big[ F_2^\gamma(x,Q^2) - 2 x F_1^\gamma(x,Q^2) \big]
- \frac{M}{2E}xy \, F_2^\gamma (x,Q^2) \bigg] \,,
\end{eqnarray}
where $M$ is the nucleon mass. The accumulated data for this process,
mostly from the HERA collider at DESY, are displayed in Fig.~\ref{fig:6.12}.
Notice that to first approximation the structure function $F_2(x,Q^2)$ 
only depends on $x$, a phenomenon known as scaling. The
parton model, to which we now turn, provides an explanation for this phenomenon.

After the interaction with the virtual boson has taken place, the charged
constituent will move in a different direction than the spectator particles.
However, during recoil it feels the influence of the binding mechanism, which
forces the constituents to recombine into a new hadronic state, such as an
excited nucleon or a nucleon with one or several pions (on a much longer time
scale than that of the primary collision). Confinement dictates 
that the nature of this final-state interaction is such that the
partons cannot be produced as isolated
particles, and that
the binding force does not interfere with the primary interaction with the
vector bosons.

\begin{figure}[htbp]
  \begin{center}
 \includegraphics[scale=1.0]{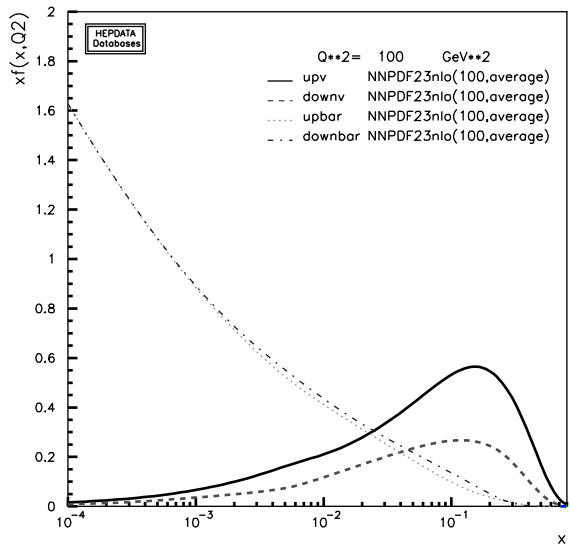}%
\caption{The distributions of $x [u(x)-\bar{u}(x)]$ and
$x [d(x)-\bar{d}(x)]$ (the valence quarks), as well as those of
the anti-up $x \bar{u}(x)$ and anti-down $x \bar{d}(x)$
in the proton. The plots correspond to the NNPDF set, version 2.3 \cite{Ball:2012cx} for a
value of $Q^2 = 100$ GeV$^2$ \cite{hepdata}.  }
\label{fig:6.23}
  \end{center}
\end{figure}

Because the spectators do not participate in the primary interaction the cross
section for inelastic lepton-nucleon scattering is given directly in terms
of the cross section for lepton-parton scattering. Assuming
that the parton is point-like and that the beam energy in the
laboratory frame is large compared
to the masses, one can compute, using the rules of QED
\begin{equation}
\label{eq:6n.70}
\Big({{\rd\sigma}\over{\rd y}}\Big)^\gamma
= {{8\pi \alpha^2 ME}\over{(Q^2)^2}}
 q^2 {{(1-y)^2 +1}\over{2}}\xi  \,.
\end{equation}
where the factor $\xi$ arises
because we have replaced the parton mass $m$ by $M\xi$. Comparing
(\ref{eq:6n.70}) to (\ref{eq:6.52}), we conclude that the contribution from elastic
parton scattering via photon exchange to the nucleon structure functions is given by
\begin{equation}
\label{eq:6n.91}
F^\gamma_2(x) = 2xF_1^\gamma(x) = q^2 x \delta(x-\xi)\,,
\end{equation}
The structure functions thus satisfy the Callan-Gross relation \cite{Callan:1969uq}
\begin{equation}
  \label{eq:6n.73}
F_2(x) = 2xF_1(x)\,,
\end{equation}
which is characteristic for (massless) spin-$\ft12$ partons
Although we have now found structure functions that depend only on
$x$, in agreement with the phenomenon of scaling discussed above,
the model is clearly unrealistic as $x$
remains restricted to a single value $\xi$. 
Therefore, to improve the situation one now assumes that the
nucleon contains many partons interacting with the intermediate photon
and carrying a fraction of the nucleon momentum according to a
probability distribution $f(\xi)$. To be precise, $f_i(\xi)d\xi$
measures the number of partons of type $i$ (e.g. a u-quark or a gluon)
in the momentum range from $\xi P_1$ to ($\xi + \rd\xi)P_1$. As the
nucleon may also contain anti-partons there is a corresponding
distribution $\bar f_i(\xi)$ to measure the number of anti-partons in
the same momentum range. In doing so we will keep ignoring the effect
of transverse parton momenta. Furthermore we assume that the
scattering on the partons is {\it incoherent} (i.e. quantum-mechanical
interference effects between scattering reactions on different partons
are ignored) so that we can simply sum and/or integrate
(\ref{eq:6n.70}) over the various (anti-)parton distributions,
\begin{equation}
  \label{eq:6n.251}
  \Big({{\rd\sigma}\over{\rd y}}\Big)^\gamma
  = {{8\pi \,\alpha^2 ME}\over{(Q^2)^2}}\,
  {{(1-y)^2 +1}\over{2}}\;
  \sum_i  q_i{\!}^2 \int_x^1\,\mathrm{d}\xi \,\xi\, f_i(\xi) \,,
\end{equation}
where the sum is over (anti-)-quark flavours $i$ having fractional
charge $q_i$ (either $\ft23$ or $-\ft13$).  Let us discuss a few more
consequences of the parton model.  Identifying the partons as
quarks\footnote{We include u,d,s and c quarks here as they can be
  treated as massless quarks in most high-energy processes. Bottom and
top quarks, being heavier, are often not treated as partons.}
we can directly derive the following parton model expression for the electromagnetic
structure functions
\begin{align}
  \label{eq:6n.74}
  x^{-1}F_2^\gamma(x)  =&\, 2F_1^\gamma(x) \\
  =&\, \ft49 [u(x) + \bar u(x) + c(x) + \bar c(x)] +
     \ft19 [d(x) + \bar d(x) + s(x) + \bar s(x)]\,. \nonumber
\end{align}
One may now also immediately state he charge sum rule
\begin{align}
  \label{eq:6n.77}
  Q^{\rm nucleon}= \int_0^1 \rd x\, \big[&\ft23 [u(x)-\bar
  u(x)+c(x)-\bar c(x)\big] \nonumber \\
   & -\ft13 [d(x)-\bar d(x)+s(x)-\bar s(x)]\big]  \,,
 \end{align}
which the parton distribution functions must obey.
 Furthermore we note that by interchange of u and d quarks a proton becomes
a neutron and vice versa (this interchange can be realized by a special
isospin transformation). Therefore all neutron quark
distributions follow from those of the proton:
$u(x)^N = d(x)^P, d(x)^N = u(x)^P$, whereas the s- and c-distributions are
equal. Henceforth we will therefore use the notation
where $u(x), d(x), s(x)$ and $c(x)$ refer to the proton
structure functions only. In Fig.~\ref{fig:6.23} we show some examples
of quark distribution functions in the proton.
Unlike  the quarks, the gluons are neutral under weak and electromagnetic
interactions, so they are not directly observed in the deep-inelastic
process. Their presence can, however, already be inferred from
the naive model discussed above, because the total fraction of the nucleon
momentum carried by the quarks (which is given by the areas under the
curves of Fig.~\ref{fig:6.23}) is roughly $\ft12$.
This is an indirect indication that gluons carry the remaining nucleon momentum.

As stated above, if the nucleon is probed at large $Q^2$ the quarks inside will behave as free
point-like objects. The QCD interactions will dissipate the momentum transfer
$Q^2$ to other quarks, and in this process gluons will be radiated which may
again interact with quarks or gluons or annihilate into quark-anti-quark pairs.
This effect becomes more sizeable if the momentum transfer $Q^2$ is shared by
many quarks and gluons, as the average momenta are then smaller so that the
effective QCD coupling grows in strength. The timescale that is relevant for
the final state interaction is therefore much larger than that for the primary
interaction. Incorporating these quark-gluon interactions into the naive quark-parton
model leads in principle to a consistent field-theoretic
set-up for calculating quantum corrections in deep-inelastic scattering,
and other processes. 

We finally remark that the universal nature of the parton distribution functions $f_i(\xi)$
should allow us to apply the parton model also to
the Drell-Yan process, in which a quark and an anti-quark
inside the nucleons collide to form a lepton-anti-lepton pair. This we
shall do further below in these notes.

\subsection{Renormalization and asymptotic freedom}
\label{sec:renorm-asympt-freed}

The examination of the quantum corrections in 
a theory can provide crucial insight into the structure of the
theory, and its consistency. For example, if they break
the gauge symmetries of a theory (so that these symmetries are anomalous), 
the theory can be inconsistent. It can also 
teach us about the predictive power of the theory. If the 
higher order corrections for some observable are so large that the very concept
of perturbation theory for this case becomes doubtful,
we have a crisis of the theory's predictive power for this observable.
Higher order corrections may contain ultraviolet divergences (we will
discuss other divergences later). Here we discuss how 
one may handle them and account for them.

\subsubsection{Regularization}
\label{sec:regularization}

In order to handle divergences one must first regularize the quantum
field theory in such a way that the infinities become temporarily 
finite (would-be infinities). If done consistently,
one can apply the renormalization procedure, upon which for appropiate quantities
the would-be infinities cancel, so that the regularization can be removed. 
A number of regularization have been invented in the past, let us review some
of them.\\[1.5ex]
\noindent \emph{Cut-off}\\[1ex]
In this method one imposes a uniform upper limit $\Lambda$ on the loop momenta
\begin{equation}
  \label{eq:151}
  \int^\Lambda \frac{dq}{q} + \ldots 
 = \ln \Lambda + \mathrm{finite\; terms}
\end{equation}
The would-be infinity is represented here by $\ln \Lambda$. When all would-be infinities
have cancelled and only $1/\Lambda^p$ terms are left, one can remove the regulator by
$\Lambda \rightarrow \infty$. The advantage of this method is that it is very intuitive,
the (serious) disadvantage is that it is very cumbersome in higher orders, in particular
for gauge theories. It is therefore mostly used in high energy physics
for didactical purposes.\\[1.5ex]
\noindent  \emph{Lattice}\\[1ex]
In this method one discretizes spacetime, and defines fields to live only on the lattice
points (or on the links between them). In this way momenta cannot be larger than $1/a$
where $a$ is the lattice spacing. A major advantage of this method is that it can
actually be used for computer simulation, so that the full path integral can be evaluated,
without need to expand it in perturbation theory. Among the drawbacks are difficulties
in maintaining continuum symmetries on the lattice. It is however a widely used method,
mostly for lower energy observables, such as hadron masses and decay constants.\\[1.5ex]
\noindent \emph{Dimensional regularization}\\[1ex]
This is the regularization that is most powerful in perturbative quantum
field theory, and therefore also most widely used. It consists of the temporary extension
of the number of dimensions in spacetime, or conversely, momentum space, from $4$
to $4+\varepsilon$
\begin{equation}
  \label{eq:152}
  \int d^4 x \mathcal{L}(x) \rightarrow \int d^{4+\varepsilon} \mathcal{L}(x)
\end{equation}
How does this method regularize ultraviolet diverences\footnote{A more 
careful treatment of dimensional regularization, including the conditions
on the complex parameter $\varepsilon$ can be found in \cite{DeWit:1986it}.}?
A careful dimensional analysis 
shows that (i) momentum space propagators continue to look like $1/(q^2+m^2)$, and
(ii) gauge couplings now get dimension $-\varepsilon/2$. Then a one-loop integral
is extended as follows
\begin{equation}
  \label{eq:153}
  \int d^4q \frac{1}{q^4} \rightarrow \int d^{4+\varepsilon}q \frac{1}{q^4}
\end{equation}
In $n=4+\varepsilon$ dimensional polar coordinates this may be written as
(introducing a lower limit $Q$ on the $q$ integral)
\begin{equation}
  \label{eq:154}
  \int d\Omega_{3+\varepsilon} \int_Q^\infty dq q^{3+\varepsilon} \frac{1}{q^4}
\end{equation}
These $d$ dimensional integrals can be carried out to yield
\begin{equation}
  \label{eq:155}
  \frac{2\pi^{2+\varepsilon/2}}{\Gamma(2+\varepsilon/2)} \, \frac{-1}{\varepsilon}Q^\varepsilon
\end{equation}
The Euler gamma function $\Gamma(2+\varepsilon/2)$ makes a frequent appearance in this regularization method.
The would-be infinity is $1/\varepsilon$. Removing the regulator would correspond to
taking the limit $\varepsilon \rightarrow 0$. 

\subsubsection{Renormalization}
\label{sec:renormalization}

Now that we know how to regularize a quantum field theory we are ready to understand 
conceptually the renormalization procedure. At its heart is the question how to have
a predictive theory when higher-order corrections 
contribute an infinite amount to various Green functions.

\begin{figure}[htbp]
  \centering
  \includegraphics[width=0.5\textwidth]{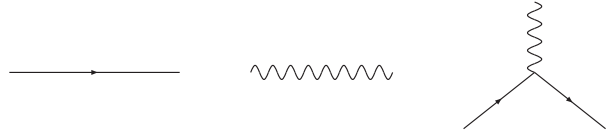}    
  \caption{Lowest order Green's functions provided by lagrangian }
  \label{fig:QEDGreenfunction_lo}
\end{figure}

\begin{figure}[htbp]
  \centering
    \includegraphics[width=0.6\textwidth]{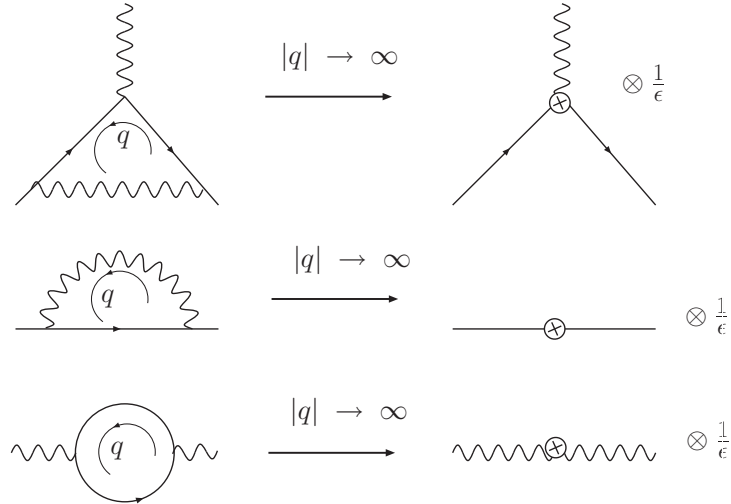}    
  \caption{One-loop corrections to lowest order Green's functions and their UV limit.}
  \label{fig:QEDGreenfunction_ho}
\end{figure}

\begin{figure}[htbp]
  \centering
      \includegraphics[width=0.5\textwidth]{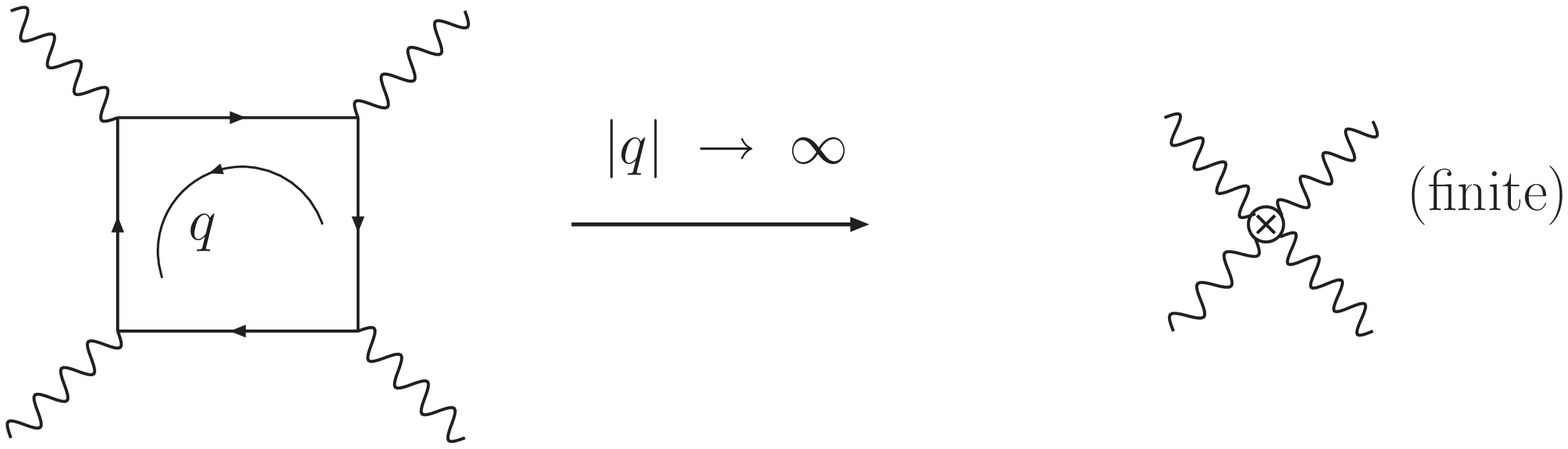}    
  \caption{UV limit of QED box graph}
  \label{fig:QEDbox_ho}
\end{figure}

Let us first form a physical picture for the case of the QED
lagrangian
\begin{equation}
  \label{eq:26}
  {\cal L}_{\mathrm{QED}} =
-\ft14\big(\partial _\mu A_\nu - \partial _\nu
  A_\mu\big)^2 - \bar \psi \,\slash{\partial} \psi - m\,\bar \psi\psi +
  \im e\, A_\mu \bar \psi\gamma^\mu\psi\,,
\end{equation}
with $A_\mu$ representation the photon field,  $\psi$ the electron
field, and $e$ the electric charge.
For QCD the conceptual points are the same, if a bit more complicated.
At lowest order, and after gauge-fixing, the lagrangian provides an electron 2-point
function (leading to the electron propagator), a photon 2-point function
(leading to the photon propagator), and a photon-electron 3-point function
(leading to the QED interaction vertex), see Fig.~\ref{fig:QEDGreenfunction_lo}.
Let us now look at some of their one-loop correction when the loop-momentum
$q$ becomes very large. In Fig.~\ref{fig:QEDGreenfunction_ho} we indicate
how these corrections may be viewed. 
Because the loop momentum becomes so large, the loop reduces to a very local 
effect, of would-be infinite strength. It should be noticed that the result
is simply a would-be infinite coefficient times the lowest order Green function.
This is an important result. For example in Fig.~\ref{fig:QEDbox_ho} we see that
the UV limit of the box graph, while leading to local 4-photon vertex which does
\emph{not} occur in the lagrangian, is also \emph{not} would-be infinite. 
This suggests that we can absorb in this case the $1/\varepsilon$'s into the couplings
and field normalizations of the lagrangian, without need to introduced new types
of interactions.
Quantum Electrodynamics is in fact a renormalizable \index{renormalizable} theory. This 
means that it is sufficient to renormalize $e,m,\psi,A_\mu$ to absorb/cancel all 
would-be infinities for any Green function in QED. Let us see how this absorption
works, using dimensional regularization.

First, let us recall that in dimensional regularization the dimension of the gauge
coupling is no longer zero, but rather 
\begin{equation}
  \label{eq:156}
  [e] = \frac{2-d}{2} = -\frac{\varepsilon}{2}
\end{equation}
To keep count of such dimensionalities, and to be able to define a dimensionless
coupling, we introduce a mass scale $\mu$, whose value is intrinsically
arbitrary, such that
\begin{equation}
  \label{eq:157}
  e  = e(\mu) \mu^{-\varepsilon/2}
\end{equation}
We now \emph{renormalize} $e$ by a factor $Z_e$ that contains would-be infinities
\begin{equation}
  \label{eq:158}
  e = e_R(\mu) \mu^{-\varepsilon/2} Z_e\left( \frac{1}{\varepsilon}, e_R(\mu) \right) =
\left( 
1 + e_R(\mu)^2 \frac{1}{\varepsilon} z_e^{(1)} +
 e_R(\mu)^4 \left[\frac{1}{\varepsilon^2} z_e^{(2)}+\frac{1}{\varepsilon} z_e^{(1,1)} \right]
+\ldots
\right)
\end{equation}
The renormalized coupling \index{renormalized coupling} $e_R(\mu)$ is finite, and is
can be directly related to an actual physical quantity like the fine-structure constant.
How this works when what is supposed to be a number actually depends on $\mu$ we will
see below.

We have not yet specified the constants $z_e^{(1)},  z_e^{(2)},  z_e^{(1,1)}$ etc. Let us
now consider an observable $O$ which we have computed to 1-loop, using the QED Feynman rules
\begin{equation}
  \label{eq:159}
  O = e\,C  + e^3\left[ A \frac{1}{\varepsilon} (Q^2)^{-\varepsilon/2} + B \right]
\end{equation}
where $Q$ is the typical energy scale of the observable.
We now renormalize the coupling according to \eqref{eq:158} and obtain, to order $e_R(\mu)^3$
and obtain
\begin{equation}
  \label{eq:160}
  O = \mu^{-\varepsilon/2}\Big\{
e_R(\mu) \, C + e_R(\mu)^3\left[ 
 A \frac{1}{\varepsilon} \left(\frac{Q^2}{\mu^2}\right)^{-\varepsilon/2}
+ C  \frac{1}{\varepsilon} z_e^{(1)} + B
\right]
\Big\}
\end{equation}
We can now choose $z_e^{(1)} = -A/C$. Then the poles in $\varepsilon$ will cancel,
and we can expand the result in $\varepsilon$
\begin{equation}
  \label{eq:161}
   O = \Big\{
e_R(\mu) \, C + e_R(\mu)^3\left[ 
 A \ln\left(\frac{Q^2}{\mu^2}\right) + B
\right]\,.
\Big\}
\end{equation}
One might think that it is not so hard to cancel divergences if one can simply choose
to do so by picking $z_e^{(1)} = -A/C$. The remarkable fact, and the
essence of the renormalizability of a theory, is however that 
this same choice works for all cases. One would always, for QED, find 
the same answer for $z_e^{(1)}$. Similarly for $Z_\psi\left( \frac{1}{\varepsilon}, e_R(\mu) \right), 
Z_\alpha\left( \frac{1}{\varepsilon}, e_R(\mu) \right), Z_m\left( \frac{1}{\varepsilon}, e_R(\mu) \right)$. 
To find their coefficients in the $e_R(\mu)$ expansion one can take some relatively simple
observables, and compute them once and for all. 

Based on the example just discussed it should not be too great a surprise to learn
that the generic structure of the observable, after renormalization, is
\begin{align}
  \label{eq:162}
  O(Q,\mu)  = &  e_R(\mu)^2 \left[O_1 \right]  + \\
          &  e_R(\mu)^4 \left[O_{10} + O_{11} \ln\left(\frac{Q^2}{\mu^2}\right) \right]  + \\
          &  e_R(\mu)^6 \left[O_{20} + O_{21} \ln\left(\frac{Q^2}{\mu^2}\right)   + 
O_{20} \ln^2\left(\frac{Q^2}{\mu^2}\right) \right]  + \ldots 
\end{align}
where the $O_{ij}$ are various constants. We note that 
(i) $O(Q,\mu)$ is finite,  and (ii) it depends on the determined scale $\mu$ both directly, via the logaritms, 
 and implicitly, via the renormalized coupling $e_R(\mu)$.
The last point is problematic: if we have consistently cancelled the divergences
only for $O$ to depend on an arbitrary scale it seems we have not gained much
predictive power. However, the $\mu$ dependence is precisely such that for
Eq.~\eqref{eq:162}
\begin{equation}
  \label{eq:163}
  \mu \frac{d}{d\mu} O(Q,\mu) = \mathcal{O}(e_R^8(\mu))
\end{equation}
i.e. one order beyond the one calculated. Should one add another order to the
result in \eqref{eq:162} the residual dependence on $\mu$ would be $\mathcal{O}(e_R^{10}(\mu))$
and therefore progressively less, and the prediction ever more precise. Some
uncertainty will however remain, and it is customary to estimate it by varying
$\mu/Q$ from $2$ to $1/2$.

\subsubsection{Running coupling, $\beta$ function}
\label{sec:runn-coupl-beta}

What is the origin of this conspirational $\mu$ dependence? It is in fact
the renormalization procedure itself. In the problem sets it was shown that from
the relation~\eqref{eq:158} one can derive (by acting with $d/d\ln \mu$ on both
side) a first order differential equation for the $\mu$ dependence of the finite renormalized
coupling
\begin{equation}
  \label{eq:187}
  \mu \frac{d}{d\mu}e_R(\mu) = \beta_0 e_R(\mu)^3 + \beta_1 e_R(\mu)^5 + \ldots
 \equiv \beta(e_R(\mu))
\end{equation}
known as the $\beta$ function equation, or sometimes also as the renormalization
group equation \index{Beta function}\index{$\beta$ function}\index{renormalization group equation}
for the running coupling.

The $\beta$-function equation plays an important role in the Standard Model. It should
be clear that its occurrence is generic. Because each coupling in the 
Standard Model requires renormalization, each will have its own $\beta$-function. 
The $\beta$ functions are only known in form of a perturbative expansion, as in 
Eq.~\eqref{eq:187}. For non-abelian gauge theory no less than the first four terms
are known (five for the case of $\SU{3}$ \cite{Baikov:2016tgj}!). The
first term
\begin{equation}
  \label{eq:188}
  \beta_0 = -\frac{11 C_A - 4 T_F N_F}{12 \pi}
\end{equation}
was calculated in the early 70's. The 2004 Nobel Prize was awarded for this calculation,
in particular for the interpretation for the fact that the term is \emph{negative}, about
which more below. From eq.~(\ref{eq:187}) we can already see that if the function has
negative coefficents, as non-abelian gauge theories such as QCD do, the coupling decreases
for a positive increment in the scale $\mu$, i.e. when $\mu \rightarrow \mu+d\mu$, leading to 
asymptotic freedom in the ultraviolet, and strong binding at low scales $\mu$. 

\subsubsection{Symmetries of QCD}
\label{sec:symmetries-qcd}

Before diving further into the perturbative aspects of QCD, let us devote now a
bit of space to considering the fundamental symmetries of QCD. We discussed some
of this already qualitatively in section \ref{sec:spectr-symm-qcd}, here we 
discuss these from a field-theoretical point of view. The defining symmetry 
of QCD is the local $\SU{3}$ symmetry, under which the quark transform as
\begin{equation}
  \label{eq:12.1}
\psi(x)\rightarrow\psi^\prime(x) = U(x) \,\psi(x),   \qquad
U=\exp(\xi^a t_a),  
\end{equation}
where the matrices $t_a$ are called the {\it generators} of the 
group defined in the representation appropriate to $\psi$, and 
the $\xi^a$ constitute a set of linearly independent 
{\it real} parameters in terms of which the group elements can be 
described. The number of generators is obviously equal to 
the number  of independent parameters $\xi^a$ and therefore to 
the dimension of the group, but is not necessarily related to the dimension of the 
matrices $U$ and $t_a$\footnote{For instance, for the three $\SU{2}$ 
generators one can choose 2,3,... dimensional matrices, corresponding
to (iso)spin $\ft12, 1,\dots$. }. Hence, $U$ is a square matrix whose
dimension is equal to the number of components in $\psi$ (3 
in the case of QCD). 
The covariant derivative should be such that when acting on 
a field that already transforms covariantly, the result will transform
covariantly also
\begin{equation}
  \label{eq:12.7}
D_\mu\psi(x)\rightarrow (D_\mu\psi(x))^\prime=U(x) 
\, D_\mu\psi(x).
\end{equation}
To this end one introduces a (set of) gauge field(s) 
\begin{equation}
  \label{eq:12.8}
D_\mu\psi \equiv \partial_\mu\psi - W_\mu \,\psi,  \qquad W_\mu = W_\mu^a \,t_a \,,
\end{equation}
so that also $W_\mu$ is matrices, and the number of gauge fields
equals the number of generators (8 in the case of $\SU{3}$). With the
property (\ref{eq:12.7}) it is easy to construct non-abelian gauge
theory. The rule (\ref{eq:12.7}) holds if $W_\mu$ transforms as
\begin{equation}
  \label{eq:12.11}
W_\mu\rightarrow W_\mu^\prime = UW_\mu U^{-1} +
 (\partial_\mu U)U^{-1}\,,
\end{equation}
i.e. inhomogeneously (the second term does not contain $W_\mu$), and
non-covariantly (the second term depends on the derivative of $U(x)$).
With the covariant derivative one can also construct the field strength tensor 
\begin{equation}
  \label{eq:12.24}
G_{\mu\nu}= - [D_\mu , D_\nu] =  \partial_\mu W_\nu -\partial_\nu W_\mu -[W_\mu,W_\nu],
\end{equation}
The field strength transforms covariantly  and homogenously
\begin{equation}
  \label{eq:12.25}
G_{\mu\nu}\rightarrow G_{\mu\nu}^\prime = UG_{\mu\nu}U^{-1}\,.
\end{equation}
Finally, the QCD coupling constant can be introduced by replacing
\begin{equation}
  \label{eq:12.35}
W^a_\mu\to g\,W_\mu^a\,,\qquad G_{\mu\nu}^a\to g\,G_{\mu\nu}^a\,.
\end{equation}
With this we can write down the QCD Lagrangian for one quark flavour, 
with mass $m$
\begin{align}
  \label{eq:3}
    {\cal L}=\,&{\cal L}_W +{\cal L}_\psi\nonumber \\
  =&\,{\textstyle{1\over4}}{\rm Tr}\,
  [G_{\mu\nu}\,G^{\mu\nu}]-\bar\psi\,\Slash{D} \psi -m\,\bar\psi\,\psi\, . 
\end{align}
Thanks to the
rules in (\ref{eq:12.1},\ref{eq:12.7}) and (\ref{eq:12.25}), it is
straightforward to check its local $\SU{3}$ invariance

We now turn to a global symmetry of the QCD Lagrangian, that is
relevant because there is more than one quark flavour.
The full QCD Lagrangian reads 
\begin{equation}
  \label{eq:4}
   {\cal L}_{\mathrm{QCD}} = -\frac{1}{4}\mathrm{Tr}(G_{\mu\nu}G^{\mu\nu})
 -\sum_{f=1}^{n_f} \overline{\psi_f}(\slashed{D} + m_f) \psi_f\,.
\end{equation}
Besides having local symmetry, this Lagrangian has an interesting
global symmetry if the masses of the quarks may be neglected. 
We can use the chiral projector $P_L = (1+\gamma_5)/2$ and 
$P_R = (1-\gamma_5)/2$ (it is easy to check that they are idempotent,
that $P_LP_R=P_RP_L=0$ and that they and sum to 1) to define left- and
righthanded quarks
\begin{equation}
  \label{eq:5}
  \psi_L = P_L \psi, \qquad   \psi_R = P_R \psi\,. 
\end{equation}
In these terms, the fermion sector of (\ref{eq:4}) reads
\begin{equation}
  \label{eq:6}
  \sum_{f=1}^{n_f} \overline{\psi_f}(\slashed{D} + m_f) \psi_f
=   \sum_{f=1}^{n_f} (\overline{\psi_{L,f}}\slashed{D}  \psi_{L,f}
+ \overline{\psi_{R,f}}\slashed{D}  \psi_{R,f}) 
  \sum_{f=1}^{n_f} m_f (\overline{\psi_{L,f}}\psi_{R,f} + \overline{\psi_{R,f}}\psi_{L,f})\,.
\end{equation}
When setting $m_f=0$ one notes that the left- and righthanded quarks
have no interactions. We may in fact mix them independently 
\begin{equation}
\label{eq:24}
  \psi_{L,i} \rightarrow   \psi'_{L,i} = U_{L,ij} \psi_{L,j} , \qquad
  \psi_{R,i} \rightarrow   \psi'_{R,i} = U_{R,ij} \psi_{R,j} \,,
\end{equation}
with $U_L$ and $U_R$ independent unitarity matrices. The dimension of
these matrices is equal to the number of quark flavours that is
(approximately) massless. This is a good a approximation for the
up and down quarks, in general still reasonable for the strange
quark, but for the heavy quarks not anymore. Note that the chiral
symmetry $U_L\otimes U_R$ is a global symmetry, we do not make an effort
to make this symmetry local. It relates many properties of 
pions and kaons. 

But this symmetry becomes especially interesting when one accounts for the
fact that the QCD nonperturbative vacuum should have the structure
\begin{equation}
  \label{eq:8}
\sum_f  \langle  \overline{\psi_{L,f}} \psi_{R,f} \rangle + (L
\leftrightarrow R)\,.
\end{equation}
In words, in the QCD groundstate left- and righthanded projections
of quark flavours are coupled, so that this chiral symmetry is
spontaneously broken. By Goldstone's theorem, the spectrum of QCD
(the set of actually realized particles) should feature massless
spinless bosons. They must however be odd under parity, so that they
are in fact pseudoscalar bosons.
The obvious candidates for these would be the pions and kaons of the
pseudoscalar meson octet.  The reason is that the groundstate
(\ref{eq:8}) is still invariant when choosing $U_L=U_R$ (so-called
vector rotations), but when transforming left- and righthanded quarks
differently, a $\gamma_5$ remains, which implies that the goldstone
bosons behaves as $\overline{\psi}\gamma_5 \psi$, i.e. as pseudoscalar
mesons.

Though it still an unsolved problem how to compute the
non-perturbative QCD spectrum fully analytically from the 
Lagrangian (\ref{eq:4}), one may set up an effective theory, 
Chiral Perturbation Theory ($\chi$PT),  for
pions (and kaons) valid for low energy scattering. However, in these
notes we shall not go further into this interesting subject.

\subsection{Evidence for colour}
\label{sec:evidence-colour}

Because the QCD colour quantum number is so central to its understanding
and functioning, it would be interesting to verify it. This is not 
straightforward, as we discussed, since colour is confined (hadrons are ``white''), 
so that its existence can only be inferred. 
Let's see how this might be done. Consider the total cross section for the production of a 
fermion-antifermion pair $f\bar{f}$ in an electron-positron collision, to 
lowest order in the electromagnetic coupling. The fermion has e.m. charge $Q_f e$
and mass $m$, and we approximate the electron to be massless. The
answer is in fact quite simple
\begin{equation}
  \label{eq:142}
  \sigma_{f}(s) 
= \frac{4\pi \alpha^2 Q_f^2}{3s}\beta \left(1+\frac{2m_f^2}{s}\right)
\theta(s-4m_f^2)
\end{equation}
where $s$ is the center-of-mass energy squared. Note that we have attached a
label $f$ to the mass the type of fermion $f$.
The factor involving the electric charges 
also depends on the fermion ``flavour''. Thus, for
an electron, muon and tau $Q_f = -1$,  for up, charm, and top quarks 
$Q_f = 2/3$, while for down, strange and bottom quarks $Q_f = -1/3$. 
The factor $\beta = \sqrt{1-4m_f^2/s}$ is a phase space volume factor; when $s$ is just
a little bit larger than $4m^2$ $\beta$ is close to zero, i.e. near threshold the
cross section is small. Far above threshold $\beta \sim 1$. The theta function
ensures that the cross section is only non-zero is the center of
mass energy is large enough to produce the quark pair.

\begin{figure}[htbp]
  \centering
          \includegraphics[width=0.5\textwidth]{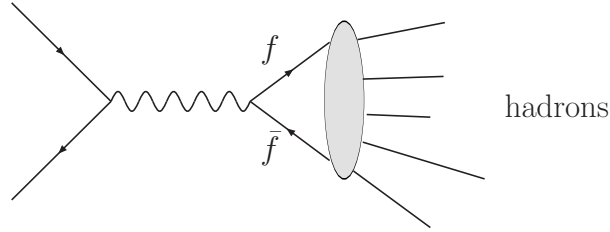}
  \caption{$e^+e^- \rightarrow$ hadrons; the blob represents the "hadronization" process. Note that
time runs from left to right in this diagram.}
  \label{fig:eehad}
\end{figure}

\begin{figure}[htbp]
  \centering
            \includegraphics[width=0.5\textwidth]{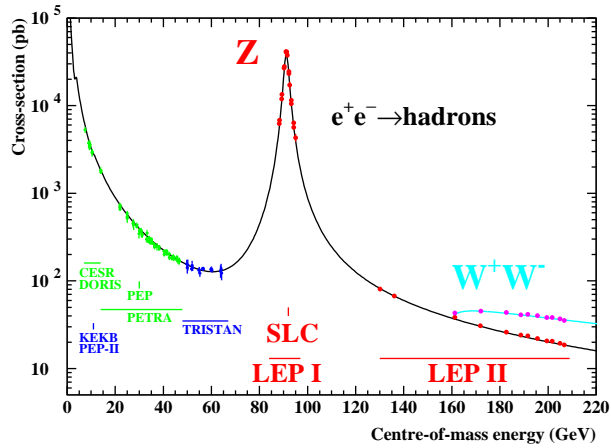}
  \caption{Total cross section for $e^+ e^-$ to fermions.}
  \label{fig:eeff}
\end{figure}

\begin{figure}[htbp]
  \centering
        \includegraphics[width=0.5\textwidth]{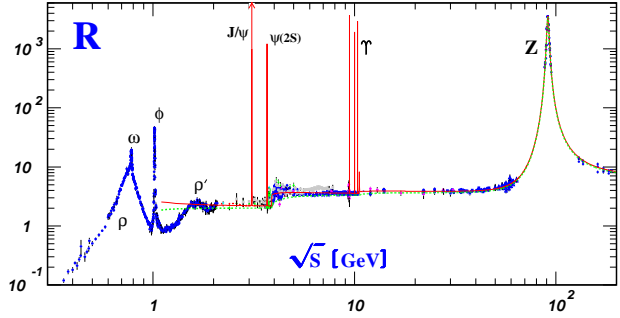}
  \caption{R-ratio vs. center of mass energy}
  \label{fig:eehad-r}
\end{figure}

How might we use this result? If the produced fermions are electrons, muons or taus
we can directly confront the result with data, and agreement is in fact very good.
There is a more interesting use of the formula in Eq.~\eqref{eq:142}. Consider
the inclusive \emph{quark} cross section
\begin{equation}
  \label{eq:144}
  \sigma_{had}(s) 
= \sum_{f=u,d,s,c,\ldots}
\frac{4\pi \alpha^2 Q_f^2}{3s}\beta \left(1+\frac{2m_f^2}{s}\right) 
\theta(s-4m_f^2) N_c
\end{equation}
The extra factor $N_c$ at the end accounts for the fact that quarks come in
$N_c=3$ colours. 
We may interpret this in fact as a prediction for the inclusive \emph{hadron}
cross section, because the quark final state must, before they reach any detector, 
make a transition to a hadronic final state, see the illustration in
Fig.~\ref{fig:eehad}.
In Fig.~\ref{fig:eeff} we see the confrontation of this result with data, and that
the agreement is very good, except that we did not anticipate the huge peak
near $\sqrt{s} \simeq 90 \mathrm{GeV}$. 
That is because we did not include in our calculation
of $\sigma_f(s)$ in eq.~(\ref{eq:142}) a second diagram in which not a photon (as in Fig.~\ref{fig:eehad}) but a $Z$-boson of mass $M_Z \simeq 90 \mathrm{GeV}$ mediates the scattering. Had we done so, we would have more 
terms in final answer for $\sigma(s)$ in Eq.~\eqref{eq:142}, with the factor $1/s$ 
replaced $1/(s-M_Z^2+\Gamma_Z^2)$, where $\Gamma_Z$ is the $Z$-boson decay width (about $2$ GeV).
The good agreement also implies that the effect of higher order corrections 
to $\sigma(s)$ should be small, and indeed they turn out
to be so, after calculation.
We can now define an observable traditionally called $R$
\begin{equation}
  \label{eq:123}
    R(s) = \frac{\sigma(e^+e^- \rightarrow \;\mathrm{hadrons})}{\sigma(e^+e^- \rightarrow \mu^+\mu^-)}
\end{equation}
The benefit of defining such a ratio is that a many common factors cancel
in the theoretical prediction, and that many experimental uncertainties cancel
in the experimental measurement. We have then
\begin{equation}
  \label{eq:122}
  R(s) = \frac{\sum_{f=u,d,s,c,\dots }\sigma(e^+e^- \rightarrow f\bar{f})}{\sigma(e^+e^- \rightarrow \mu^+ \mu^-)}
\end{equation}
For large center-of-mass energy $\sqrt{s}$ we can derive from \eqref{eq:144}
that
\begin{equation}
  \label{eq:124}
  R(s) \stackrel{s\rightarrow \infty}{\longrightarrow}
  N_c \sum_{f=u,d,s,c,\dots} Q^2_f \theta(s-4m_f^2)
\end{equation}
In Fig.~\ref{fig:eehad-r} we confront this result with experiment. 
We can draw the
conclusions that (i) there is again fairly good agreement between prediction and
measurement; (ii) we see the effects of new quark flavour $f$ being ``turned on'' 
as the energy increases beyond $2m_f$ ($m_c \simeq 1.5$ GeV,
$m_b \simeq 5 $ GeV);
(iii) the larger step at charm than at bottom (proportional 
 to $Q_c^2 = 4/9$ and $Q_b^2 = 1/9$, respectively) is 
well-predicted;
(iv) the value of $R(s)$, say beyond the bottom quark
threshold 
\begin{equation}
  \label{eq:143}
  R(s) =   N_c \sum_{f=u,d,s,c,b} Q^2_f \theta(s-4m_f^2)
 = 3 \left(\tfrac{4}{9}+ \tfrac{1}{9}+\tfrac{1}{9}+\tfrac{4}{9}+\tfrac{1}{9} \right)
= \tfrac{11}{3}
\end{equation}
agrees with experiment, and indicated that quarks come indeed in 3 colours.

\section{Higher orders}
\label{sec:higher-orders}

In this section we discuss a number of key aspects relevant for computing
higher-order effects in QCD. These are crucial to present-day applications of QCD
for collider physics, so we provide a fair amount of detail.

\subsection{Parton distribution functions (PDFs)}
\label{sec:part-distr-funct}

Before discussing how to compute higher-order partonic cross sections, let us
discuss the quantities that form the interface of these to the hadronic cross
sections: parton distribution functions. A recent, excellent review can be found
in \cite{Forte:2013wc}. We already encountered the PDFs
in section \ref{sec:parton-mode} in the context of the parton model,
where they were functions of the momentum fraction variable (``$x$'')
only. However, in the context of higher-order calculations they
play a central role in the cancellation of initial state collinear
divergences, and in that process also acquire (logarithmic)
factorization scale dependence. It should be clear that, being the
interface between hadronic and partonic cross sections, they play a
crucial role at the hadron colliders such as the HERA, Tevatron and
LHC, and the quality of theoretical predictions is directly tied to
knowing the PDFs well. Thus, we need to understand how to determine
the PDFs and their uncertainties. To be precise, 11 of them: 5 quark,
5 anti-quark and 1 gluon PDF, which we denote by
$\phi_{i/P}(\xi,\mu_F)$, the number of partons of type $i$ in the proton
with momentum fraction $\xi$, at factorization scale $\mu_F$. 

Key to this determination is their universality: the QCD factorization theorems
\cite{Collins:1989gx} ensure that it is same set PDFs that occurs in 
all well-defined partonic cross sections to any fixed
order. Therefore, one may choose (with care) a set of observables
(e.g.  DIS structure functions, certain hadron collider cross
sections) to infer the PDFs from. Since each is described as 
a combination of PDFs and partonic cross sections we have the set of
equations
\begin{equation}
  \label{eq:9}
  \left[O_n + \Delta O_n\right]^{\mathrm{exp}} = \sum_{i=1}^{11} 
\phi_{i/P} \otimes \left[\hat\sigma_{n,i} \pm \delta \hat\sigma_{n,i}  \right]^{\mathrm{theory}}\,,
\end{equation}
where also the experimental and theoretical uncertainties are
indicated. From this set of equations the PDFs may be inferred. 
Notice that if the calculations on the rhs are all of order $\mathrm{N^kLO}$,
then the PDFs inferred are also labelled $\mathrm{N^kLO}$, even though there
are intrinsically non-perturbative functions. The $\mathrm{N^kLO}$ PDFs can
then be consistently used for other $\mathrm{N^kLO}$ (and of course
also for lower $k$) calculations.

The determination of the PDFs is not a trivial matter, and is
performed by various groups, each taking different approachs. The
groups are known by acronyms of various lengths: MSTW, CTEQ, NNPDF,
GJR, HERAPDF, ABKM, etc. Below we shall discuss briefly some features
and results of some of these approaches. 

First, a brief aside on the formal aspects of a PDF. Although they
cannot yet be computed from first principles, it is possible to give a
precise definition of PDFs in terms of operators. In essence, it is the
expectation value of a parton counting operator (think of $a^\dagger a$ for a harmonic
oscillator in quantum mechanices) in a proton state. For the quark case it is
\begin{equation}
  \label{eq:10}
  \phi_{q/P}(\xi,\mu)
= \frac{1}{4\pi} \int_{-\infty}^{+\infty}dy^- e^{-\im p^+ y^-}
\langle p|\bar{q}(0,y^-,0_T) \gamma^+ q(0,0,0_T)  |p \rangle\,.
\end{equation}
We have introduced here also lightcone notation for 4-vectors
\begin{equation}
  \label{eq:11}
  p^\pm = \frac{p^0\pm p^3}{\sqrt{2}}, \qquad   p\cdot q = -p^+ q^- - p^- q^+ + p_T\cdot q_T\,.
\end{equation}
so that $\gamma^+ = (\gamma^0+\gamma^3)/\sqrt{2}$ in \eqref{eq:10}.
The benefit of having a definition such as (\ref{eq:10}) is that one can compute
now higher-order corrections to the operator, renormalize it, and then have 
a renormalization group equation for it. This is in fact then precisely the
DGLAP equation. Note this can all be done in QCD perturbation theory. For the purposes
of such  calculations one can replace the proton states with parton states. 
Of course, the non-perturbative aspect comes in when evaluating the operator in a proton state. 
The DGLAP evolution equation reads
\begin{equation}
  \label{eq:12}
  \mu \frac{d}{d\mu}  \phi_{i/P}(\xi,\mu)
 = \sum_j\int_\xi^1 \frac{dz}{z} P_{ij}(z,\alpha_s(\mu)) 
\phi_{j/P}\left(\frac{\xi}{z},\mu\right)\,,
\end{equation}
where the $P_{ij}$ are the Altarelli-Parisi splitting functions, which act here as
evolution kernels. With evolution is meant the change in form of the function as 
the energy scale $\mu$ evolves. They are now known to NNLO (3-loop) \cite{Moch:2004pa,Vogt:2004mw}.
The logic is thus not unlike that of the running coupling, but now we have ``running functions''.

Returning now to how to extract the actual functional form of the PDFs from the equation (\ref{eq:9}),
we see how the DGLAP equation is very useful. The data, on the lhs of (\ref{eq:9}), are 
taken at various energy scales. The theoretical description can for each observable be 
computed at the same energy scale because the scale evolution of the PDFs is known, so 
that meaningful comparison can be made.

The selection of observables to be used in eq. (\ref{eq:9}) must be done with care. 
Different observables should be sensitive in different ways to the various PDFs, so
that a reliable extraction of a PDF is possible for each parton type. For instance, 
in DIS the most important partonic subprocess is $\gamma^* q \rightarrow q + X$ (where the
off-shell photon is exchanged with the initial electron), so that associated observables
are particularly sensitive to light quark PDFs. A nice overview of the main processes
involved can be found in table \ref{tab:processes}, 
taken from Reference~\cite{Martin:2009iq},
which lists the processes that are included in a typical present-day
global fit (MSTW08), and the PDFs they constrain.
\begin{table}%
  \def~{\hphantom{0}}
  \begin{center}
    \begin{tabular}{llll}
      \hline
      Process & Subprocess & Partons & $x$ range \\ \hline
      $\ell^\pm\,\{p,n\}\to\ell^\pm\,X$ & $\gamma^*q\to q$ & $q,\bar{q},g$ & $x\gtrsim 0.01$ \\
      $\ell^\pm\,n/p\to\ell^\pm\,X$ & $\gamma^*\,d/u\to d/u$ & $d/u$ & $x\gtrsim 0.01$ \\
      $pp\to \mu^+\mu^-\,X$ & $u\bar{u},d\bar{d}\to\gamma^*$ & $\bar{q}$ & $0.015\lesssim x\lesssim 0.35$ \\
      $pn/pp\to \mu^+\mu^-\,X$ & $(u\bar{d})/(u\bar{u})\to \gamma^*$ & $\bar{d}/\bar{u}$ & $0.015\lesssim x\lesssim 0.35$ \\
      $\nu (\bar{\nu})\,N \to \mu^-(\mu^+)\,X$ & $W^*q\to q^\prime$ & $q,\bar{q}$ & $0.01 \lesssim x \lesssim 0.5$ \\
      $\nu\,N \to \mu^-\mu^+\,X$ & $W^*s\to c$ & $s$ & $0.01\lesssim x\lesssim 0.2$ \\
      $\bar{\nu}\,N \to \mu^+\mu^-\,X$ & $W^*\bar{s}\to\bar{c}$ & $\bar{s}$ & $0.01\lesssim x\lesssim 0.2$ \\ \hline
      $e^\pm\,p \to e^\pm\,X$ & $\gamma^*q\to q$ & $g,q,\bar{q}$ & $0.0001\lesssim x\lesssim 0.1$ \\
      $e^+\,p \to \bar{\nu}\,X$ & $W^+\,\{d,s\}\to \{u,c\}$ & $d,s$ & $x\gtrsim 0.01$ \\
      $e^\pm p\to e^\pm\,c\bar{c}\,X$ & $\gamma^*c\to c$, $\gamma^* g\to c\bar{c}$ & $c$, $g$ & $0.0001\lesssim x\lesssim 0.01$ \\
      $e^\pm p\to\text{jet}+X$ & $\gamma^*g\to q\bar{q}$ & $g$ & $0.01\lesssim x\lesssim 0.1$ \\ \hline
      $p\bar{p}\to \text{jet}+X$ & $gg,qg,qq\to 2j$ & $g,q$ & $0.01\lesssim x\lesssim 0.5$ \\
      $p\bar{p}\to (W^\pm\to\ell^{\pm}\nu)\,X$ & $ud\to W,\bar{u}\bar{d}\to W$ & $u,d,\bar{u},\bar{d}$ & $x\gtrsim 0.05$ \\
      $p\bar{p}\to (Z\to\ell^+\ell^-)\,X$ & $uu,dd\to Z$ & $d$ & $x\gtrsim 0.05$
      \\ \hline
    \end{tabular}
  \end{center}
  \caption{The main processes, their dominant subprocesses and the parton types they
mostly affect, and the relevant $x$ range. that are included in the
MSTW 2008 global PDF analysis. They are partitioned  into fixed-target experiments, HERA and the Tevatron. }
  \label{tab:processes}
\end{table}
A priori, the space of functions is too large to be constrained through a global fit implied
by solving eq.~(\ref{eq:9}) for the PDFs using a finite amount of data, so some assumptions must be made. 
The various groups differ in their approaches to this issue to varying degrees, and, related
to this, also in their determination of the errors of the extracted PDFs. 

A few constraints are taken along. First, charm and bottom PDF's can be 
determined from the light flavour ones, assuming that such heavy quark
arise from gluon splittings in the proton. This can be done
in different ways, known as variable flavour number schemes, 
see also \cite{Forte:2013wc} for further comments and references.
Also, the already mentioned charge and momentum sum rules must be obeyed
precisely.  

The most common approach is to take a physically motivated form for the
PDFs at a low fixed scale $Q_0$ such as
\begin{equation}
  \label{eq:13}
  \phi_{i/P}(x,Q_0^2 = x^{\alpha_i}(1-x)^{\beta_i}g_i(x)\,,
\end{equation}
with the choice of function $g_i(x)$ differing per group (polynomials, exponentials etc).
The form at other scales is found by solving the DGLAP evolution equation
(\ref{eq:12}). Typically about 20-30 parameters are then to be fitted using 
$\chi^2$ as goodness-of-fit
\begin{equation}
  \label{eq:14}
  \chi^2 = \sum_{i,j=1}^{N_{data}} \left(D_i - T_i\right)\left(V^{-1}\right)_{ij} \left(D_j - T_j\right)\,,
\end{equation}
where $D_i, T_i$ are data and theoretical prediction, respectively, and $V$ is the 
experimental covariance matrix. The uncertainties are determined by varying the 
parameters such that per variation along certain directions in parameter space
(determined by the Hessian matrix) the $\chi^2$ increased by a fixed amount.
In this way, one generates a best-fit PDF set and a collection of one-sigma
error sets, from which then uncertainties for physical observables may be determined.

Another approach is to use, instead of the fixed forms in (\ref{eq:13}), an approach
that does not include theoretical bias at the outset, using neural networks. The
number of free (architecture) parameters in this case is of order 200-300 and very
redundant, but that's ok. The probability distribution in the space of function
is modelled by a Monte Carlo sample of replica's, so that averages and standard deviations
can be easily computed using sums over replicas. 

\begin{figure}[htbp]
  \centering
        \includegraphics[width=0.6\textwidth]{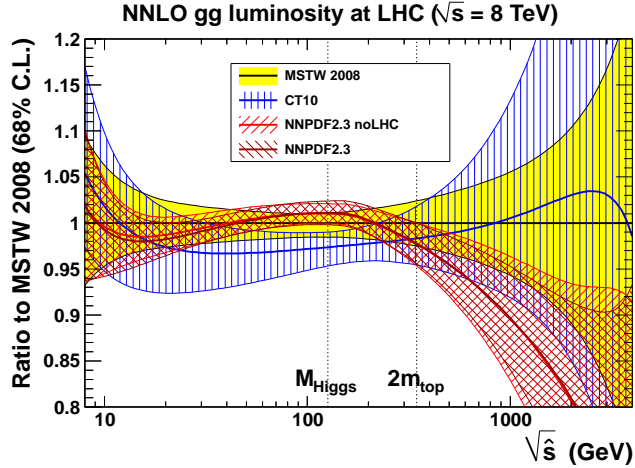}
  \caption{Comparison of gluon-gluon luminosity functions for various PDFsets, relative
to the MSTW08 set. Taken from \cite{Forte:2013wc}.}
\label{fig:pdfcomp}
\end{figure}

Both approaches agree quite well, and differences are very instructive. A comparison
of various recent sets for the LHC at 8 TeV is shown in Fig.~\ref{fig:pdfcomp}
Clearly, the topic of PDF determination is highly important, and progress
continues as data accumulate and understanding of subtle bias effects improves.
A recent review \cite{Rojo:2015xta} describes the state of affairs at the start
of LHC run 2. 

\subsection{$e^+e^-$ collisions and event shapes}
\label{sec:event-shapes}

Before continuing with aspects of QCD at hadron colliders, let us first have a look at
issues in QCD at $e^+e^-$ colliders, such as the former LEP collider at CERN.
Such colliders are, in a sense, the cleanest place to study QCD, due to the
pointline, non-strongly interacting initial state particles. We already saw
how the number of colours, and the masses of heavy quarks can be seen in 
in $R$-ratio. However, the $R$-ratio involves the total cross section, and is not sensitive
to the particular shape or structure of the final state. 

There are other observables or variables that can be, and were, measured, experimentally.  
and are also theoretically consistent (that is, they are infrared-safe, which we discuss in section
\ref{sec:higher-orders}), and are sensitive to the geometry or structure of the
final state. These \textit{event shape} observables describe properties of
final state configurations differently from the
total cross section in $e^+e^-$ collisions (they have also
been generalized to other collision types).  

A well-known example of such an infrared safe event shape is 
the maximum directed momentum, or thrust $T$, 
in $e^+e^-$ collisions. The thrust of an event is defined by 
\begin{equation}
  \label{eq:17.202}
T = \mathrm{max}_{\hat{n}}
\frac{\sum_i\vert \vec p_i \cdot \vec{\hat{n}}\vert}
{ \sum_i \vert \vec p_i \vert}\,,
\end{equation}
where the $p_i$ are the momenta of the particles
and the unit three-vector
$\vec{\hat{n}}$ is varied until a maximum value of $T$ is
obtained. It varies
between  $T=\ft12$ for a spherical energy flow and $T=1$
for a pencil-like linear energy flow of two very narrow, back-to-back jets.
\begin{figure}[htbp]
  \begin{center}
\includegraphics[scale=0.7]{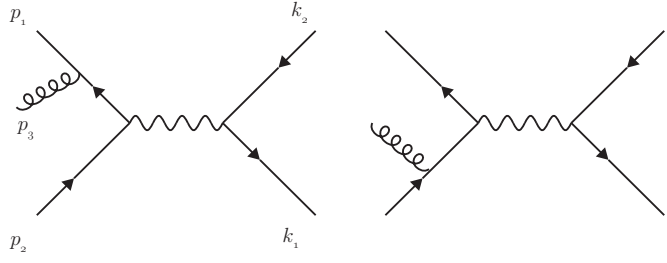}
    \caption{Feynman diagrams for $e^+ e^- \rightarrow \gamma
\rightarrow q\bar{q} g$ at lowest order in the QCD coupling}
    \label{fig:17.3}
  \end{center}
\end{figure}
Let us illustrate this discussion by the calculation of the thrust
distribution for the reaction 
\begin{equation}
  \label{eq:17.203}
e^+(k_1) + e^- (k_2) \rightarrow \gamma(q) \rightarrow q(p_1) + \bar{q}(p_2) + g(p_3) \,.  
\end{equation}
The Feynman diagrams are shown in Fig.~\ref{fig:17.3}.  The kinematical
situation is that of an off-shell photon decaying into three massless
partons, which allows us to use Dalitz plot variables to describe the final state. 
The Dalitz plot for a decay into 3-particle final state is a scatterplot of events 
in a plane spanned by two of the final particle energies. The reason for plotting
events this way is that the phase space measure is ``flat'' in those variables, so
that any clustering of events represents an intermediate resonance in the decay.

\begin{figure}[htbp]
  \begin{center}
\includegraphics[scale=0.6]{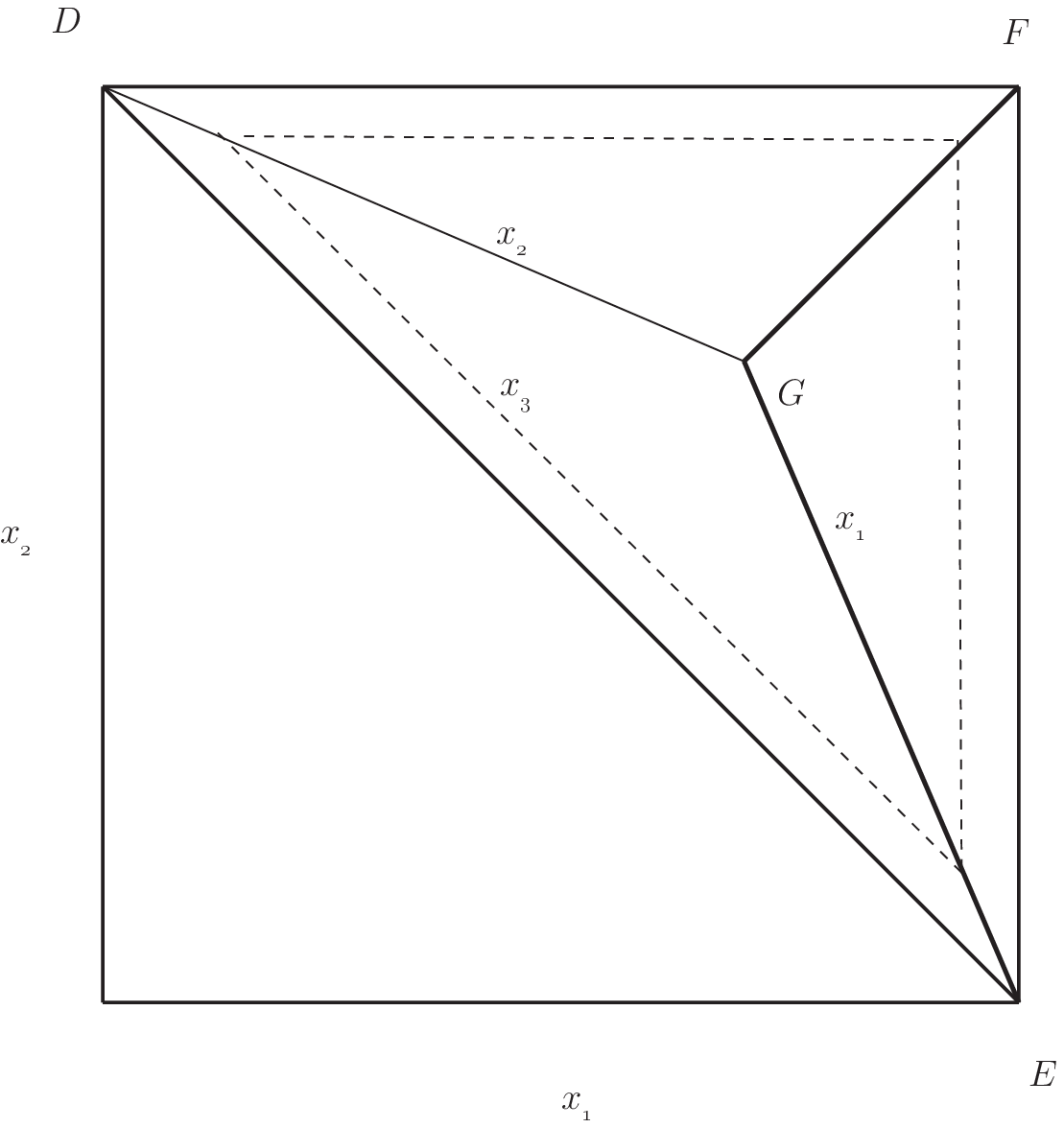}
\caption{Contributing regions in $x_{1,2}$ for a given value of thrust
  $T$. On the line DE $x_3=1$ (recall that $x_1+x_2+x_3=2$). At the
  point $G$ all three $x_i$ are equal to $2/3$. In each of the three
  triangles that join at $G$ one of the three $x_i$ is largest, as
  indicated. On the dashed line the value of $T$ is constant.}
    \label{fig:17.4}
  \end{center}
\end{figure}

Let us then choose two energies to specify the allowed region in a Dalitz plot,
or, more conveniently and equivalently, choose invariant mass variables. 
Hence, using the particle name to represent its four
momentum, we introduce $s_{13} = - (p_1+p_3)^2$, $s_{23} = -
(p_2+p_3)^2$ and $s_{12}= -(p_1+p_2)^2$.  Since the three final particles are coplanar, the
whole kinematics is specified by $s_{13}, s_{23}$ and $3$ angular
variables.  One angular variable $\theta$ specifies the polar angle
between the beam axis and a line in the three-particle plane. 
Another azimuthal angular variable $\phi$ specifies the orientation
of the plane will respect to this line and finally there is an overall
azimuthal angle $\chi$. The phase space for the final three particles therefore becomes
\begin{equation}
  \label{eq:17.33}
{{1}\over {(2\pi)^5}} \int {{d^3 p_1}\over{2E_1}} \int
{{d^3p_2}\over{2E_2}} \int {{d^3 p_3}\over{2E_3}} =
{{1}\over{(2\pi)^5}} \int \frac{1}{32 q^2} ds_{13} ds_{23} d \phi d \sin
\theta d \chi \,.
\end{equation}
The expression for the cross-section after squaring the matrix element
for $e^+e^-\rightarrow q \bar{q} g$ and integrating over $\phi$  and
$\chi$ turns out to be
\begin{equation}
  \label{eq:17.34}
{{d^3 \s}\over{ds_{13} ds_{23} d \sin \theta}} = {{\a^2_e}\over{8}}
{{\a_s}\over{q^2}} \left( x_1{}^2 + x_2{}^2 \right) \left( 2+\cos^2
\theta \right) {{1}\over{s_{13}s_{23}}} \,, 
\end{equation}
where the variables $x_i = E_i/E $, with $E=\sqrt{q^2}/2$, are related to the invariant
mass variables by $s_{13} = q^2 (1-x_2), s_{23}= q^2
(1-x_1), s_{12} = q^2 (1-x_3)$.  Note that $x_1 + x_2 + x_3= 2$.
From (\ref{eq:17.34}) we see that the
angular distribution of the normal to the plane with respect to the
beam line is given by $2+\cos^2 \theta$.  A final integration over $\sin\theta$ 
yields the two equivalent expressions.
\begin{equation}
  \label{eq:17.35}
\s^{-1}_T {{d^2 \s}\over{ds_{13} ds_{23}}} = {{2}\over{3\pi}} \a_s {{x_1^2 +
x_2^2}\over{s_{13} s_{23}}} \,, 
\end{equation}
or
\begin{equation}
  \label{eq:17.36}
\s^{-1}_{T} {{d^2 \s}\over{dx_1 dx_2}} = {{2}\over{3\pi}} \a_s {{x_1^2 +
x_2^2}\over{(1-x_1)(1-x_2)}} \,,
\end{equation}
where we have divided both sides of the equation by $\s_T = {{4}\over{3}} \pi\a_e^2 /s $,
the $e^+e^- \rightarrow \m^+\m^-$ cross section.
These distributions diverge for small invariant masses, or, equivalently
as the scaled energies $x_{1,2}$ of the quark and antiquark go to one. 
It is not very difficult to show that the thrust
variable $T$ for the present case is equal to $\mathrm{max}(x_1, x_2, x_3)$ for each event, 
with ${{2}\over{3}} \leq T \leq 1$. 
For fixed $T$ the allowed region in $x_1,x_2$ is then shown  in
Fig.~\ref{fig:17.4}.
%%%%%%%%%%%%%%%%%%%%%%%%%%%%%%%%%%%%%%%%%%%%%%%%%%%%%
%%%%%%%%%%%%%%%%%%%%%%%%%%%%%%%%%%%%%%%%%%%%%%%%%%%%%
The lines EF, FD and DE are the lines $x_1=1$, $x_2=1$ and $x_3 =1$,
respectively.   The figure shows the
subdivision of the final phase space into three regions depending on
which particle has the largest $x$ value. Consider first the case
$T=x_2$. Then we have
\begin{eqnarray}
  \label{eq:17.37}
 \s^{-1}_T {{d\s}\over{dT}}& &=  \frac{2\alpha_s}{3\pi} \int dx_1
 dx_2 \delta(T-x_2)\theta(T-x_1)\theta(T-x_3)
\nonumber \\&&
\times \frac{x_1^2+x_2^2}{(1-x_1)(1-x_2)}
\nonumber \\
&&= \frac{2\alpha_s}{3\pi} \int^T_{2(1-T)} dx
{{T^2+x^2}\over{(1-T)(1-x)}} 
\nonumber \\
&&= \frac{2\alpha_s}{3\pi} \Big\{ {{1+T^2}\over{1-T}} \ln
{{2T-1}\over{1-T}} + {{3T^2-14T+8}\over{2(1-T)}}  \Big\}\,,
\end{eqnarray}
with an identical result for $T=x_1$.  The $T=x_3$ case is slightly
different, and corresponds to integrating over the dashed line
parallel to DE in Fig. \ref{fig:17.4}. On this line we have that 
$x_1=2-T-x_2$, while $x_2$ ranges from $2(1-T)$ to $T$. One then
finds
\begin{equation}
  \label{eq:17.38}
\s^{-1}_T {{d\s}\over{dT}} = {{4\alpha_s}\over{3\pi}} \Big\{ {{1+(1-T)^2}\over{T}}
\ln {{2T-1}\over{1-T}} + 2 - 3T \Big\} \,.
\end{equation}
Clearly there is different, interesting dependence on $T$ for the various cases.

Note that the thrust distributions for the quark and antiquark are
singular as $T\rightarrow 1$, signifying the appearance of soft and/or
collinear singularities, where either the gluon is very soft, or
the (anti-)quark-gluon splitting is essentially collinear (in the
$T=x_3$ case \eqref{eq:17.38} the distribution is singular but integrable).
In these infared and collinear regions of phase space  non-perturbative effects
must start playing a role in order to cure this apparent problem in perturbative QCD.

The different expressions for the thrust dependence for different regions in 
Fig.~\ref{fig:17.4} allow us to make an interesting observation.
Since the integral of the gluon $T$ distribution in (\ref{eq:17.38})
is integrable at $T=1$, we
can integrate it from $T=2/3$ to $T=1$ to find the probability
that the gluon is the most energetic particle.  This yields
\begin{equation}
  \label{eq:17.39}
\s^{-1}_{T} \int^1_{2/3} {{d\s}\over{dT}} dT = 0.61 {{\alpha_s}\over{\pi}} \,.
\end{equation}
We can reasonably \footnote{A higher-order calculation of the thrust distribution
  \cite{Fabricius:1981sx,Vermaseren:1980qz,Kunszt:1980vt,Ellis:1980wv},
  which requires  renormalization of the QCD coupling, confirms this.} assume that $\alpha_s$ is a function of $q^2$, 
the center-of-mass energy squared, so that
the total probability that the gluon is the most energetic particle decreases with increasing $q^2$. 
The probability that the quark or the antiquark is the most energetic
particle is then given by $(1-0.61 \alpha_s / \pi)$.

The total thrust distribution is twice the result (\ref{eq:17.37})
(accounting for the cases $T=x_1$ and $T=x_2$)
added to the result (\ref{eq:17.38}), which yields
\begin{equation}
  \label{eq:17.40}
\s^{-1}_{T} {{d\s}\over {dT}} = {{2\alpha_s}\over{3\pi}} \Big[
{{2(3T^2-3T+2)}\over{T(1-T)}} \ln {{2T-1}\over{1-T}} - 
{{3(3T-2)(2-T)}\over{1-T}} \Big] \,.
\end{equation}
Because the integrand is integrable at $T=1$ we can also compute
the average value of $(1-T)$ from this formula
\begin{equation}
  \label{eq:17.41}
\langle 1-T \rangle \equiv \s^{-1}_T \int dT  {{d\s}\over{dT}} (1-T) = 1.05 {{\alpha_s (q^2)}\over{\pi}} \,.
\end{equation}
We see that this average value of $(1-T)$ decreases with
increasing $q^2$.

\begin{figure}[htbp]
  \centering
\includegraphics[scale=0.75,angle=270]{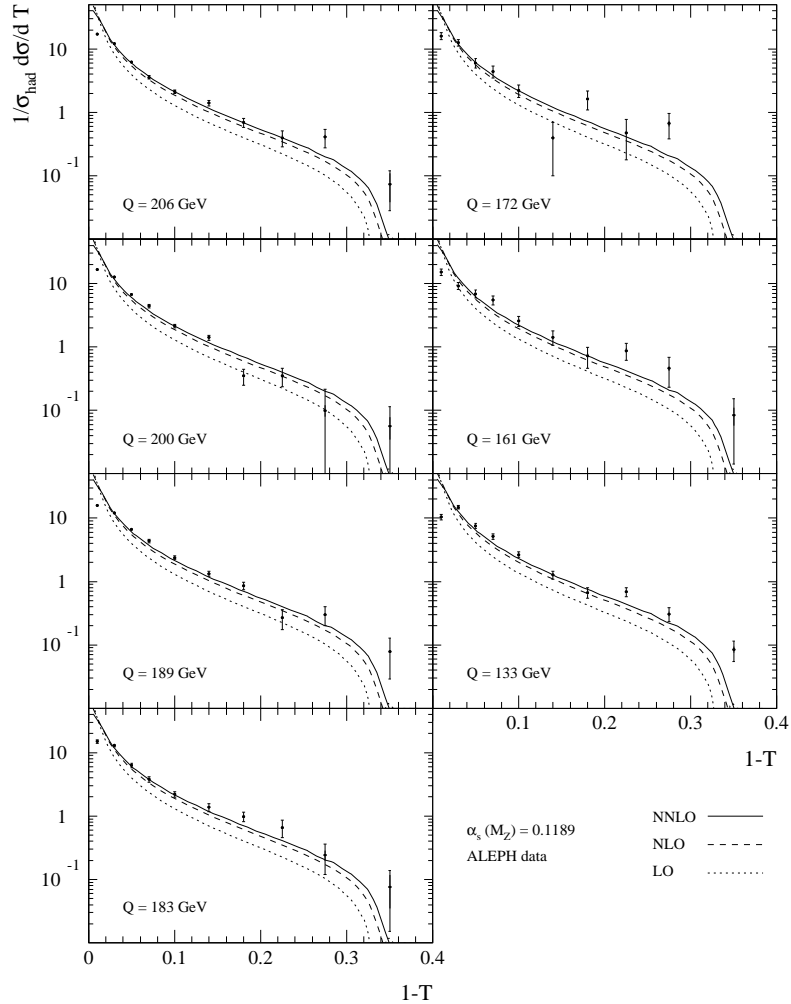}
  \caption{A comparison of thrust data from the ALEPH collaboration,
with LO, NLO and NNLO calculations for various LEP cm energies. 
Figure taken from \cite{GehrmannDeRidder:2007hr}.}
\label{fig:alepht}
\end{figure}

A comparison of higher-order calculations for thrust with data
is shown in Fig.~\ref{fig:alepht}, showing the good quality of
(but also the need for) the NNLO approximation.

This concludes our rather detailed look at event-shape variables, 
where we already saw the appearance of infrared and collinear singularities. 
We now turn to a more detailed discussion of these, as this is a
central issue in the application of perturbative QCD for colliders. 

\subsection{More on $e^+e^-$ cross sections, IR divergences, KLN theorem}
\label{sec:more-e+e-}

The ratio $R= \sigma(e^+ e^- 
\rightarrow {\rm hadrons}) /\sigma (e^+ e^- \rightarrow \mu^+ 
\mu^-)$  was already discussed in section \ref{sec:evidence-colour} when discussing
evidence for colour. In order to compute higher-order QCD corrections to this ratio,
i.e. to the $e^+ e^-$ total cross section, we must deal with infrared and collinear
divergences (often collectively called ``infrared''). This can be seen if one
would integrate the expression (\ref{eq:17.40}) over $T$; it would produce
a divergence due to the $1/(1-T)$ behaviour. 

The Kinoshita-Lee-Nauenberg theorem \cite{Kinoshita:1962ur, Lee:1964is} (KLN)
now states that, when summing over all contributions to the observable at that order (i.e. also
the loop-corrections) the divergences cancel if the sum is over a sufficiently degenerate
set of states. 

This is a very powerful result, so let us discuss it a bit further.
One of the crucial aspects of massless particles is indeed that one is dealing 
with degenerate states. For instance, in quantum electrodynamics, it is not meaningful to
distinguish between a single electron and an electron accompanied by
any number of zero-momentum photons, as the corresponding states carry
the same electric charge, energy and momentum. An (infinite)
degeneracy of states implies that a naive application of perturbation
theory may run into difficulties, a phenomenon that is, for instance,
also known from applications in quantum mechanics. According to the
\index{Kinoshita-Lee-Nauenberg theorem} Kinoshita-Lee-Nauenberg (KLN)
theorem \cite{Kinoshita:1962ur,Lee:1964is} the divergences that are in
principle present in partial transition probabilites, must cancel when
averaging over a suitable set of degenerate states.  This theorem
encompasses in fact the older Bloch-Nordsieck \cite{Bloch:1937pw}
theorem. 

Observables sufficiently inclusive to allow a sum over a sufficiently
large ensemble of degenerate states
for the KLN cancellations to occur, are known as \emph{infrared safe}.
How large an ensemble should be depends on the experimental
process that one is considering. Of course in electron-positron
annihilation at high energies, the {\em total} cross section (where
one sums over {\em all} possible finite states) certainly constitutes an
infrared safe quantity, so reliable predictions
in perturbative QCD should be possible.

Let us, then, examine the first three terms in 
the perturbation series for $R$ \cite{Baikov:2012zn}, in the limit of zero fermion 
mass (owing to the KLN theorem the quark mass could be 
suppressed without encountering infrared divergences)
\begin{eqnarray}
  \label{eq:16.31}
R(t) &=& \left({\textstyle\sum_f Q_f^2}\right) 
\Bigg\{ 1 + \frac{\a_s(t)}{4\pi}3C_2(R) \nonumber \\ 
&& + \left( {\a_s(t) \over 4\pi} \right)^2 \left[ -\ft12 C_2^2(R) 
+\left(\ft{123}2 - 44\zeta(3)\right) \,C_2(G)\,C_2(R)   
\right.\nonumber \\ 
&&\left. \quad + n_{\rm f} \left(-22+16\zeta(3)\right) \,C_2^2(R)\,{\textstyle{\frac{3}{8}}}   \right] \Bigg\}\,, 
%(14.d12)
\end{eqnarray}
where $({\textstyle\sum_f Q_f^2})$ denotes the square of the electric charges of 
the fermions and $C_2(G), C_2(R)$ are colour factors of SU(3),
As before, 
$t$ represents the logarithm of the ratio of two energy 
scales, one being the total center-of-mass energy $q^2$ of the incoming 
electron-positron pair and the other some 
reference scale. Obviously, the running coupling constant should be 
evaluated to the same order as the cross section. Through $\a_s(t)$ 
this result thus depends on the number of  quark flavours $n_{\rm f}$. 

The Riemann zeta function invariably appears in higher-loop calculations. It is  
defined by $\zeta(z) = \sum_{n=0}^\infty {1/ n^z}$. The 
specific value encountered above is $\zeta(3) \approx  1.2020569$.
Using these values, the numerical coefficients 
for $\SU{3}$ with $n_{\rm f}=5$ quark flavours yield 
\begin{equation}
  \label{eq:16.32}
R(t) = \left({\textstyle\sum_f Q_f^2}\right)  \left( 1 +  {\a_s(t) \over \pi}
+1.409 \Big( {\a_s(t) \over \pi} \Big)^2  \right)\; .
%\eqno(14.d13)
\end{equation}
We see that the coefficient in front of the  $(\a_s(t)/\pi)^2$ is not
too large, and the perturbative description is well-behaved. 
As is clear from Fig.~\ref{fig:eehad-r} the result above should be used with great caution in the vicinity of
heavy flavour thresholds, because bound states
appear in $R(t)$, which are not describable in finite order perturbation theory. 

Let us next discuss two important, technical but generic issues that arise in the derivation of
results such as (\ref{eq:16.32})  in the context 
of dimensional regularization. The first concerns the integration
over phase space. Because the quantities that one 
calculates are infrared safe, infrared divergences must cancel at 
the end of the calculation. This requires one to determine the full cross  
section in $n$ dimensions and take the limit $n\to 4$ only at
the end. In particular, also the phase-space integrals should be 
evaluated in $n$ dimensions. The second issue is that 
that the mass shell for massless particles poses problems in perturbative
calculations. On-shell massless particles may split into perfectly collinear
massless particles which then remain on their mass shells. This makes
the mass shell ill-defined, and singularities appear, as we will see below.
Moreover, there is a conceptual problem in that
on-shell massless particles should correspond to asymptotic states.
But in a confined theory such as QCD the massless partons do not 
correspond to \emph{physical} states, which consist of massive hadrons. 
For infrared safe observables this is in fact not fatal to predictive
power, but for calculations this is at least at an intermediate level a cumbersome feature.
We now discuss  these two aspects in turn. 

In $n$ dimensions the two-particle phase-space integral is 
defined by 
\begin{equation}
  \label{eq:16.33}
I^{(n)}(s, m_3^2, m_4^2) =  \int \; 
{{{\rm d}^{n-1}p_3 }\over{ (2\pi)^{n-1} 2 \omega_3}}
{{{\rm d}^{n-1}p_4 }\over{ (2\pi)^{n-1} 2 \omega_4}}
\,(2\pi)^n\, \delta^{(n)}(p_1 + p_2 - p_3 -p_4)\,,
%\eqno(14.d14)
\end{equation}
where $s=-(p_1+p_2)^2$ and $\o_{3,4}=({\bf p}_{3,4}^2 
+m^2_{3,4})^{1/2}$. We choose 
the centre-of-mass frame and decompose the full integral in one  
over the $n-2$  angular variables (which we leave unperformed) and 
one over the 
length of the $(n-1)$-dimensional momentum vector ${\bf p}_3=-{\bf 
p}_4$, which contains a delta function, and which we do perform. It
yields 
\begin{equation}
  \label{eq:16.34}
I^{(n)}(s, m_3^2, m_4^2) =   
{1\over 8\pi\sqrt s}\, \Big[{ \lambda(s, m_3^2, m_4^2)\over 16\pi^2 
s}\Big]^{{n-3\over 2}} \, \int {\rm d}\O_{CM}\,,
%\eqno(14.d15)
\end{equation}
where $\lambda(x,y,z) = x^2+y^2+z^2-2xy-2xz-2yz$.
As usual this expression must be combined with the square of 
the invariant amplitude to yield a cross section or decay rate. 
Assuming that the invariant amplitude depends only on the 
deflection angle $\theta_{CM}$ between ${\bf p}_1$ and ${\bf 
p}_3$, we can integrate over the remaining $n-3$ angles,
using the formula 
\begin{equation}
  \label{eq:16.35}
\int {\rm d}\O_{CM} =  {2\pi^{-1+\ft12n}\over \G(\ft12n-1)} 
\int_0^\pi  \;{\rm d}\theta_{CM}\, \big[ \sin 
\theta_{CM}\big]^{n-3} \,,
\end{equation}
where the integral on the left-hand side runs over all $n-2$ 
angles, while the integral on the right-hand side contains only 
the deflection angle. Typically one needs the integral 
for $m_4 = 0$, reflecting emission of a massless particle. Replacing $m_3$ by $m$ and introducing the variables 
$x = m^2/s$ and $y = \ft12 (1 + \cos \theta)$ the combined result 
for the phase-space integral reads 
\begin{equation}
  \label{eq:16.36}
I^{(n)}(s, m^2,0)   = {{1}\over{8\pi}}
\Big({m^2\over 4\pi}\Big)^{-2+\ft12 n} \,
{{x^{2-\ft12 n} (1-x)^{n-3}}\over
{\Gamma(\ft12n-1)}}\, 
\int_0^1 {\rm d}y \;[y(1-y)]^{-2+\ft12 n} \,,
%\eqno(14.d16)
\end{equation}
which has the correct dimension of a mass to the power $n-4$. 
For $n>2$ this expression is free of 
singularities. Corresponding expressions can be derived for 
multi-particle phase-space integrals. Formulae like (\ref{eq:16.36}) are 
obviously needed for calculating decay rates and cross sections 
in arbitrary dimension, the squared invariant amplitudes for which
are then functions of $y$. 
The $y$-integral can be evaluated by using the relation
\begin{equation}
  \label{eq:Eulerbeta}
  \int_0^1 dy \, y^{p-1}(1-y)^{q-1} = B(p,q) = \frac{\Gamma(p)\Gamma(q)}{\Gamma(p+q)}\,,
\end{equation}
where $B(p,q)$ is the Euler beta-function. Depending on the invariant
amplitude of the process, infrared divergences can then appear as
poles in the Gamma function, just as for virtual corrections. 

The second issue involves the definition of the mass shell for
massless particles in dimensional regularization. To this end we 
let us turn to the evaluation of a typical one-loop self-energy 
diagram
\begin{figure}[htbp]
  \centering
\includegraphics[scale=0.75]{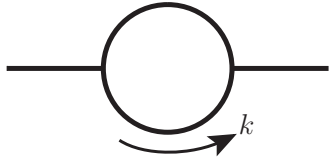}
  \caption{A typical one-loop self-energy diagram}
\label{fig:prop}
\end{figure}
This involves an integral of the type 
\begin{equation}
  \label{eq:16.37}
I(k^2,m_1^2,m_2^2) = {1\over (2\pi)^n}\int {{\rm d}^nq \over 
\big((q+\ft12 k)^2 +m_1^2\big) \big((q-\ft12 k)^2 +m_2^2\big)}\,.
\end{equation}
We imagine $k$ to be the momentum of a massless particle that is off-shell and put $m_1=m_2=0$. 
Using Feynman parameters one obtains 
\begin{equation}
  \label{eq:16.38}
I(k^2,0, 0)= {\im\over 16\pi^2} \, \Gamma(2-\ft12 n) \Big({k^2\over 
4\pi}\Big)^{-2+\ft12 n}\int_0^1 {\rm d}x \; [x(1-x)]^{-2+\ft12 n}\,.
\end{equation}
The $x$ integral can be evaluated using (\ref{eq:Eulerbeta}) and becomes
$ (\Gamma(\ft12 n-1))^2 / \Gamma(n-2)$, so that 
\begin{equation}
  \label{eq:16.39}
I(k^2,0, 0)= {\im\over 16\pi^2}  \,{ \Gamma(2-\ft12 n)  \Gamma(\ft12 
n-1)^2\over  \Gamma(n-2)}  \Big({k^2\over 4\pi}\Big)^{-2+\ft12 n}\,.
\end{equation}
This expression exhibits poles for both large and small values 
of $n$, signaling ultraviolet and infrared singularities 
respectively. The last factor in (\ref{eq:16.39}) shows that the result is in fact 
ambiguous when approaching 
the mass shell, $k^2 \to 0$. When considering infrared divergences one
assumes $n>4$ so that the integral is in fact zero on the mass
shell. Hence, one can omit self-energy loop corrections for massless
external particles. In fact, their zero contribution can be shown to 
be due to a perfect cancellation between a UV divergence and a collinear 
divergence.  However, one must still include the UV
counterterms on the external lines. The sum of the two is then in fact the
collinear divergence, which in turn will cancel in a calculation of infrared
safe quantities.

\subsection{Jets}
\label{sec:jets}

\begin{figure}[htbp]
  \begin{center}
\includegraphics[scale=0.5]{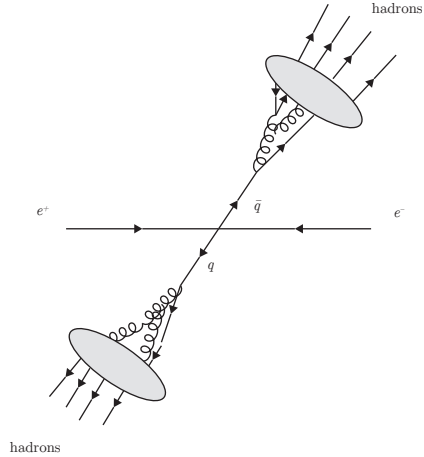}
\caption{Two jets of hadrons arising from quark-antiquark production
  in an $e^+e^-$ collision. The blobs represent the hadronization process.}
\label{fig:17.1}
  \end{center}
\end{figure}

Besides event shapes there is another important class of infrared
safe observables that uses
the notion of a\index{jet} jet.  In high-energy 
$e^+e^-$ collisions the photon couples
directly to a quark-antiquark pair, and the latter are then produced back-to-back 
in the $e^+e^-$ cm frame, with high momentum.  
As the quarks begin to fly apart they undergo the
complicated fragmentation or hadronization process that leads to
colourless hadronic final states.  One could therefore expect the
final hadrons to follow the line of flight of
the quarks to produce two streams of back-to-back particles, as
depicted in Fig.~\ref{fig:17.1}. 
The angular distribution of the quarks is $ 1 + \cos^2 \theta$, where
$\theta$ is the polar angle between a quark and the beam direction.
The angular distribution of the hadrons
should then have roughly the same dependence on $\cos \theta$. Indeed this
is the case.

\begin{figure}[htbp]
  \begin{center}
\includegraphics[scale=0.5]{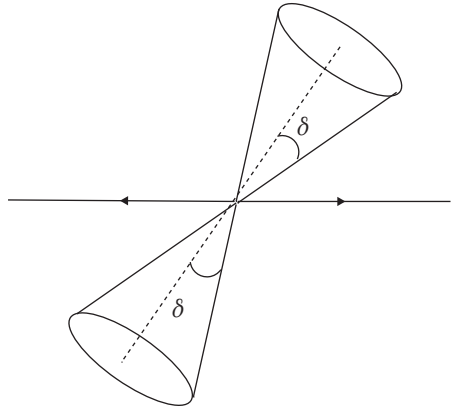}
    \caption{Two jets defined by an opening angle $\delta$}
    \label{fig:17.2}
  \end{center}
\end{figure}

In the context of considering the effect of higher orders,
one may ask how this angular distribution
is changed by the emission of an additional gluon. 
Remember that the contribution due to 
gluon emission contains soft and collinear
divergences, to be cancelled via the KLN theorem.
Intuitively, one would think that one does not need
to integrate over \emph{all} possible gluon emission energies and angles;
if we only integrate the gluon emission rate over a small
angle close to the quark or antiquark direction the  
collinear divergence should already cancel with the divergence
contained in the virtual contribution, leaving a contribution 
depending on the size of the angular range.
Also, if we allow a (very) soft gluon to be emitted and add this 
contribution to that from the virtual contribution
we expect that the infrared divergences will
cancel too. The result will then still depend on one angle and one energy.
One could therefore define a two-jet event as one where 
almost all of the energy, namely $(1-\eps)\sqrt{s}$,
is contained in two small cones of semi-angle $\delta$, where $\eps$ and $\delta$
are fixed, and can be reasonably large, as shown in
Fig.~\ref{fig:17.2}. 
%%%%%%%%%%%%%%%%%%%%%%%%%%%%%%%%%%%%%%%%%%%%%%%%
%%%%%%%%%%%%%%%%%%%%%%%%%%%%%%%%%%%%%%%%%%%%%%%%%
An explicit calculation of the corrections to the
(anti)quark angular distribution shows that the
angular distribution is still proportional to $1+\cos^2\theta$ but the
coefficient in front is modified by the factor 
\begin{equation}
  \label{eq:17.201}
1 - \frac{\alpha_s(q^2)C_2(R)}{\pi} \Big[ (4 \ln 2 \eps + 3) \ln
\d + \pi^2/3 - 5/2 + 0 (\eps) + 0 (\d) \Big]\,.  
\end{equation}
As one would expect, if one would take the limit $\eps$ and $\delta \rightarrow 0$
divergences the divergences show up again in this factor,
so one must be careful to choose $\epsilon$ or $\delta$ small but
large enough that the $\alpha_s$ correction in (\ref{eq:17.201}) is still relatively small.
In this way, due to Sterman and Weinberg \cite{Sterman:1977wj}, 
the jet angular distribution is well-defined
and has been successfully compared with experiment.  

One may now generalize the definition of a jet such that singularities
still cancel, but that their definitions are more easily implemented
in both experimental measurements and theoretical calculations, the latter
in the form of a Monte Carlo program.
To this end one constructs an iterative algorithm for combining
the measured hadrons (or computed partons)  into jets. The starting
point of such an algorithm is a list of
particles (hadrons or partons) with their energies and angles.
For example, in one algorithm, for all particle pairs $i$ and $j$ 
one then calculates the quantity $y_{ij}= 2E_iE_j(1-\cos\theta_{ij})/s$.
All $y_{ij}$'s are now compared with a chosen value $y_{\rm cut}$. For
each $y_{ij}$ that is smaller than $y_{\rm cut}$
the two momenta of particles $i$ and $j$ are combined according 
some chosen prescription, for instance ``add the four-momenta''.
Particles $i$ and $j$ are then removed from the list, but their
combination is returned to the list as a new 'pseudoparticle'.  
The procedure is repeated until no two (pseudo)particles have an
$y_{ij}$  that is smaller than $y_{\rm cut}$. 
This subdivides the experimentally measured or theoretically simulated 
events into a number of clustered jets, of which one can study the
properties. One should be aware however 
that not all algorithms are infrared safe for all collider types.

This concludes our discussion of higher orders for 
collider processes with quarks and gluons only in the final
state. We now turn to the case where there are strongly 
interacting particles in the initial state as well.

\subsection{The Drell-Yan process}
\label{sec:drell-yan:-history}

The Drell-Yan  process \cite{Drell:1970yt} is, and has been,
an important reaction in particle physics.  It
involves the production of a lepton-antilepton pair in
proton-(anti)proton collisions,
\begin{displaymath}
  p + \bar{p}/p \rightarrow l  + \bar l + X
\end{displaymath}
where $X$ denotes the rest of the final state. Leptons are relatively
\index{charm quark discovery} \index{bottom quark discovery} easy to
detect and through this reaction a number of important discoveries
such as of the $J/\Psi$ and the $\Upsilon$ mesons (and therefore
of the charm and bottom quarks), and of the $W$ and $Z$ vector bosons
were made.  From a theoretical point of view, its QCD corrections are
prototypical for any high-energy cross section with initial state
hadrons, and it is from this perspective that we shall discuss these
corrections here.  For simplicity we will only examine the QCD
corrections to the single differential cross section in the lepton
pair invariant mass $Q$, i.e. $d\sigma/dQ^2$. Therefore the process is
inclusive in all the hadron final states, which renders the KLN
theorem for the QCD corrections in principle operative (as we will
see, only for the final state).  To calculate the
corrections we consider the reaction at the partonic level, where the
lowest order approximation only involves only quark-antiquark
annihilation into a virtual photon, which then couples to the
lepton-antilepton pair.  The total cross
section for quark-antiquark annihilation in the reaction $q(p_1) +
\bar q(p_2) \rightarrow l(q_1) + \bar l (q_2)$ can be computed as
\begin{equation}
  \label{eq:17.1}
  \sigma_{q\bar q}^{(0)}(\hat s)   =
  \frac{1}{4 N_c^2} \frac{1}{2 \hat s}
  \int {{d^3q_1}\over{(2\pi)^3 2\omega_1}}
  \int {{d^3q_2}\over{(2\pi)^3 2\omega_2}}
  (2\pi)^4 \delta(p_1+p_2 -q_1 -q_2) 
  \sum \vert {\cal M} \vert^2 \,,   
\end{equation}
where $\hat{s}=-(p_1+p_2)^2=-(q_1+q_2)^2 = Q^2$, the sum is over all initial and final
spin and colour indices, and initial state spins and colours are
averaged over.

We remind the reader that the hadronic cross section follows by
convoluting this result with partonic densities in the incoming
hadrons, as in sections \ref{sec:parton-mode} and \ref{sec:part-distr-funct}.
We write $\hat s = \xi_1 \xi_2 s $, where
$\xi_{1,2}$ are parton momentum fractions and $s$ is the collider
cm energy squared. Also we introduce
the variable $\tau = Q^2/ s$ so that
\begin{equation}
  \label{eq:17.5}
  {d\sigma_{pp}^{(0)}(\tau)\over{dQ^2}}  =
  \sum_{i,j}
  \int^1_{\xi_{1,{\rm min}}} 
  d \xi_1
  \int^1_{\xi_{2,{\rm min}}}
  d \xi_2\;
  f_{i/p}(\xi_1) f_{j/p}(\xi_2)  
  {d\sigma_{ij}^{(0)}(\xi_1,\xi_2)\over{dQ^2}}  \,,
\end{equation}
where the $f$'s are the parton distribution functions for the quarks
and antiquarks in the proton and antiproton.  The sum runs over all
quarks and antiquarks in both the proton and antiproton, while
$\xi_{1,{\rm min}}= \tau$, and $\xi_{2,{\rm min}} = \tau/\xi_1$. 
We return to this formula towards the end of this section.

 The only Feynman diagram to compute is the one photon exchange
 diagram for which the square of the amplitude yields
 \begin{equation}
   \label{eq:17.2}
   \sum \vert {\cal M} \vert^2
   =e^4 Q_f^2 {\rm Tr}(\gamma_\mu \slash{p_1} \gamma_\nu \slash{p_2})
   {\rm Tr}(\gamma_\mu \slash{q_2} \gamma_\nu \slash{q_1}) 
   {{1}\over{\hat s^2}} \,.
\end{equation}
The charge of quark flavour $f$ is $Q_f e$, and the final trace is over
 the unit $N_c$-dimensional matrix labelled by the colour indices. It
 is not difficult to work out (\ref{eq:17.2}). Moverover the integration over the lepton trace
in (\ref{eq:17.1}) can be done by applying the so-called Lenard identity (here given in $n$ dimensions)
\begin{multline}
  \int \,\rd^nq_1 \,\delta(q_1^2)\int\,\rd^nq_2\,
  \delta(q_2^2)\,\delta^n(q -q_1-q_2) \,
  q_1^{\mu}q_2^{\nu} = \\
\left(\frac{-q^2}{4} \right)^{(n-4)/2} \frac{\pi^{(n-1)/2}}{\Gamma((n+1)/2)}\frac{1}{32}
(q^2\eta^{\mu\nu}+ 2q^\mu q^\nu)\,.  \label{eq:16}
\end{multline}
The final result, in 4 dimensions,
 \begin{equation}
   \label{eq:17.3}
   \sigma^{(0)}_{q\bar q}(\hat s) 
   = {{4\pi\alpha^2}\over{3N_c\hat s}} \,,
 \end{equation}
 is only a function of $\hat s$.  The differential cross
 section with respect to $Q^2=-(q_1+q_2)^2$ can now be derived using 
 \begin{equation}
   \label{eq:17.4}  
   {d \sigma^{(0)}_{q\bar q}(Q^2)\over{dQ^2}} =
   \left[ \frac{4\pi\alpha^2}{3N_c(Q^2)^2} \right] 
   \, \delta\left(1-\frac{Q^2}{\hat{s}} \right)\,.
 \end{equation}
Indeed, as a check
 \begin{equation}
   \label{eq:17.401}
   \sigma^{(0)}_{q\bar q}(\hat{s}) 
   = \int {{d \sigma^{(0)}_{q\bar q}}(Q^2)\over{dQ^2}} dQ^2 
   = \frac{4\pi\alpha^2}{3N_c} \int \frac{dQ^2}{(Q^2)^2} 
   \delta\left(1-\frac{Q^2}{\hat{s}} \right)
   = \frac{4\pi\alpha^2}{3N_c \hat{s}} \,. 
 \end{equation}
 Note that (\ref{eq:17.4}) is no longer a function but a distribution
 as it is proportional to a $\delta$-function with argument
 proportional to $\hat{s}-Q^2$. 
Therefore as far as the lowest order formula is concerned
 we can write either $d \sigma^{(0)}_{q\bar q}/ dQ^2$ or $d
 \sigma^{(0)}_{q\bar q}/ d\hat{s}$.  The expression in square brackets
 in (\ref{eq:17.4}) we will refer to as $\sigma^{(0)}_\gamma$.

 For the calculation of the QCD corrections we would prefer not 
 to include the part of the diagram where the photon decays
 into leptons, which is common to all diagrams to any order in QCD 
perturbation theory.  One can account for that by computing the
ratio $K$ of the squared amplitude for
 the process $q(p_1)+\bar{q}(p_2) \rightarrow \gamma^\ast(q)$ ($q^2=-Q^2$) and 
 the $q(p_1)+\bar{q}(p_2) \rightarrow l \bar{l}(q)$ at lowest order as follows
 \begin{equation}
   \label{eq:17.701}
   \sigma^{(0)}(l \bar{l}) = K \sigma^{(0)}(\gamma^\ast) \,.
 \end{equation}
The factor $K$ can be computed in dimensional
 regularization. It is valid to all orders in perturbative QCD,
 because it only involves the electroweak final state.
We shall not give the
expression here, but thanks to (\ref{eq:17.701}) we can now suffice with
computing the cross section for $\gamma^*$ production.

\begin{figure}[htbp]
  \centering
\includegraphics[scale=0.75]{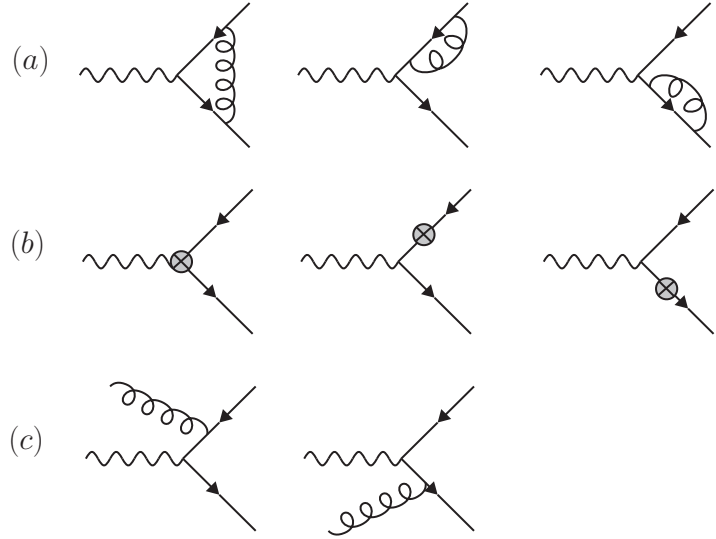}
  \caption{The Feynman diagrams for the first order QCD corrections to the
partonic Drell-Yan reaction in the quark-antiquark collisions producing an off-shell photon. 
Shown are (a) loop contributions (b) counterterm contributions, and (c) radiative graphs.
The leptons into which the photon decays are not shown. Time runs from right to left
in this figure.} 
  \label{fig:17.6}
\end{figure}

Let us now evaluate the next order corrections to
 (\ref{eq:17.5}).  We consider the quark-antiquark channel and calculate the
 processes involving the virtual corrections to the Born reaction,
and the counterterm contributions. In Fig.~\ref{fig:17.6} we show these, and
also contributions to the quark-gluon channel, which are typically smaller.
The Feynman diagrams 
must be evaluated in $n$-dimensions and a colour 
matrix must be added at the quark gluon vertex. We split the correction as follows
\begin{equation}
  \label{eq:17.7}
{{d\sigma^{(1)}_{q \bar q}}\over{d Q^2}} =
{{d\sigma^{(1)}_{q \bar q}}\over{d Q^2}} \vert_{\rm virtual}+ 
{{d\sigma^{(1)}_{q \bar q}}\over{d Q^2}} \vert_{\rm real} \,.  
\end{equation}
Incidentally, we use $n$-dimensional regularization also for infrared divergences,
and consider the quark and anti-quark
to be massless and on-shell. To see how this affects the loop
integrals, let present the result for
$J(t,0,0)$, the scalar vertex function, where $t = -Q^2$. It 
occurs in the first diagram in Fig.~\ref{fig:17.6}a and is 
defined by 
\begin{equation}
  \label{eq:8.29}
  J(t,0,0) = {{1}\over{(2\pi)^n}}
  \int \, {{\rd^nq}\over{((p+q)^2) ((p'+q)^2) (q^2) }} \,,
\end{equation}
where $p^2 = p^{\prime 2}=0$. The integral may be computed using standard
methods in dimensional regularization. The result is
\begin{equation}
  \label{eq:17.8}
J(t,0,0) = \im (4\pi)^{-\frac{n}{2}}\Big(\frac{-t}{\mu^2}\Big)^{\frac{n-6}{2}}
(\mu^2)^{\frac{n-6}{2}}
\frac{\Gamma(3-n/2)\Gamma^2(n/2-1)}{\Gamma\left(n-3\right)} 
\times\frac{4}{(n-4)^2} \,,
\end{equation}
where we have inserted a mass scale $\mu$ to make the integral
have the correct dimension in $n$ space-time dimensions.
Notice that the last factor shows a double pole in $n-4$, arising from the
overlap of an infrared and a collinear singularity, when the virtual
gluon both becomes soft and collinear to either the incoming
quark or anti-quark.
Also a two-denominator integral occurs when including numerator factors
in the leftmost graph of Fig.~\ref{fig:17.6}a. It reads
\begin{equation}
  \label{eq:8.23}
  I(k^2,0,0) = {{1}\over{(2\pi)^n}}
  \int \, {{\rd^nq}\over{((q+\ft12 k)^2)
      ((q-\ft12 k)^2) }}\,,
\end{equation}
with $k^2=Q^2$, and the result of doing the integral is 
\begin{equation}
  \label{eq:17.9}
I(-t,0,0) =  \im (4\pi)^{-\frac{n}{2}}\Big(\frac{-t}{\mu^2}\Big)^{\frac{n-4}{2}}
(\mu^2)^{n/2-2} 
\frac{\Gamma(3-n/2)\Gamma^2(n/2-1)}{\Gamma(n-2)}
\times \frac{2}{4-n} \,.
\end{equation}
Again it features a pole in $n-4$. Note that it is not always obvious from superficial inspection to see
whether a $1/(n-4)$ pole has an ultraviolet, infrared or collinear
origin. However, in general, a UV divergence occurs after the $n$-dimensional integral over
the loop momentum, while the IR and collinear singularities arise from the
integrations over the Feynman parameters. The full result for the vertex graph in Fig.~\ref{fig:17.6}a
including numerator factors, reads, after substituting $n=4+\varepsilon$ 
\begin{eqnarray}
  \label{eq:17.15}
\Lambda(p',p) & = & e^3\gamma_\mu  
\im (4\pi)^{-2}\Big({{-t}\over{4\pi \mu^2}}\Big)^{\varepsilon/2}
(\mu^2)^{\varepsilon/2} 
{{\Gamma(1-\varepsilon/2)\Gamma^2(1+\varepsilon/2)}\over{\Gamma
(2+\varepsilon)}}
\nonumber \\ &\ & \quad\quad
\times\left[{{8}\over{\varepsilon^2}}
+ {{2}\over{\varepsilon}}
+ 1 \right]\,.
\end{eqnarray}
Besides the vertex graph we should also consider other virtual
contributions, namely the one-loop gluon self-energy corrections to the 
incoming on-shell quark and anti-quark, as well as contributions from 
counterterms. However, none of these contribute to
the present calculation. To see this, consider first the self-energy 
contribution for an on-shell massless fermion
\begin{equation}
  \label{eq:17.601}
  \Sigma(p) = -\im \slash{p} - g^2 \int \frac{d^nq}{(2\pi)^n}
\frac{(-\im \slash{p})(-\im \slash{p} - 
\im \slash{q})(-\im \slash{p})}{(p+q)^2 q^2} \,.
\end{equation}
Using $p^2 = 0$ this reduces to 
\begin{equation}
  \label{eq:17.602}
  \Sigma(p) = -\im \slash{p} - \im \slash{p} \,
g^2 \int \frac{d^nq}{(2\pi)^n}
\frac{2p\cdot q}{(p+q)^2 q^2}\,.
\end{equation}
Writing
\begin{equation}
  \label{eq:17.603}
2p.q = (q+p)^2 - q^2 \,,
\end{equation}
we see that the $\mathcal{O}(g^2)$ correction vanishes, by the
rules of dimensional regularization, in which scaleless 
loop integrals may be consistently set to zero.

Besides the loop diagrams also the ${\cal O}(\alpha_s)$ counterterms in
the Lagrangian must be included in the virtual contributions, shown 
in Fig.~\ref{fig:17.6}b.
This is so even when a loop graph itself is zero, such as for the quark
and antiquark self energy corrections.  There
are in fact three counterterm diagrams in the Lagrangian, indicated
in Fig.~\ref{fig:17.6}b.  The quark colours,
when including initial quark colour averaging, here merely lead to a common factor
$C_2(R)/N_c$ for all three contributions. 
The counterterm contributions for the self energy corrections must be
included with a factor $1/2$ due to the need to normalize the scattering
amplitude using the residue at the pole. When one does this,
the counterterm contributions cancel agains each other.

So, remarkably, in the end only the triangle diagram contributes to the virtual
contribution, and we have the result 
\begin{eqnarray}
  \label{eq:17.18}
  {{d\sigma^{(1)}_{q \bar q}}\over{d \hat s}} \vert_{\rm virtual }  
  & = & \sigma^{(0)}_\gamma Q_f^2 {{1}\over{2\pi}} C_2(R) 
  \Big( {{4\pi\mu^2}\over{\hat s}}\Big)^{-\varepsilon/2}
  {{\Gamma(1+\varepsilon/2)}\over{\Gamma(1+\varepsilon)}}
  \nonumber \\ &\ & 
  \times \left[-{{8}\over{\varepsilon^2}} +
  {{6}\over{\varepsilon}} 
  -8 + {{2\pi^2}\over{3}} 
  + O(\varepsilon) \right] \delta\left(1-x \right)\,,
\end{eqnarray}
where $x = \hat{s}/s$. We have used the expansion
$\Re{(-1)^{\varepsilon/2}} = \Re{\exp(\varepsilon\im \pi/2)} \simeq 1 - \pi^2\varepsilon^2/8$. (We
dropped the imaginary part $\varepsilon\im \pi/2$ since we only need the real part of the virtual
contributions in the interference with the Born diagram.)
The other $\pi^2$ terms in (\ref{eq:17.18}) follow from expansion of
the Gamma functions
\begin{equation}
  \label{eq:17.17}
  \Gamma(1-\varepsilon/2)\Gamma(1+\varepsilon/2) = 1 + 
  {{\pi^2}\over{6}} {{\varepsilon^2}\over{4}} + O(\varepsilon^3) \,.
\end{equation}

Next we must consider the real gluon bremsstrahlung
graphs which as far as the partonic channel is concerned contribute to
the two-to-two body scattering cross section for $q(p_1) + \bar q(p_2)
\rightarrow \gamma(q) + g (k)$. Now since $\hat{s} = -(p_1 + p_2)^2$ is the
square of the total centre-of-mass energy and $Q^2=-q^2 =
(q_1+q_2)^2$ is  the invariant mass of the dilepton pair then $\hat{s}
\ne Q^2$.   It is convenient to rewrite the Mandelstam invariants
in terms of the two variables $x=  Q^2 / s$ 
and $y = (1+\cos \theta)/2$.  This yields the relations
$s = Q^2 /x$,  $\hat t =  - Q^2 (1-x)(1-y)/x$, and  
$\hat u =  - Q^2 (1-x)y/x$.
In this bremsstrahlung correction we will 
have contributions from the region $x = Q^2 /s < 1$,
whereas the virtual and counterterm diagrams only contribute
at $x=1$.

Let us introduce a
convenient shorthand notation for an $l$-particle $n$-dimensional
phase-space measure
\begin{equation}
  \label{eq:17.501}
\int_{\rm PSl} dq_1 \cdots dq_l = 
(2\pi)^{n+l(1-n)}
\int \frac{d^{n-1}q_1}{2q_1^0} \cdots
 \frac{d^{n-1}q_l}{2q_l^0} 
\, \delta^{(n)}(P-\sum_i^l q_i) \,,
\end{equation}
where we have left the masses of each particle unspecified.
We need to evaluate
\begin{equation}
  \label{eq:17.19}
{{d\sigma^{(1)}_{q \bar q}}\over{d Q^2}} \vert_{\rm real}  
= {{1}\over{8N_c^2\hat s}}
\int_{\rm PS3} dk dq_1 dq_2 
\delta^{(n)}(p_1+p_2-k-q_1-q_2)
\sum\vert{\cal M}\vert^2 \,,
\end{equation}
where ${\cal M}$ is the matrix element of the
two-to-three body reaction $q(p_1) + \bar q(p_2) \rightarrow
l(q_1) + \bar l(q_2) + g(k)$.
The three-body phase-space integral can be factorized
into two two-body phase-space integrals by inserting
\begin{equation}
  \label{eq:17.20}
1 = \int \frac{dQ^2}{2\pi} \int d^n q \delta^{(n)}(q-q_1-q_2) 
(2\pi) \delta(q^2-Q^2)\,,  
\end{equation}
into the integral. Then we write the integral over $dq$ as
a $n-1$ dimensional integral using $\delta(q^2 - Q^2)$. If we use
the notation $p=p_1+p_2$ then the integrals can be written as
\begin{equation}
  \label{eq:17.21}
{{d\sigma^{(1)}_{q \bar q}}\over{d Q^2}} \vert_{\rm real}  
= {{1}\over{16 \pi N_c^2\hat s}} 
\int_{\rm PS2} dk dq \int_{\rm PS2} dq_1 dq_2 
\sum\vert{\cal M}\vert^2 \,,
\end{equation}
where the first phase space-integral has a $\delta$-function
$\delta^{(n)}(p-q-k)$ and the second one
$\delta^{(n)}(q-q_1-q_2)$.
This enables us again to factor off the decay of the $\gamma^*$ into
the lepton-antilepton pair, leading to the equation (\ref{eq:17.701})
but now also for the real emission contribution.

The square of the partonic
matrix element summed over all initial and final spins
and polarizations can then be written in terms of the Mandelstam
invariants for the reaction $q(p_1) + \bar q(p_2)
\rightarrow \gamma (q) + g(k)$. These we will call  
$  s = -(p_1+p_2)^2 $ , 
$  t = -(p_1 - k)^2 $, and  
$  u = -(p_2 - k)^2 $, which
satisfy $  s +   t +   u = -Q^2$
Note that a term involving a new mass scale 
$\mu$ will be required because the
QCD coupling constant $g$ has mass dimension $(4-n)/2$
in $n$-dimensions. 
There is no need to write an $n$-dimensional 
generalization for the QED coupling constant so we
can keep $e$ in four dimensions.
The answer in terms of $n=4+ \varepsilon$ reads
\begin{eqnarray}
  \label{eq:17.25}
{{d\sigma^{(1)}_{q \bar q}}\over{d \hat{s} }} \vert_{\rm real}  
& = & \sigma^{(0)}_\gamma Q_f^2 {{1}\over{2\pi}} C_2(R)
\Big( {{4\pi\mu^2}\over{\hat s }}\Big)^{-\varepsilon/2}
{{\Gamma(1+\varepsilon/2)}\over{\Gamma(1+\varepsilon)}}
{{4}\over{\varepsilon}} 
\nonumber \\ &\ & \quad\quad
\times \left[ 2x^{1-\varepsilon/2} (1-x)^{-1+\varepsilon} +
x^{-\varepsilon/2}(1-x)^{1+\varepsilon}
\right] \,.  
\end{eqnarray}
A collinear pole in $\varepsilon$ resulting from
the angular integral is now explicit. 
If we integrate over the variable $x$, which we must to
form the hadronic cross section, 
a second pole will appear from the region $x \rightarrow 1$.
That is the infrared pole. After integration there are therefore
double pole terms from overlapping divergences and single pole
terms from the either soft or the collinear singularities.
Using the KLN theorem to cancel these pole terms against the contributions
from the virtual graphs, which only exist for $x = 1$ 
would be convenient, especially 
before doing the integration over $x$.  So, we would need a
way of combining the contributions from
the virtual and bremsstrahlung graphs as functions of $x$.

One way to do this is to split off a
small piece in (\ref{eq:17.25}) between $x=1-\delta$ and $x=1$
and call this the "soft" bremsstrahlung piece. 
In this small range near unity one can 
substitute $x=1$ whenever this is allowed and simply do the $x$
integral yielding terms in $\ln \delta$ as well as poles in
$\varepsilon$. These pieces can then be added to the contributions
from the virtual graphs.  The remaining "hard" bremsstrahlung
integral over the range $0$ to $1-\delta$ is finite, and can be take
in $n=4$ dimensions., 
Integration will then yield a term involving $\ln\delta$ which
should cancel with the corresponding $\ln\delta$ term in the
virtual graphs. This method is called the phase-space
slicing method.

We will employ another method.
We would like a
relation that expresses the double pole terms immediately in
terms of $\delta(1-x)$. Such a relation does exist but in the
sense of distributions, namely when multiplied by a smooth 
function $F(x)$ and integrated between $0$ and $1$ (like the
$\delta$-function itself). 
Assume the function $F(x)$ has a Taylor expansion near $x=1$
so we can write $F(x) = F(1) + F(x) - F(1)$, where the difference
between the last two terms is proportional 
to the finite derivative of $F(x)$ at $x=1$.
Let us consider therefore
\begin{eqnarray}
  \label{eq:17.26}
\int_0^1 \, dx\,  
{{F(x)}\over{(1-x)^{1-\varepsilon}}}
&=&
F(1) \int_0^1 \, dx\, 
{{1}\over{(1-x)^{1-\varepsilon}}}   
+ \int_0^1 \, dx\,  
{{F(x)-F(1)}\over{(1-x)^{1-\varepsilon}}} \,.
\nonumber \\ &\ & 
\end{eqnarray}
The first integral yields $F(1)\varepsilon^{-1}$. We can rewrite
this again as an integral over $dx$ with the argument $\delta(1-x)$.
In the second integral we can expand the denominator so it yields 
\begin{eqnarray}
  \label{eq:17.27}
\int_0^1 \, dx\,  
{{F(x)-F(1)}\over{(1-x)^{1-\varepsilon}}}
& = &
\int_0^1 \, dx\,  
{{F(x)-F(1)}\over{(1-x)}} 
\nonumber \\ &\ & \quad 
+ \varepsilon \int_0^1 \, dx \, 
[F(x)-F(1)]{{\ln(1-x)}\over{(1-x)}} + O(\varepsilon^2) \,,
\end{eqnarray}
near $\varepsilon=0$. 
Therefore we have the identity
\begin{eqnarray}
  \label{eq:17.28}
\int_0^1 \, dx \, {{F(x)}\over{(1-x)^{1-\varepsilon}}}
 & = &
{{1}\over{\varepsilon}} \int_0^1 \, dx \, F(x) \delta(1-x)
+
\int_0^1 \, dx \, {{F(x)-F(1)}\over{1-x}}
 \nonumber \\ &\ & \quad + \varepsilon 
\int_0^1 \, dx [\, F(x)-F(1)]{{\ln(1-x)}\over{1-x}} 
+O(\varepsilon^2)\,.
\end{eqnarray}
This we will write in shorthand notation as
\begin{equation}
  \label{eq:17.29}
{{1}\over{(1-x)^{1-\varepsilon}}}
 = {{1}\over{\varepsilon}} \delta(1-x) +
\left[{{1}\over{1-x}}\right]_+
+ \varepsilon \left[{{\ln(1-x)}\over{1-x}}\right]_+ 
+ O(\varepsilon^2) \,,
\end{equation}
where on the right hand side we see so-called "plus"
distributions. Note that this result is exact for a lower integration limit
$x=0$. If the lower limit is not zero then there are additional finite
terms 
involving logarithms of this lower limit.

Our final result for the gluon radiation graphs therefore
follows by expanding the terms in the square bracket
in (\ref{eq:17.25}) in powers of $\varepsilon$ and using (\ref{eq:17.29}).
We find
\begin{eqnarray}
  \label{eq:17.30}
{{d\sigma^{(1)}_{q \bar q}}\over{d \hat{s} }} \vert_{\rm real}  
& = & \sigma^{(0)}_\gamma Q_f^2 {{1}\over{2\pi}} 
\Big( {{4\pi\mu^2}\over{\hat s }}\Big)^{-\varepsilon/2}
{{\Gamma(1+\varepsilon/2)}\over{\Gamma(1+\varepsilon)}} 
\left[ {{8}\over{\varepsilon^2}} \delta(1-x)   
\right.\nonumber \\ &\ &\quad \left.
+{{4}\over{\varepsilon}} (1+x^2)\left[{{1}\over{1-x}}\right]_+
+ 4 (1+x^2) \left[{{\ln 1-x}\over{1-x}}\right]_+
\right.\nonumber \\ &\ &\quad \left.
-2(1+x^2) {{\ln x }\over{ 1-x}}  
+ O(\varepsilon)\right]\,.
\end{eqnarray}

Now we have isolated the term in $\delta(1-x)$ containing the 
double pole we see that it cancels the corresponding term from 
the virtual graphs in (\ref{eq:17.18}). 
These are the overlap terms containing both soft and
collinear divergences and they cancel as expected from the
KLN theorem. The single pole term in $\varepsilon$
however cannot possibly cancel against a virtual contribution, 
as it is not purely a $\delta(1-x)$ term. Therefore we are left with
an uncancelled collinear singularity.

Finally we can finally sum (\ref{eq:17.18}) and (\ref{eq:17.30}) and find
\begin{eqnarray}
    \label{eq:17.31}
{{d\sigma^{(1)}_{q \bar q}}\over{d \hat{s}}}  
& = & \sigma^{(0)}_\gamma Q_f^2 {{1}\over{2\pi}} C_2(R) 
\Big( {{4\pi\mu^2}\over{\hat s}}\Big)^{-\varepsilon/2}
{{\Gamma(1+\varepsilon/2)}\over{\Gamma(1+\varepsilon)}}
\nonumber \\ &\ & \times \Big\{ 
{{4}\over{\varepsilon}} \Big((1+x^2)\left[{{1}\over{1-x}}\right]_+
+{{3}\over{2}}\delta(1-x) \Big) +
4 (1+x^2) \left[{{\ln (1-x)}\over{1-x}}\right]_+ \nonumber \\ &\ & \quad  
-2(1+x^2) {{\ln x }\over{ 1-x}} + 
(4 \zeta(2) -8) \delta(1-x)
+ O(\varepsilon) \Big\} \,,
\end{eqnarray}
with $\zeta(2) = \pi^2/6$.
The remaining pole term in $\varepsilon$  implies that the
KLN theorem is inoperable when there are collinear 
singularities in the initial partonic state. How are we then
going to make sense of this result?

First, let us observe that if one expands all functions in 
(\ref{eq:17.31}) in $\varepsilon$ one finds
\begin{align}
  \label{eq:17.300}
{{d\sigma^{(1)}_{q \bar q}}\over{d \hat{s}}} & = 
\sigma^{(0)}_\gamma Q_f^2 {{1}\over{2\pi}} C_2(R) 
2 \left({2\over{\varepsilon}} - \ln 4\pi + \gamma_E \right)
 \Big((1+x^2)\left[{{1}\over{1-x}}\right]_+
\nonumber \\ &\hspace{2mm} +{{3}\over{2}}\delta(1-x) \Big) 
+ \mathcal{O}(\varepsilon^0) \nonumber \\
& = \sigma^{(0)}_\gamma Q_f^2 {{1}\over{2\pi}} C_2(R) 
2 \left({{2}\over{\varepsilon}} - \ln 4\pi + \gamma_E \right)
\left[\frac{1+x^2}{1-x}\right]_+ + \mathcal{O}(\varepsilon^0) \,. 
\end{align}
Next, we realize that this expression should be substituted
into the convolution (\ref{eq:17.5}). At this point one may, in a sense,
\emph{renormalize} (or rather: \emph{factorize})
the parton distributions in (\ref{eq:17.5}) as 
\begin{equation}
  \label{eq:17.301}
  f_{q/A}(\xi) = \int_0^1 dz \int_0^1 dy
  f_{q/A}(y,\mu_F) \Phi^{-1}_{qq}(z,\mu_F) \delta(\xi-zy) \,,
\end{equation}
with $\mu_F$ the factorization scale, introduced in the previous
section, and $\Phi_{qq}$ a transition function. This fucntion is
analogous to the $Z$-factors for UV renormalization in section \ref{sec:renormalization}.

To first order, 
the above relation can be written as
\begin{multline}
  \label{eq:17.302}
  f_{q/A}(\xi) = f_{q/A}(\xi,\mu_F) -
\int_{\xi}^1 \frac{dz}{z} f_{q/A}\left(\frac{\xi}{z},\mu_F\right) \\
\times \left\{{\alpha_s(\mu)C_2(R)\over 2 \pi}{1\over \varepsilon}
\left(\frac{4\pi\mu^2}{\mu_F^2}\right)^{-\varepsilon/2}
\left[{1+z^2\over 1-z}\right]_+ \right\} \,.
\end{multline}
Collecting terms we see indeed, as we announced, the 
collinear singularities cancel after renormalization of
the parton distribution by the transition functions, leaving a finite
remainder. The final result is
\begin{eqnarray}
    \label{eq:17.303}
{{d\sigma^{(1)}_{q \bar q}}\over{d \hat{s}}}  
& = & \sigma^{(0)}_\gamma Q_f^2 {{1}\over{2\pi}} C_2(R) 
\nonumber \\ &\ & \times \Big\{ 
2\ln\left(\frac{Q^2}{\mu_F^2} \right) \left[{1+z^2\over 1-z}\right]_+ +
4 (1+x^2) \left[{{\ln (1-x)}\over{1-x}}\right]_+ \nonumber \\ &\ & \quad  
-2(1+x^2) {{\ln x }\over{ 1-x}} + 
(4 \zeta(2) -8) \delta(1-x)
\Big\} \,.
\end{eqnarray}
This result we can now insert into (\ref{eq:17.5}), use NLO PDF's and predict 
the Drell-Yan cross section.

\subsection{Factorization}
\label{sec:factorization}

The fact that the initial state divergences cancel through a
renormalization/factorization of the PDFs, as in (\ref{eq:17.301}) is
a one-loop manifestation
of the QCD factorization theorem \cite{Collins:1989gx}. This is the full QCD generalization of
the parton model formula, and states that for IR safe cross sections,
the initital state collinear divergences can be consistently removed
in this way. The consistency lies in the fact that this
factorization does not depend on the process, i.e. that it is \emph{always the same
set} of $\Phi_{ij}$ functions, computed to the appropriate order. 
This aspect is the one that preserves predictive power: indeed if we
devote certain set of observables to infer the PDF's, as we discussed extensively in
section \ref{sec:part-distr-funct}, we can use these for any other reaction
and predict the outcome. To cover the details of the factorization proof for the
inclusive Drell-Yan cross section would take us too far. However, it is 
worthwhile to point out that factorization proofs for other observables (differential
cross sections, cross sections near phase space edges, or with vetoed phase space
regions) are an active and important area of research \cite{Forshaw:2012bi,Catani:2012iw}.

\section{Modern methods}
\label{sec:modern-methods}

In this section I discuss a number of modern methods in the application
of perturbative QCD, focussing mostly on spinor helicity techniques,
and the essence of the recent ``NLO revolution''. For lack of space 
I shall not discuss the enormous strides made in Monte Carlo methods and
applications in recent years.

\subsection{Spinor methods, recursion relations}
\label{sec:spin-meth-recurs}
\def\half{{\scriptstyle\frac{1}{2}}}
\def\upl{u_+}
\def\umi{u_-}
\def\vpl{v_+}
\def\vmi{v_-}
\def\ucpl{u^c_+}
\def\ucmi{u^c_-}
\def\vcpl{v^c_+}
\def\vcmi{v^c_-}
\def\barup{\overline{u_+}}
\def\barum{\overline{u_-}}
\def\barvp{\overline{v_+}}
\def\barvm{\overline{v_-}}
\newcommand\hlp[1]{\langle#1\!+\! |}
\newcommand\hlm[1]{\langle#1\! -\! |}
\newcommand\hrp[1]{ #1 + \rangle}
\newcommand\hrm[1]{ #1 - \rangle}
\newcommand\spm[1]{\langle #1 \rangle}
\newcommand\spp[1]{[ #1 ]}
%%%%%%%%%%%%%%%%%%%%%%%%%%%%%%%%%%%%%%%%%%%%%%%%%%%%%%%%%%%%
\noindent %%%%%%%%%%%%%%%%%%%%%%%%%%%%%%%%%%%%%%%%%%%%%%%%%%%%%
At high center-of-mass energies, final states produced
in particle colliders usually contain many more than two
particles. Calculations of such processes are long and complicated
because one must write down the individual amplitudes for the Feynman
graphs and then square the result, which involves all the cross
products between them. In this section we describe methods to shorten
these calculations by using clever choices for external line
polarizations and simplifications owing to the masslessness of the
particles. 
We also note that at high energies most of the final state particles
can be considered massless, so that in order to represent fermions we
may make use of a chiral spinor basis because at large momenta
chirality and helicity are related. In
that case many external helicities configurations are in fact simply not allowed by
parity invariance. There are moreover many relations among the amplitudes
so the number of amplitudes to compute is not overly large.
An interesting thing to note is that by specifying all external line quantum numbers, the
expression for each helicity
amplitude is simply a complex number.
This can then obviate the need for analytically spin-summing over the absolute value squared of the invariant
amplitudes, and allow this task to be handled by a computer, reducing
the amount of laborious computation further.

Let us see how the use of spinors of definite
chirality or helicity can significantly simplify the calculation of
Feynman diagrams with massless fermions and gauge bosons. We will
also use the freedom of gauge choice for external gauge fields to
maximal advantage. I try to give a reasonably explicit and self-contained presentation
of these helicity spinor methods. We shall need  
the Dirac gamma matrices $\gamma^\mu$, $\gamma^5$ and the charge
conjugation matrix $C$  in the Weyl basis:
\begin{align}
  \label{eq:7.2.1}
  \gamma^k = &\,
  \begin{pmatrix}
  0 &   \sigma_k \\[2mm]
   \sigma_k & 0
\end{pmatrix}
\,, \quad k=1,2,3; \qquad
\gamma^0 =
\begin{pmatrix}
0  &  \mathbbm{1} \\[2mm]
 - \mathbbm{1} & 0
\end{pmatrix}\,, \nonumber\\[3mm]
  \gamma^5= &\,\gamma_5 = - \mathrm{i}\gamma^0
  \gamma^1 \gamma^2 \gamma^3
  = \begin{pmatrix}
    \mathbbm{1} & 0 \\[2mm]
    0 & - \mathbbm {1}
\end{pmatrix} \,,   C = {\rm i}\gamma^1 \gamma^3 = \begin{pmatrix}
       \sigma_2  & 0 \\[2mm]
       0 & \sigma_2
     \end{pmatrix}
 \,.
\end{align}
The explicit form of the $u$ and $v$ spinors in this basis is
\begin{align}
  \label{eq:weyl-u-v}
  u(\bm{P},\xi) =
  \frac{\mathrm{e}^{\mathrm{i}\pi/4}}{ \sqrt{2(m+\omega(\bm{P})) }}
  \begin{pmatrix}
     \big[(m+\omega(\bm{P}))\mathbbm{1} -
     \bm{P}\cdot\bm{\sigma}\big] \xi \\[4mm]
         -\mathrm{i} \big[(m+\omega(\bm{P}))\mathbbm{1} +
     \bm{P}\cdot\bm{\sigma}\big]\xi
    \end{pmatrix} \,, \nonumber\\[4mm]
%%%%%%
  v(\bm{P},\bar\xi) = \frac{\mathrm{e}^{\mathrm{i}\pi/4}}{\sqrt{2(m+\omega(\bm{P}))}}
  \begin{pmatrix}
        - \big[(m+\omega(\bm{P}))\mathbbm{1} -
     \bm{P}\cdot\bm{\sigma}\big] \bar\xi \\[4mm]
         - \mathrm{i}\big[(m+\omega(\bm{P}))\mathbbm{1} +
     \bm{P}\cdot\bm{\sigma}\big] \bar\xi
   \end{pmatrix}
   \,,
\end{align}
with $\bar\xi=\mathrm{i}\sigma_2 \xi^\ast$. The momentum $P^\mu
=\omega(\bm{P}),\bm{P})$ is the on-shell momentum of the fermion,
and the charge conjugation matrix is used to define the charge
conjugate spinor
\begin{equation}
  \label{eq:17}
  \psi^{\rm c} \equiv C^{-1} \bar \psi^{\rm T} \,.
\end{equation}

Having this explicit form will allow
us to derive a number of useful identities which make calculations with massless
particle must more efficient. The following identity,
\begin{equation}
  \label{eq:square}
  \big[(m+\omega(\bm{P}))\mathbbm{1} \pm
     \bm{P}\cdot\bm{\sigma}\big]^2 = 2\big(m+\omega(\bm{P})\big)\,
     \big[\omega(\bm{P})\,\mathbbm{1} \pm \bm{P}\cdot\bm{\sigma}\big] 
\end{equation}
suggests that there is a more systematic way to write these spinors. To
see this let us define $\sigma_\mu$ and $\bar{\sigma}_\mu$ as
four-vector arrays of $2\times2$ hermitian matrices,
\begin{equation}
  \label{eq:6n.3}
  \sigma_\mu = (-\mathbbm{1},\bm{\sigma}), \quad
  \bar\sigma_\mu = (-\mathbbm{1},-\bm{\sigma}), \quad  (P\cdot\sigma)(P\cdot
  \bar{\sigma}) = -P^2 = m^2\,.
\end{equation}
In terms of these matrices one has the identities
\begin{align}
  \label{eq:6n.4}
    (\im \Slash{P} \pm m)=&\, \begin{pmatrix}
      \pm m \mathbbm{1} & \im P\cdot \sigma \\[2mm]
     -\im P\cdot \bar\sigma & \pm m \mathbbm{1}
    \end{pmatrix}
    \,,  \nonumber\\
    - P^\mu \sigma_\mu=&\, \omega(\bm{P})\,\mathbbm{1} -
    \bm{P}\cdot\bm{\sigma} \,,    \nonumber \\
    - P^\mu\bar\sigma_\mu=&\, \omega(\bm{P})\,\mathbbm{1} +
    \bm{P}\cdot\bm{\sigma} \,.
\end{align}
Observe that $-P\cdot\sigma$ and $-P\cdot \bar\sigma$ are hermitian
positive definite matrices with eigenvalues equal to $\omega(\bm{P})
\pm \vert\bm{P}\vert$.

Let us now consider the case of massless spinors. In that case the
matrices $-P\cdot \sigma$ and $-P\cdot \bar\sigma$ have one zero
eigenvalue and become equal to $2\,\omega(\bm{P})$ times a projection
operator, as follows from (\ref{eq:square}). Indeed, the massless
limit of (\ref{eq:weyl-u-v}) equals
\begin{align}
  \label{eq:massless-weyl-u-v}
  u(\bm{P},\xi) = \frac{\mathrm{e}^{\mathrm{i}\pi/4}}{\sqrt{2\omega(\bm{P})}}
  \begin{pmatrix}
      (-P\cdot\sigma)  \,\xi \\[4mm]
         -\mathrm{i}(-P\cdot\bar\sigma) \,\xi
    \end{pmatrix} \,, \nonumber\\[4mm]
%%%%%%
  v(\bm{P},\bar\xi) = \frac{\mathrm{e}^{\mathrm{i}\pi/4}}{\sqrt{2\omega(\bm{P})}}
  \begin{pmatrix}
         -(-P\cdot \sigma)\, \bar\xi \\[4mm]
         - \mathrm{i}(- P\cdot\bar\sigma)\,  \bar\xi
   \end{pmatrix}
   \,.
\end{align}
Before proceeding, let us introduce the \index{light-cone basis}
light-cone basis for a generic massless momentum $p^\mu$.
In this basis the components $p^0$ and $p^3$ are replaced by
\begin{equation}
  \label{eq:7.42}
  p^+ = {p^0+p^3\over \sqrt{2}},\qquad p^- = {p^0-p^3\over \sqrt{2}},
\end{equation}
where the two remaining components are denoted by the two-component
vector $p_\perp=(p^1,p^2)$. In this basis
\begin{equation}
  \label{eq:7.43}
  p^2 = -2\,p^+ p^- + p_\perp{}^2\,.
\end{equation}
The advantage of this basis is clear when considering a massless
particle moving along the $3$-axis. In the standard basis the momentum
four-vector has two non-zero components, namely $p^0$ and $p^3$, but
in the light-cone basis there is only one non-vanishing component
(i.e. either $p^+$ or $p^-$), which helps with the calculations as
we will see below.

The positive frequency solution is degenerate and can be further
classified using the chirality projectors $P_\mathrm{L} =
(1+\gamma_5)/2$ and $P_\mathrm{R} = (1-\gamma_5)/2$, which project
onto the upper and lower two components of the spinor,
respectively. Thus we have the left- and right-handed solutions
\begin{equation}
  \label{eq:6n.8}
      u_\mathrm{L}(\bm{P},\xi) =  \frac{\mathrm{e}^{\mathrm{i}\pi/4}}{\sqrt{2\omega(\bm{P})}}  \left(\begin{matrix}
    (-P\cdot \sigma)\,\xi \\0 \\ \end{matrix}
  \right), \qquad
      u_R(\bm{P},\xi) = \frac{\mathrm{e}^{-\mathrm{i}\pi/4}}{\sqrt{2\omega(\bm{P})}}   \left(\begin{matrix}
   0\\   (-P\cdot \bar\sigma) \, \xi \\ \end{matrix}
  \right)\,,
\end{equation}
and likewise for the spinors $v_\mathrm{L}$ and $v_\mathrm{R}$.
We can specify further the two-component spinors $\xi$. We
note that $-P\cdot \sigma$ and $-P\cdot \bar\sigma$ project onto
negative and positive helicity eigenstates, respectively. For instance, from
\begin{equation}
  \label{eq:6n.8.1}
  (-P\cdot \sigma) \xi = (|\vec{P}|-\vec{P}\cdot \vec{\sigma})\xi\,,
\end{equation}
we see that the right hand side is only non-zero for $\xi$
a negative helicity, $\xi_-$. Because $-P\cdot \sigma$ and
$-P\cdot \bar\sigma$ are projectors, we can, without loss of
generality, choose a convenient basis for the $\xi_{\pm}$ spinors
independent of momentum.
We choose $\xi_-$ ($\xi_+$) such that,
in the frame where $\vec{P}$ is along the $z$-axis, $u_L$ ($u_R$) has
$j_3$ eigenvalue $-\frac{1}{2}$ ($+\frac{1}{2}$),
in correspondence with the helicity-chirality relation \index{helicity vs. chirality}
 $2h = -\gamma_5$.
 We thus choose
\begin{equation}
  \label{eq:6n.9}
  \xi_- =
\left(\begin{matrix}
   0\\ 1\\\end{matrix}
  \right),
\qquad
  \xi_+ =
\left(\begin{matrix}
   1\\ 0\\\end{matrix}
  \right)\,.
\end{equation}
In this case we have for $u_L(P,\xi)$ and  $u_R(P,\xi)$
\begin{equation}
  \label{eq:6n.10}
      u_L(\bm{P},\xi) = \frac{e^{\im \pi/4}}{\sqrt{2P^0}} \left(\begin{matrix}
    -P_T^\ast \\\sqrt{2} P^+ \\0\\0\\ \end{matrix}
  \right)\,,
\qquad
      u_R(\bm{P},\xi) = \frac{e^{-\im \pi/4}}{\sqrt{2P^0}} \left(\begin{matrix}
 0\\0\\ \sqrt{2} P^+  \\P_T\end{matrix}
  \right)
\,.
\end{equation}
For the rest of this section we change from chirality to helicity
labels, and allow for a change in normalization
\begin{equation}
  \label{eq:6n.12}
  u_L \equiv  \frac{1}{c_-} u_-\,, \quad
  u_R \equiv \frac{1}{c_+} u_+\,.
\end{equation}

Once can show that in order to have $u_\pm(P)^\dagger u_\pm(P) = 2P^0$
one must, up to phases, choose $c_- = 2^{1/4} \sqrt{P^0/P^+}$ and
$c_+ = 2^{1/4} \sqrt{P^0/P^-}$.
We choose the phases of $u_{\pm}$ now such that
\begin{equation}
  \label{eq:6n.13}
      u_-(\bm{P},\xi) = 2^{1/4} \left(\begin{matrix}
    -\sqrt{P^-}e^{-\im \phi_p} \\ \sqrt{P^+} \\0\\0\\ \end{matrix}
  \right)\,, \qquad
      u_+(\bm{P},\xi) = 2^{1/4} \left(\begin{matrix}
   0\\0\\ \sqrt{P^+} \\ \sqrt{P^-} e^{\im \phi_p}\\ \end{matrix}
  \right)\,.
\end{equation}
where the phase $\phi_p$ is defined through
\begin{equation}
  \label{eq:6n.11}
  P_T = e^{\im \phi_p}\sqrt{2P^+P^-}\,.
\end{equation}
Having constructed quite explicit forms for helicity spinors we now use
them to derive useful computational rules. Arguments of spinors we
now indicate with lower-case four-momenta.
To begin, we define \emph{spinor products}
together with bra-ket notation, as follows
\begin{equation}
  \label{eq:6n.14}
\im\overline{u_-}(k) u_+(p) \equiv  \hlm{k} \hrp{p} \equiv \spm{kp} \,,
\end{equation}
and
\begin{equation}
  \label{eq:6n.15}
\im  \overline{u_+}(k) u_-(p) \equiv  \hlp{k} \hrm{p} \equiv \spp{kp}\,.
\end{equation}
One may show that
\begin{equation}
  \label{eq:p.exer.6.5.1}
\langle kp \rangle = (e^{\im \phi_k} \sqrt{2k^-p^+}
-e^{\im \phi_p} \sqrt{2k^+p^-})\,,
\end{equation}
and
\begin{equation}
  \label{eq:p.exer.6.5.2}
[kp ] =
\langle kp \rangle = (e^{-\im \phi_k} \sqrt{2k^-p^+}
-e^{-\im \phi_p} \sqrt{2k^+p^-})\,,
\end{equation}
so that
\begin{equation}
  \label{eq:p.exer.6.5.3}
\langle kp \rangle = - \langle pk \rangle, \quad
[ kp ] = - [ pk ], \quad \langle kp \rangle^* = [ kp ]\,,
\end{equation}
and
\begin{equation}
  \label{eq:p.exer.6.5.4}
  \langle kp \rangle[ kp ] = -2k \cdot p \,.
\end{equation}
The real benefits of working with helicity spinors come to the fore
when also the polarization vectors $\varepsilon^\mu(k,\lambda)$ of massless spin-1
particles with momentum $k$ and helicity $\lambda$ are
expressed in terms of them.  To see how this works, we first choose
the frame in which the on-shell massless gauge boson momentum has
only a $+$ component.
In this frame only the $\varepsilon^1$ and $\varepsilon^2$ components
are meaningful, corresponding to the two helicity states
of the massless vector field.
For the chosen frame the third component of the spin is identical to
the helicity, and the transversality condition $k\cdot
\varepsilon(k)=0$ becomes
\begin{equation}
  \label{eq:7.69}
  k^+ \varepsilon^- (k,\lambda)=0\,,
\end{equation}
which implies that $\varepsilon^0=\varepsilon^3$.
From this explicit solution one observes
that $(\varepsilon_\mu(k,+))^*$ has negative helicity.
Our normalization is such that
\begin{equation}
  \label{eq:7.75}
(\varepsilon_\mu(k,+))^* =\varepsilon_\mu(k,-)\,.
\end{equation}
and
\begin{equation}
  \label{eq:7.76}
  \varepsilon(k,+)\cdot \varepsilon(k,-) = -1  \,.
\end{equation}
We will also use the notation
\begin{equation}
  \label{eq:7.77}
  \varepsilon^\mu(k,\pm) = \varepsilon_\pm^\mu(k) \,.
\end{equation}
Let us now demonstrate that we may write the polarization vector
indeed in terms of spinors of fixed helicity, as follows
\begin{equation}
  \label{eq:7.78}
  \varepsilon_+^{\mu}(k,p)
= A_+ \barup(k,+) \gamma^\mu u(p,+)
\equiv -\im A_+ \hlp{k} \gamma^\mu |\hrp{p}\,,
\end{equation}
and similarly for negative helicity.
Note the extra momentum $p$, called the \index{reference momentum}
\emph{reference momentum},  of the second $u$ spinor.
It is in fact arbitrary, with $p^2=0$, we will discuss it further below.
From the explicit form of the helicity spinors in (\ref{eq:6n.13})
and the form of the solutions one may derive
\begin{equation}
  \label{eq:p.exer.6.6.1}
A_+ = \frac{-\im}{\sqrt{2}\langle kp \rangle}\,, \quad
A_- = \frac{-\im}{\sqrt{2}[kp]}\,.
\end{equation}
Recall that any multiple of $k^\mu$ may be added to the expressions
for the photon polarizations without changing the amplitude, as this is just a gauge transformation.

From the explicit form of the $u$ and $v$ spinors (\ref{eq:6n.13}),
one can prove the following series of identities
\begin{align}
  \label{eq:p.exer.6.7.1}
& \hlp{k} \hrp{p}   = 0\,,\\
& \hlp{k} \gamma^\mu | \hrm{p} = \hlp{k} \gamma_5 | \hrp{p}  = 0\,, \\
& \hlp{k} \gamma^\mu | \hrp{k} = 2 k^\mu\,, \\
& \hlp{k} \gamma^\mu | \hrp{p} =  \hlm{p} \gamma^\mu | \hrm{k}\,,
\end{align}
and similarly with all helicities reversed.
These identities are remarkably useful in practical calculations with
helicity spinors. Another very important property for this is Fierz
reordering, with which one may ``recouple'' the spinors.
Consider the following expression
\begin{equation}
  \label{eq:7.89}
\hlp{1} \gamma^\mu | \hrp{2} \hlm{3}  \gamma_\mu| \hrm{4}\,,
\end{equation}
where we have abbreviated $|\hrp{k_1} = |\hrp{1}$ etc. This is in fact
the most general form for such a contraction of spinor products.
Let us define the following complete set of 16 matrices 
\begin{equation}
  \label{eq:7.90}
  O_I = \left\{ 1,\gamma_5,\gamma^\mu, \im\gamma^\mu\gamma_5,
    \sigma^{\mu\nu}\right\}\,,
\end{equation}
with the orthogonality property
\begin{equation}
  \label{eq:6n.16}
{1\over 4} {\rm Tr}\left[ O_I O_J\right] = \delta_{IJ}
 \,,\;\; I=1\ldots 5 \,.
\end{equation}
We can insert this relation into (\ref{eq:7.89}), which may then be written as
\begin{equation}
  \label{eq:6n.17}
\frac{1}{4}\sum_I \hlp{1} \gamma^\mu O_I \gamma_\mu |\hrm{4} \hlm{3}
O_I| \hrp{2} \,.
\end{equation}
Because of the chirality properties of the bra's and kets, only the two diagonal $O_I$ yield a non-zero
result, $1$ and $\gamma_5$, and they moreover yield the same result. 
Hence the Fierz recoupling identity reads simply
\begin{equation}
\label{eq:6n.18}
\hlp{1} \gamma^\mu | \hrp{2} \hlm{3}  \gamma_\mu| \hrm{4}
= 2 \hlp{1}\hrm{4}\hlm{3}\hrp{2}\,,
\end{equation}
where e.g. $\hlp{1}$ has been recoupled to $|\hrm{4}$ in the spinor product.
With this identity we can now check the normalization of the
polarization vectors and find
\begin{equation}
  \label{eq:7.88}
  \varepsilon^+(k,p)\cdot \varepsilon^-(k,p)
= -A_+ A_-
\hlp{k}\gamma^\mu |\hrp{p}
\hlm{k} \gamma_\mu | \hrm{p}   = 1 \,,
\end{equation}
and similarly that $\varepsilon_\pm(k,p)\cdot \varepsilon_\pm(k,p)=0$.
The identities involving the sum over spin polarizations
read in terms of helicity spinors
\begin{equation}
  \label{eq:7.97}
  \slash{k}= |\hrp{k} \hlp{k} + |\hrm{k} \hlm{k}\,.
\end{equation}
One can derive the completeness relation for the
polarization vectors in the representation (\ref{eq:7.78})
\begin{equation}
  \label{eq:p.pro.6.6.2}
  \sum_{\lambda=\pm} \varepsilon^\mu_\lambda(k,p)
  \left(\varepsilon^\nu_\lambda(k,p) \right)^* =
\eta^{\mu\nu}-\frac{p^\mu k^\nu + p^\nu k^\mu}{p\cdot k}\,,
\end{equation}
and that a change in reference momentum amounts to a different gauge choice
\begin{equation}
  \label{eq:7.98}
  \varepsilon_+^\mu(k,p)-  \varepsilon_+^\mu(k,q)
 = \frac{\sqrt{2}\langle pq \rangle}{\langle kp \rangle \langle kq \rangle}
 k^\mu \,.
\end{equation}
We have now sufficient ingredients to demonstrate the efficiency
of using helicity spinors in computing invariant amplitudes for a
few examples. For each amplitude we shall discuss the result
for various sets of helicities for the external particles. We shall
also take each external particle as massless so that helicity
is a conserved quantum number. For convenience we choose momenta
of the external particles outgoing, and express possible anti-fermion
spinors in terms of $u$ spinors using $v_+=u_-$ and $v_-=u_+$.

\subsubsection*{The reaction ${\rm e}^+{\rm e}^- \rightarrow \mu^+ \mu^-$  }
\label{sec:rm-e+rm-e-1}

We first consider the reaction
\begin{equation}
\label{eq:6n.116}
  e^-(k_1) +   e^+(k_2) \rightarrow   \mu^-(k_3) +   \mu^+(k_4) \,,
\end{equation}
mediated via a photon. The invariant amplitude may be represented as
\begin{equation}
  \label{eq:6n.117}
  {\cal M}(1^{\lambda_1},2^{\lambda_2},3^{\lambda_3},4^{\lambda_4})\,,
\end{equation}
where we have indicated only the label of each external
line momentum, and the associated helicity. Using the rules derived in
this section we have
\begin{equation}
  \label{eq:6n.118}
    {\cal M}(1^+,2^-,3^+,4^-) = (\im e)^2 \hlm{2} \gamma^\mu | \hrm{1}
    \frac{-\im}{s_{12}} \hlp{3} \gamma_\mu |\hrp{4}\,,
\end{equation}
where we used the notation $s_{ij} = -(k_i+k_j)^2$. Using the Fierz
identity (\ref{eq:6n.18}) and the shorthand notation of
(\ref{eq:6n.14}) and (\ref{eq:6n.15}) this can be written as
\begin{equation}
  \label{eq:6n.119}
      {\cal M}(1^+,2^-,3^+,4^-) = 2\im e^2 \frac{\spp{24}\spm{31}}{\spm{12}\spp{12}}\,.
\end{equation}
Using momentum conservation this may be further rewritten as
\begin{equation}
  \label{eq:6n.120}
      {\cal M}(1^+,2^-,3^+,4^-) = 2\im e^2 \frac{\spp{13}^2} {\spp{12}\spp{34}}\,.
\end{equation}
It may be readily verified that
\begin{equation}
  \label{eq:6n.121}
      {\cal M}(1^-,2^+,3^-,4^+) = 2\im e^2 \frac{\spm{13}^2} {\spm{12}\spm{34}}\,.
\end{equation}
The expressions in (\ref{eq:6n.120}) and (\ref{eq:6n.121}) are quite
compact, and can be transformed into each other by either a parity
transformation or a charge conjugation.
For any other helicity configuration the amplitude actually vanishes.

\subsubsection*{The reaction ${\rm e}^+{\rm e}^- \rightarrow \mu^+\mu^- \gamma$  }
\label{sec:rm-e+rm-e-2}

\begin{figure}[htbp]
  \begin{center}
 \includegraphics[scale=0.6]{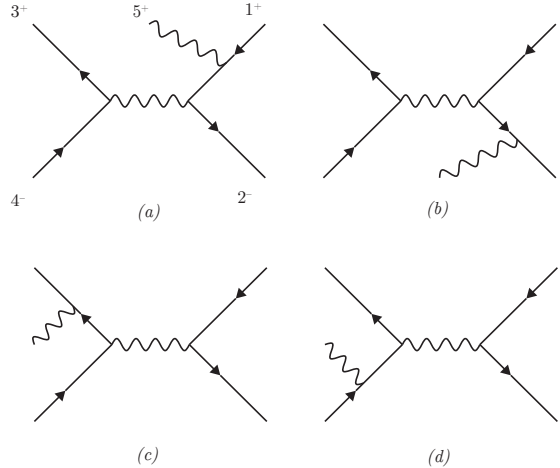}%
   \caption{Feynman diagrams contributing a particular helicity
     amplitude, indicated in diagram (a), for ${\rm e}^+{\rm e}^-
     \rightarrow \mu^+\mu^- \gamma$ at lowest order. All momenta are outgoing, and
time runs from right to left.}
\label{fig:6.16}
  \end{center}
\end{figure}

In this second example we study the production of a muon pair together
with a photon in electron positron annihilation
\begin{equation}
  \label{eq:6nn.17}
    e^-(k_1) +   e^+(k_2) \rightarrow
   \mu^-(k_3) +   \mu^+(k_4) +   \gamma(k_5)
\end{equation}
The photon can be radiated off any of the four external fermion lines,
leading to the four diagrams shown in Fig.~\ref{fig:6.16}.
%%%%%%%%%%%%%%%%%%%%%%%%%%%%%%%%%%%%%%%%%%%%%%%%%%%%%%%%
%%%%%%%%%%%%%%%%%%%%%%%%%%%%%%%%%%%%%%%%%
Let us list another useful identity, not difficult to prove, for a positive helicity massless vector boson
with polarization $\varepsilon_+^\mu(k,p)$ emitted from a fermion by
\begin{equation}
    \label{eq:p.exer.6.9.1}
\slash{\varepsilon}_+(k,p)=  \frac{\im\sqrt{2}}{\langle  k p\rangle} \left(|\hrp{p}\hlp{k}+|\hrm{k}\hlm{p}  \right)\,.
\end{equation}
For the negative helicity case one has, in analogy
\begin{equation}
    \label{eq:p.exer.6.9.2}
\slash{\varepsilon}_-(k,p)=  \frac{-\im \sqrt{2}}{[ k p ]} \left(|\hrp{k}\hlp{p}+|\hrm{p}\hlm{k}  \right)\,.
\end{equation}
The invariant amplitude reads
\begin{equation}
  \label{eq:6n.122}
  {\cal M}(1^{\lambda_1},2^{\lambda_2},3^{\lambda_3},4^{\lambda_4}, 5^{\lambda_5})\,.
\end{equation}
Given that each external line can have two helicity values, it might seem
that this process allows thirty-two different independent helicity amplitudes.
However, helicity conservation and invariance under charge
conjugation and parity transformation ensure that there is in fact only one
independent amplitude. We thus consider the helicity amplitude
\begin{equation}
  \label{eq:6n.123}
    {\cal M}(1^+,2^-,3^+,4^-,5^+) \,,
\end{equation}
and choose as $k_4$ as reference momentum for the outgoing
photon. Diagram (a) then reads
\begin{equation}
  \label{eq:6n.124}
      {\cal M}_a(1^+,2^-,3^+,4^-,5^+)  = (\im e)^3 \hlm{2}\gamma^\mu
\frac{-\im(\slash{1}+\slash{5})}{-s_{15}}
      \slash{\varepsilon}_+(k_5,k_4) |\hrm{1} 
\times \frac{-\im}{s_{34}} \hlp{3} \gamma_\mu | \hrp{4}\,.
\end{equation}
Using the results in
eqs.~(\ref{eq:p.exer.6.9.1}) and (\ref{eq:p.exer.6.9.2})  we find
\begin{equation}
  \label{eq:6n.125}
        {\cal M}_a(1^+,2^-,3^+,4^-,5^+)  = - 2\sqrt{2} e^3
        \frac{\spm{24}^2 \spp{23}}{\spm{15}\spm{45}\spm{34}\spp{34}}\,.
\end{equation}
For diagram (b) we find similarly
\begin{equation}
  \label{eq:6n.126}
        {\cal M}_b(1^+,2^-,3^+,4^-,5^+)  = 2\sqrt{2} e^3
        \frac{\spm{24}^2 \spp{13}}{\spm{25}\spm{45}\spm{34}\spp{34}}\,,
\end{equation}
while for (c) we have
\begin{equation}
  \label{eq:6n.127}
        {\cal M}_c(1^+,2^-,3^+,4^-,5^+)  = 2\sqrt{2} e^3
        \frac{\spm{24}^2}{\spm{12}\spm{35}\spm{45}}\,.
\end{equation}
Notice that with our choice of reference momentum we have
\begin{equation}
  \label{eq:6n.128}
        {\cal M}_d(1^+,2^-,3^+,4^-,5^+)  = 0\,.
\end{equation}
Adding up the contributions we find
\begin{equation}
  \label{eq:6n.129}
        {\cal M}(1^+,2^-,3^+,4^-,5^+)  = 2\sqrt{2} e^3
        \frac{\spm{24}^2}{\spm{12}} \left(
\frac{\spp{34}}{\spm{15}\spm{45}\spp{12}} +
\frac{1}{\spm{35}\spm{45}} \right)\,.
\end{equation}
Again this is a nice, compact result, a complex number fully expressed in terms of
helicity spinors.

Without further proof we can list what is perhaps the most famous
result in tree-level QCD amplitudes calculations \cite{Parke:1986gb}:
the so-called maximal helicity violating (MHV) amplitudes
(aka. Parke-Taylor amplitudes) for $n$-gluon scattering. One
may first organize the full tree-level invariant amplitude in the colour
quantum number as
\begin{equation}
\label{eq:p11.101.1}
 M_n(p_i,\lambda_i,a_i) = g^{n-2} \sum_{\sigma\in S_n/Z_n} 
 \left(T_{\sigma(a_1)} \dots T_{\sigma(a_n)}\right) A_n\left (\sigma(p_1^{\lambda_1}),\ldots, \sigma(p_n^{\lambda_n}) \right)\,.
\end{equation}
where the sum is over all permutations $\sigma$ modulo the cyclic
ones. The amplitudes $A_n$ are called ``colour-ordered''.
Such amplitudes \cite{Berends:1988yn,Berends:1987me,Mangano:1988kk} 
 are considerably easier to calculate. First, if all gluons have the
 same helicity, say $+$, then the amplitude is zero. The same holds if 
one of them has helicity $-$. With two helicities $-$, we have the 
MHV amplitude. The stunningly simple expression (a result of 
millions of Feynman diagrams if $n$ is large enough ) for the
colour-ordered  amplitude reads
\begin{equation}
    \label{eq:p14.102.2}
  A_n(1^+,\ldots,i^-,\ldots,j^-,\ldots n^+) = \im
  \frac{\spm{ij}^4}{\spm{12}\spm{23}\ldots \spm{n-1,n}\spm{n1}}\,.
\end{equation}
When flipping all $-$ to $+$ and vice versa, all one has to do is replace
the angled brackets by squared ones.

Helicity spinor methods are now a standard tool in the computation
of QCD scattering amplitudes for the LHC. It is worth mentioning
that among the very interesting
developments in QCD in recent years has been the realization of 
recursion relations among these amplitudes \cite{Britto:2004ap,Britto:2005fq},
after new insights where gained after phrasing them in terms of 
so-called twistors \cite{Witten:2003nn,Cachazo:2004kj}. Such recursion
relations, besides the still very powerful, and often faster
\cite{Duhr:2006iq,Dinsdale:2006sq}, earlier ones by 
Berends and Gield \cite{Berends:1987me} have been important in 
extending analytical and numerical computational power to
high-multiplicity amplitudes.

\subsection{The NLO revolution}
\label{sec:nlo-revolution}

I will here briefly touch upon recent ideas that have spurred what is
sometimes referred to as the NLO revolution. An extensive and clear 
review by some of the instigators is Ref.~\cite{Bern:2007dw}.
For many years the bottleneck in computing NLO cross sections for
many external lines were the one-loop diagrams for the virtual part of
the cross section. They become increasingly hard to calculate when
the number of external lines grows from 4 to 5, 6 etc.
Because of similar arguments 
just mentioned for the case of high-multiplicity tree-level amplitudes
one can restrict oneself to a smaller set of diagrams having
a particular colour order.  The objects to compute have, besides the
denominator factors due to the propagators in the loop, a 
numerator containing dot products among external momenta, polarization
vectors, and the loop momentum. Hence, the integral over the loop
momentum has possibly a number of loop momenta in the numerator, with
open Lorentz- indices. 

Such tensor integrals can be reduced to scalar integrals in a well-defined
procedure \cite{Passarino:1978jh}.  New stable and efficient reduction techniques for tensor integrals have
been proposed in Refs.\cite{Denner:2005nn,Binoth:2005ff}, and have found much use.

\begin{figure}
  \begin{center}
    \includegraphics[width=0.8\textwidth]{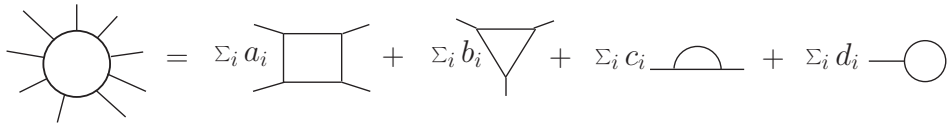}
  \end{center}
  \caption{Expansion of $n$-leg one-loop amplitude in sum of tadpoles, bubbles, triangles
and boxes.}\label{Fig:sumbox}
\end{figure}

One may also express external vectors in terms of a basis set of four. 
In this procedure also denominators are cancelled, reducing the $n$-point function to lower-point
ones. This leads to an expansion of the amplitude in terms of scalar functions down 
from $n$-point ones. Furthermore, up to (here irrelevant) $\mathcal{O}(\epsilon)$ terms, five- and
higher point functions can be expressed in terms of four-point functions and lower
\cite{vanNeerven:1983vr,Bern:1992em,Bern:1993kr}.
The price to pay is that for these lower point functions the external momenta are not
subsets but rather
combinations of the original, massless external momenta. These combinations
then are not massless. The upshot is that one has, schematically
\begin{equation}
  \label{eq:7}
A_n^{\mathrm{one-loop}}  = \sum_{j\in B} c_j \mathcal{I}_j
\end{equation}
where $B$ is a basis set that consists of a certain set of box-, triangle and bubble integrals
with or without massive external legs \cite{Ellis:2007qk},
and the $c_j$ are rational functions of dot products of
external momenta and polarization vectors, see Fig.~\ref{Fig:sumbox}.
With a generic representation (\ref{eq:7}) 
in hand, the task of calculating $A_n^{\mathrm{one-loop}}$ is
then mapped to the task of find the coefficients $c_j$. 

For this one may use unitarity methods \cite{Bern:1994zx}. In
Eq.~(\ref{eq:7}) the elements of the basis set on the \emph{right hand side}
may have branchcuts in the invariants on which the logarithms and dilogarithms in the $\mathcal{I}_j$
depend. For instance, a particular integral may have terms of the type
$\ln(-s_{ij}/\mu^2)$, with $s_{ij} = p_{ij}^2=(p_i+p_j)^2 = 2p_i\cdot
p_j$, which clearly has a branchcut in the $s_{ij}$ variable.

On the other hand, one can also examine a particular discontinuity
across a particular branch cut for a particular
invariant, or channel, for \emph{the left-hand side} in Eq.~(\ref{eq:7}), 
which is done by cutting the amplitude and replacing cut propagators
in the loop by delta functions. 
\begin{equation}
\label{eq:25}
  \frac{1}{p^2+i\epsilon}\rightarrow -i 2\pi\,\delta(p^2)
\end{equation}
This amounts to taking the imaginary part. From the comparison of both
sides the coefficients 
$c_j$ can then be determined. Essentially, one thus determines the function 
$A_n^{\mathrm{one-loop}}$ from its poles and cuts. This is vastly more
efficient than computing every term fully by itself, and is the key
insight that spurred the NLO revolution.

\begin{figure}
  \begin{center}
    \includegraphics[width=0.6\textwidth]{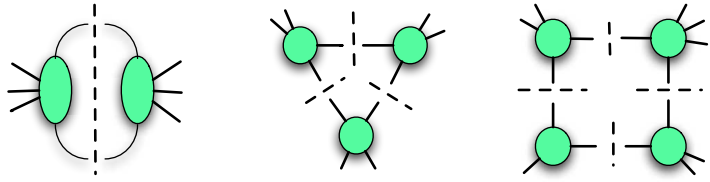}
  \end{center}
  \caption{Generalized unitarity}\label{Fig:unitarity}
\end{figure}

However, there are important subtleties. Using four-dimensional momenta
in the cuts leaves an ambiguity in the form of a rational function. Using
a $n=4+\varepsilon$ version of the unitary method \cite{Bern:1996ja,Bern:1995db} avoids this, 
but this is  somewhat more cumbersome to use. A number of other methods have been devised
to fix this ambiguity, such as using recursion relations \cite{Britto:2004ap,Britto:2005fq},
or using $D$-dimensional unitarity \cite{Anastasiou:2006gt,Brandhuber:2005jw}.
Particularly fruitful is the use of complex kinematics, which allows non-vanishing, 
non-trivial three-point amplitudes. This allows taking multiple cuts of a box integral,
such as in Fig.~\ref{Fig:unitarity}, which goes under the name ``generalized unitarity''.
By so doing, one may determine the coefficients $c_j$ purely algebraically \cite{Britto:2004nc},
since the four delta-functions fix the loop momentum. 

An effective way of solving Eq.~(\ref{eq:7}) was proposed in
Ref.\cite{Ossola:2006us}, and is known as the OPP method.
Writing the equivalent of Eq.~(\ref{eq:7}) at the \emph{integrand} level, the
coefficients of the box etc integral can then be extracted by choosing different values of
the loop momentum, and perform the inversion numerically.

Furthermore, numerical \cite{Nagy:2006xy,Anastasiou:2007qb} and semi-numerical
\cite{Ellis:2005zh} techniques for loop integrals have progressed
to the level where much work is taken care of for the user
through programs like Blackhat \cite{Berger:2008ag}, Cuttools \cite{Ossola:2007ax},
or Rocket \cite{Giele:2008bc} and MCFM \cite{Campbell:2000bg}. 
The level of automation, including the
matching
to parton showers, has now been stepped up tremendously, with the
POWHEG Box \cite{Alioli:2010xd} and aMC@NLO-MadGraph5
\cite{Alwall:2014hca} framework. They have brought NLO calculations 
now to the general user.

As this snaphot of the NLO revolution suggests, the
area of NLO calculations was a very lively marketplace
of ideas and methods. Although it is still a bustling place, attention
is now shifting to exporting the revolution to NNLO.

\subsection{Aspects of NNLO}
\label{sec:aspects-nnlo}

Here I will not say much, as this falls out of the scope of the
lectures. Many of the conceptual issues in earlier sections play
a role here as well. The accounting of singularities in a flexible
way is much harder at this order. An equation as 
(\ref{eq:7}) does not yet fully exist for this order, though impressive
progress is being made \cite{Johansson:2012sf}.
Nevertheless,
results were obtained first 
already many years ago for
DIS \cite{Zijlstra:1992qd}, Drell-Yan \cite{Hamberg:1991np}
and some time later for 
Higgs production \cite{Anastasiou:2002yz,Harlander:2002wh,Ravindran:2004mb}.
Recently the latter was even computed to NNNLO using powerful
and clever methods involving threshold expansions \cite{Anastasiou:2015ema}.

Essential for any NNLO calculation for hadron colliders are the NNLO (3-loop)
Altarelli-Parisi splitting functions. These were calculated some time ago 
\cite{Moch:2004pa,Vogt:2004mw} thanks also to the powers of the 
computer algebra program FORM \cite{Vermaseren:2008kw,Kuipers:2012rf}.

For top quark pair production \cite{Czakon:2013goa} the first full
two-to-two QCD process calculated to NNLO was completed
recently (more about this below). 
Many other results are now appearing (see e.g. \cite{Ridder:2013mf}
for NNLO results on jet cross sections), the review of which
would take us too far afield, and would anyway be out of date in very
short order. 

This concludes our discussions of finite order QCD methods and
results. Let us now turn to aspects of QCD resummation, and all-order
results. 

\section{All orders}
\label{sec:all-orders}

``Resummation'' is shorthand for all-order summation of classes
of potentially large terms in quantum field perturbation theory. 
To review status and progress in a field defined so generally is an
impossibly wide scope, and I will restrict myself
to certain types in QCD, related
of course to observables at high-energy hadron colliders.

Let us first form an impression of what resummation is and what it does.
Let $d\sigma$ be a (differential) cross section with the schematic
perturbative expansion
\begin{equation}
  \label{eq:1}
  d\sigma = 1+ \alpha_s (L^2 + L + 1) + \alpha_s^2 (L^4 + L^3 + L^2 + L + 1) + \ldots
\end{equation}
where $\alpha_s$ is the coupling, also serving as
expansion parameter, $L$ is some logarithm that is potentially large. 
In our discussion we focus on gauge theories, and on 
the case with at most two extra powers of $L$ per order, as Eq.~(\ref{eq:1})
illustrates. 
An extra order corresponds to an extra emission of 
a gauge boson, the two (``Sudakov'') logs resulting from the situation where
the emission is simultaneously soft and collinear to the parent particle direction.

Denoting $L = \ln A$, we can next ask what $A$ is.
In fact, $A$ will in general depend on the cross section at hand.
For example, for a thrust
($T$) distribution $A = 1-T$, while for $d\sigma(p\bar{p}\rightarrow
Z+X)/dp_T^Z$ $A = M_Z/p_T^Z$. It should be pointed out already here
that $A$ is not
\textit{necessarily} constructed out of measured variables but can also
be a function of unobservable partonic momenta that are to be integrated over.
E.g. for inclusive heavy quark production $A$ could be $1-4m^2/(x_1
x_2 S)$ in hadron collisions with energy $\sqrt{S}$,
where $x_1,x_2$ are partonic momentum fractions.  When $L$ is
numerically large so that even for small $\alpha_s$, the convergent
behaviour of the series is endangered, resummation of the problematic terms
into an analytic form might provide a remedy, and thereby extend the theory's predictive power to the range of large $L$.
In general the resummed form of $d\sigma$ may be written schematically as
\begin{eqnarray}
  \label{eq:2}
d\sigma_{res}  = C(\alpha_s)\, \exp\left[Lg_1(\alpha_s L) + g_2(\alpha_sL) + \alpha_s 
g_3( \alpha_sL)+\ldots  \right]   + R(\alpha_s)
\end{eqnarray}
where $g_{1,2,\ldots}$ are computable functions. The series
$C(\alpha_s)$ multiplies the exponential, and $R(\alpha_s)$ 
denotes the remainder.

The key aspect of resummation is finding the functions $g_i$. With
only $g_1$ one has leading logarithmic resummation (LL), with also
$g_2$ NLL etc. For NNLL resummation the matching function $C$ must 
also be known to next order in $\alpha_s$. 

\subsection{Resummation basics, eikonal approximation, webs}
\label{sec:resumm-basics-eikon}

Well-developed arguments exist for the exponentation
properties of the Drell-Yan cross section
near threshold~\cite{Sterman:1987aj,Catani:1989ne}.
Some are based on identifying further evolutions equations
\cite{Contopanagos:1997nh,Sterman:1987aj} based on 
refactorizations of the cross section into different 
regions only sensitive to either collinear, soft or hard corrections.
This has been made into a formidable systematic programme 
based on effective field theory \cite{Bauer:2000yr,Bauer:2001yt,Beneke:2002ph,Becher:2014oda}.

The connection between refactorization and resummation is
already illustrated by
perturbative renormalization, in
which the general relation of unrenormalized and
renormalized Green functions of fields $\phi_i$
carrying momenta $p_i$ is
\begin{equation}
  \label{eq:18}
G_{\rm un}(p_i,M,g_0)=
\prod_i Z_i^{1/2} (\mu/M,g(\mu))\; G_{\rm ren}(p_i,\mu,g(\mu))\, .
\end{equation}
$M$ is an  ultraviolet cutoff, and $g(\mu)$
and $g_0$ are the renormalized and bare couplings respectively.
The independence of $G_{\rm un}$
from $\mu$ and $G_{\rm ren}$ from $M$ may be used to derive
renormalization group equations,
\begin{equation}
  \label{eq:20}
\mu{d\ \ln\; G_{\rm ren}\over d\mu}=-\sum_i \gamma_i(g(\mu))\, ,  
\end{equation}
in which the anomalous dimensions $\gamma_i=(1/2)(\mu d/d\mu)\ln Z_i$
appear as constants in the separation of variables, free of explicit dependence on
either $\mu$ or $M$. The solution to (\ref{eq:20}) 
\begin{equation}
  \label{eq:21}
G_{\rm ren}(p_i,1,g(M))= G_{\rm ren}(p_i,M/\mu_0,g(\mu_0))
\exp\left[-\sum_i\int_{\mu_0}^M \frac{d\mu}{\mu} \gamma_i(g(\mu)) \right]\,,
\end{equation}
is clearly an exponential. While in this example the factorization
involves separation of UV modes from finite energy ones, for the
resummation we discuss in this section one (re)factorizes collinear
modes from soft-, anticollinear and hard modes. In a manner similar
to this example differential equations may be set up whose solution,
in terms of appropiate exponentials of (double) integrals over
anomalous dimensions, is the resummed cross section \cite{Contopanagos:1997nh}.

To see the appearance of exponentials in a different way
we can observe that in the refactorization
approach the soft or eikonal part of the observable is isolated in a well-defined way.
One may then use the property that moments of the eikonal DY cross
section exponentiate  at the level of integrands
\cite{Sterman:1981jc,Gatheral:1983cz,Frenkel:1984pz,Catani:1989ne}, with exponents
consisting of so-called \textit{webs}. These are selections of cut diagrams
under criteria defined by graphical topology 
(irreducibility under cuts of the eikonal lines) and with possibly
modified colour weights.  Each web is a cut diagram, 
and can be integrated over the momentum $k$ that it contributes 
to the final state. 

\begin{figure}
  \begin{center}
  \includegraphics[scale=0.6]{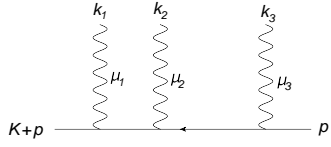}
  \caption{Soft photon emission from an energetic line} 
  \label{fig_basic_line}
  \end{center}
\end{figure}

To see how webs work, let us first consider the abelian
case\footnote{This text is derived from section 2 in \cite{Laenen:2010uz}.}.
Webs are phrased in terms of eikonal Feynman rules. 
In order to derive these one may
consider a single hard massless external line of final on-shell 
momentum $p$, originating from some unspecified hard interaction
described by ${\cal M}_0(p)$. The hard line 
may emit a number $n$ of soft photons with momenta $k_i$, as 
depicted in Fig.~\ref{fig_basic_line}, where $k_1$ is emitted closest to the hard interaction.
We shall take the emitting particle to be a scalar (the
argument for fermions is very similar). For this case
the hard interaction is dressed according to
\begin{equation}
  \label{eq_spin0_string}
  {\cal M}^{\mu_1 \ldots \mu_n} (p, k_i)  = {\cal M}_0 (p) \,
  \frac1{(p + K_1)^2}(2 p + K_1 + K_2)^{\mu_1} \ldots
  \frac1{(p + K_n)^2}(2 p + K_n)^{\mu_n} \, ,
\end{equation}
where we have introduced the partial momentum sums 
$K_i = \sum_{m = i}^n k_m$.

The eikonal approximation in this case can simply be defined as
the leading-power contribution to the amplitude when the photon
momenta $k_i^{\mu_i} \to 0$, $\forall i$, in both numerator and denominator. In this limit,
eq.~(\ref{eq_spin0_string}) becomes
\begin{equation}
  {\cal M}^{\mu_1 \ldots \mu_n} (p, k_i)  = {\cal M}_0 (p) \,
  \frac{p^{\mu_1} \ldots p^{\mu_n}}{(p \cdot K_1) 
  \ldots  (p \cdot K_n)} \, .
\label{eq_spin0_most}
\end{equation}
The eikonal factor is insensitive 
to the spin of the emitting particles.  
One may also notice that 
the eikonal factor does not depend on the energy of the
emitter, since it is invariant under rescalings of the hard 
momentum $p^\mu$: at leading power in the soft momenta,
one is effectively neglecting the recoil of the hard particle 
against soft radiation.

The eikonal factor can be further simplified by employing Bose
symmetry. Indeed for the physical quantity 
depending on the amplitude ${\cal M}^{\mu_1 \ldots \mu_n} 
(p, k_i)$, one must sum over all diagrams corresponding
to permutations of the emitted photons along the hard line. 
Having done this, the eikonal factor multiplying ${\cal M}_0 
(p)$ on the {\it r.h.s.} of (\ref{eq_spin0_most}) will be replaced 
by the symmetrized expression
\begin{equation}
  E^{\mu_1 \ldots \mu_n} (p, k_i) \, = \, \frac{1}{n!} \, 
  p^{\mu_1} \ldots p^{\mu_n} \, \sum_\pi
  \frac{1}{p \cdot k_{\pi_1}} \frac{1}{p \cdot (k_{\pi_1} + 
  k_{\pi_2})} \, \ldots \, \frac{1}{p \cdot(k_{\pi_1} + 
  \ldots + k_{\pi_n})} \, ,
\label{eikfact2}
\end{equation}
where the sum is over all permutations of the photon momenta, 
and $k_{\pi_i}$ is the $i^\text{th}$ momentum in a given 
permutation. There are $n!$ permutations, and each gives the 
same contribution to any physical observable. This becomes 
manifest using the {\it eikonal identity}
\begin{equation}
  \sum_\pi \frac{1}{p \cdot k_{\pi_1}} \frac{1}{p \cdot 
  (k_{\pi_1} + k_{\pi_2})} \ldots \frac{1}{p \cdot (k_{\pi_1}
  + \ldots k_{\pi_n})} = \prod_i \frac{1}{p \cdot k_i} \, .
\label{eikonalid}
\end{equation}
Using (\ref{eikonalid}), the eikonal factor $E^{\mu_1 \ldots 
\mu_n} (p, k_i)$ arising from $n$ soft emissions on an external 
hard line becomes simply
\begin{equation}
  E^{\mu_1 \ldots \mu_n} (p, k_i) \, = \, \prod_i \, 
  \frac{p^{\mu_i}}{p\cdot k_i} \, ,
\label{eikfactfin}
\end{equation}
which is manifestly Bose symmetric and invariant under rescalings
of the momenta $\{p_i\}$.
In practice, each eikonal emission can then be expressed by the 
effective Feynman rule
\begin{equation}
\label{eikrule}
  \includegraphics[width=2cm]{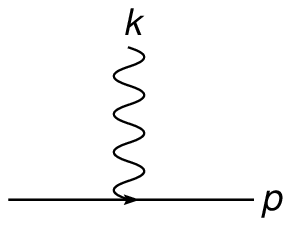}   \qquad
  {\raisebox{3ex}{\ensuremath{{\, = \, \, \,
  \Large\frac{p^\mu}{p \cdot k}}}}}
\end{equation}
These Feynman rules can be obtained by 
replacing the hard external line with a Wilson line along the 
classical trajectory of the charged particle. In abelian quantum 
field theories this is given by the expression
\begin{equation}
  \Phi_\beta (0, \infty) =
  \exp \left[\, {\rm i} e \int_{0}^{\infty} d \lambda \,
  \beta \cdot A (\lambda \beta) \, \right]~,
\end{equation}
where $\beta$ is the dimensionless four-velocity corresponding 
to the momentum $p$. (For non-abelian gauge theory the gauge
field is a matrix $A_\mu = A^a_\mu T_a$ with $T_a$ matrices that
represent the generators of the group. Because the exponent is an 
integral over a matrix-valued function, the exponential is 
a path-ordered expression. )
 This expresses the fact that soft emissions 
affect the hard particle only by dressing it with a gauge phase.
\begin{figure}
  \begin{center}
  \includegraphics[scale=0.8]{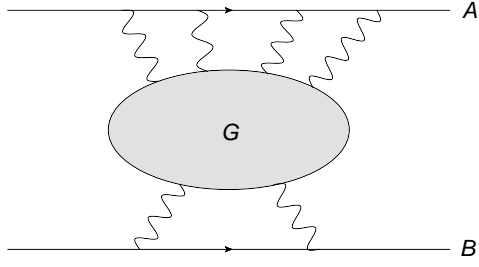}
  \caption{A process involving two eikonal lines A and B, interacting
   through the exchange of soft gluons forming diagram 
   $G$.}\label{fig_eik_exp}
  \end{center}
\end{figure}
Having constructed the effective Feynman rules, one may proceed 
to demonstrate the exponentiation of soft photon corrections
as follows. As an example, we consider graphs of the form shown 
in Fig.~\ref{fig_eik_exp}, at a fixed order in the perturbative expansion.
Fig.~\ref{fig_eik_exp} consists of two eikonal lines, labelled $A$ and 
$B$, each of which emits a number of soft photons.
One may envisage lines $A$ and $B$ as emerging from a hard 
interaction, and one may consider the graph $G$ either as a 
contribution to an amplitude, or to a squared amplitude (in which 
case some of the propagators in $G$ will be cut). 

Diagram $G$ can be taken as consisting only of soft photons 
and fermion loops. Photons originating from one of the 
two eikonal lines must land on the other one, or on a fermion loop 
inside $G$. Indeed, a photon cannot land on the same eikonal line, 
as in that case the diagram is proportional to $p^\mu p_\mu = 0$. 

Using eikonal Feynman rules, one finds that graphs of the form of
Fig.~\ref{fig_eik_exp} contribute to the corresponding (squared) 
amplitude a factor
\begin{equation}
  {\cal F}_{AB} \, = \, \sum_G \, \left[ \prod_i 
  \frac{p_A^{\mu_i}}{p_A \cdot k_i} \right]
  \left[\prod_j \frac{p_B^{\nu_j}}{p_B \cdot l_j} \right]
  G_{\mu_1 \ldots \mu_n ; \nu_1\ldots \nu_m} (k_i, l_j) \, ,
\label{amp3}
\end{equation}
where $k_i$, $l_j$ are the momenta of the photons emitted 
from lines $A$ and $B$ respectively, with $i = 1, \ldots, n$ 
and $j = 1, \ldots,  m$. 

\begin{figure}
  \begin{center}
  \scalebox{0.6}{\includegraphics{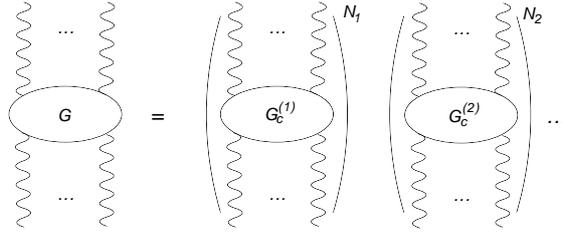}}
  \caption{Decomposition of a soft photon graph into 
  connected subdiagrams $G_c^{(i)}$, each of which 
  occurs $N_i$ times.}
  \label{decompcon}
  \end{center}
\end{figure}

Given that we have already summed over permutations in order 
to obtain the eikonal Feynman rules, each diagram $G$ can be 
uniquely specified by the set of connected subdiagrams it 
contains, as indicated schematically in Fig.~\ref{decompcon}, where 
each possible connected subdiagram $G_c^{(i)}$ occurs $N_i$ 
times. According to the standard rules of perturbation theory, 
diagram $G$ has a symmetry factor corresponding to the number 
of permutations of internal lines which leave the diagram invariant. 
This symmetry factor is given by
\begin{equation}
  S_G = \prod_i \, S_i^{N_i} \, (N_i)! \, ,
\label{sym}
\end{equation}
where $S_i$ is the symmetry factor associated with each 
connected subdiagram $G_c^{(i)}$, and the factorials account 
for permutations of identical connected subdiagrams, which 
must be divided out. Contracting Lorentz indices as in 
(\ref{amp3}), the eikonal factor ${\cal F}_{AB}$ may be 
written as
\begin{equation}
  {\cal F}_{AB} \, = \, \sum_{\{N_i\}}\, \prod_i \, 
  \frac{1}{N_i!} \, \left[ {\cal F}_c^{(i)} \right]^{N_i} \, ,
\label{amp4}
\end{equation}
where 
\begin{equation}
  {\cal F}_c^{(i)} \, = \, \frac{1}{S_i} \, \left( \prod_q 
  \frac{p_A^{\mu_q}}{p_A \cdot k_q} \right)
  \left(\prod_r \frac{p_B^{\nu_r}}{p_B \cdot l_r} \right)
  G^{(i)}_{\mu_1 \ldots \mu_{n_q} ; \nu_1\ldots \nu_{m_r}} 
  (k_q, l_r) \,,
\label{Gcdef}
\end{equation}
is the expression for each connected subdiagram, including the 
appropriate symmetry factor. Recognising (\ref{amp4})
as an exponential series, it follows that
\begin{equation}
  {\cal F}_{AB} \, = \, \exp\left[\sum_i {\cal F}_c^{(i)} \right] \, .
\label{ampexp}
\end{equation}
We conclude that soft photon corrections exponentiate in the 
eikonal approximation, and the exponent is given by the sum of 
all connected subdiagrams.
Having seen the abelian case, the non-abelian case for the eikonal
cross section is not all that
much more difficult, though we shall not treat it here.

Following arguments similar to that for the abelian case
\cite{Gatheral:1983cz}\cite{Frenkel:1984pz,Sterman:1981jc,Laenen:2008gt,Gardi:2010rn,Mitov:2010rp}
one may in fact arrive at the result that the eikonal cross section is a sum
over eikonal diagrams $D$ 
\begin{equation}
  \sigma^{\rm (eik)} =\exp\left[\sum_W\sum_{D,D'}{\cal F}(D)R^{(W)}_{DD'}C(D')\right],
\label{Sstruc}
\end{equation}
where a mixing matrix $R$ connects the momentum space parts of the
diagrams ${\cal F}(D)$ and their colour factors $C(D)$ in an
interesting way.  A very recent, pedagogical review of this and other
aspects of webs can be found in \cite{White:2015wha}.

The functions $g_i$ that constitute the resummation are usually
not only defined by the eikonal cross section. Hard collinear modes
in the higher-order corrections can be resummed in different ways, 
through so-called jet functions. These are in fact also universal, 
so that by now constructing a resummed cross section is often
a matter of putting the right set of all-order functions together. 
Some automation of this has already been undertaken \cite{Banfi:2004yd,Becher:2014aya}
and is at present being worked on further by various groups.

\subsection{Some applications in threshold resummation: heavy quark production, Higgs production}
\label{sec:appl-heavy-quark}

\begin{figure}[htbp]
\begin{center}
 \includegraphics[scale=0.5]{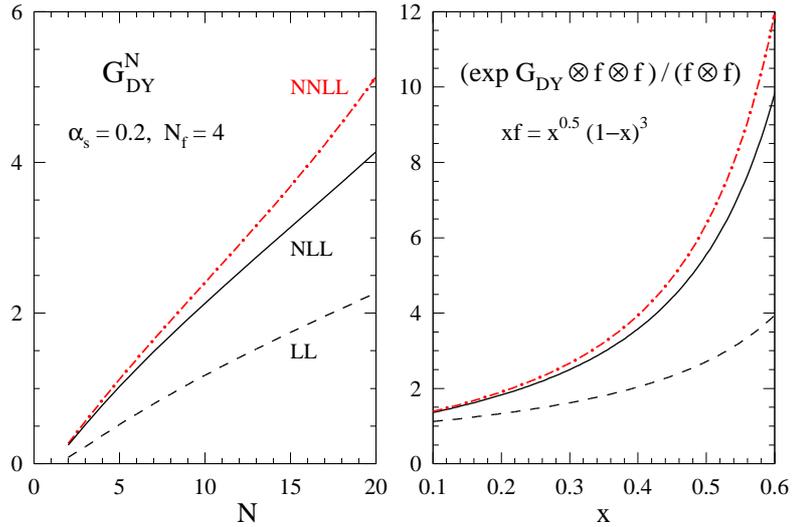}%
  \caption{Convergence behavior of Drell-Yan partonic and hadronic
    cross section \cite{Vogt:2000ci}}
\label{fig:dyall}
  \end{center}
\end{figure}

The very general arguments in the previous section can be applied to
transverse momentum resummation and threshold resummation. Here we
focus on the latter. 
An illuminating study of the effects of threshold resummation was given in Ref.\cite{Vogt:2000ci}.
One can representing the partonic resummed cross section in moment space as
\begin{eqnarray}
\label{eq:15}
  \sigma_{DY}(N,Q^2) &=& H (Q) \, \exp\left[G_{DY}(N,Q)\right] \\
G_{DY} &=& 2\ln N g_1(2\lambda) + g_2(2\lambda) + \alpha_s g_3(2\lambda)+\ldots \\
    \lambda &=& \beta_0 \alpha_s \ln N \,,
\end{eqnarray}
which was already more schematically given in (\ref{eq:2}), and where
\begin{equation}
  \label{eq:19}
  g_1(\lambda) = \frac{C_F}{\beta_0 \lambda}\left[
\lambda + (1-\lambda) \ln(1-\lambda)\right]\,.
\end{equation}
In Fig.~\ref{fig:dyall} \cite{Vogt:2000ci} convergence properties for both the exponent
and the resummed cross section are shown (for toy PDF's) 
when increasing the logarithmic accuracy of the exponent, and of the
hadronic $K$ factor. 
One observes good convergence as the logarithmic accuracy of the resummation is
increased. For the inverse Mellin transform, required to compute
the hadronic cross section in momentum space, one may use the so-called minimal prescription 
\cite{Catani:1996yz}. 

\begin{figure}[htbp]
\begin{center}
 \includegraphics[scale=1.2]{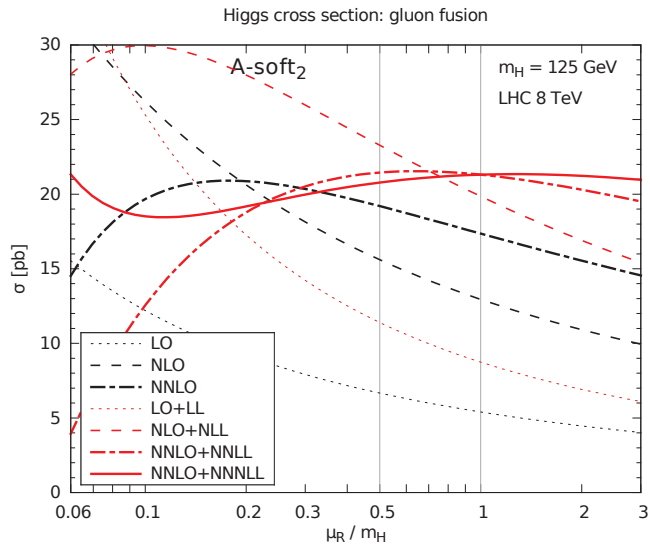}%
  \caption{A recent result \cite{Bonvini:2014joa} for the N$^3$LL cross section for Higgs
    production. One observes the notably less dependence of the result
on the renormalization scale $\mu_R$. }
\label{fig:BM}
  \end{center}
\end{figure}

Very similar to Drell-Yan is Higgs production, in the large top mass
limit where there is essentially a pointlike gluon-gluon-Higgs
coupling. A recent N$^3$LL threshold resummed result
\cite{Bonvini:2014joa} is shown in Fig.~\ref{fig:BM}.

\begin{figure}[htbp]
\begin{center}
 \includegraphics[scale=0.7]{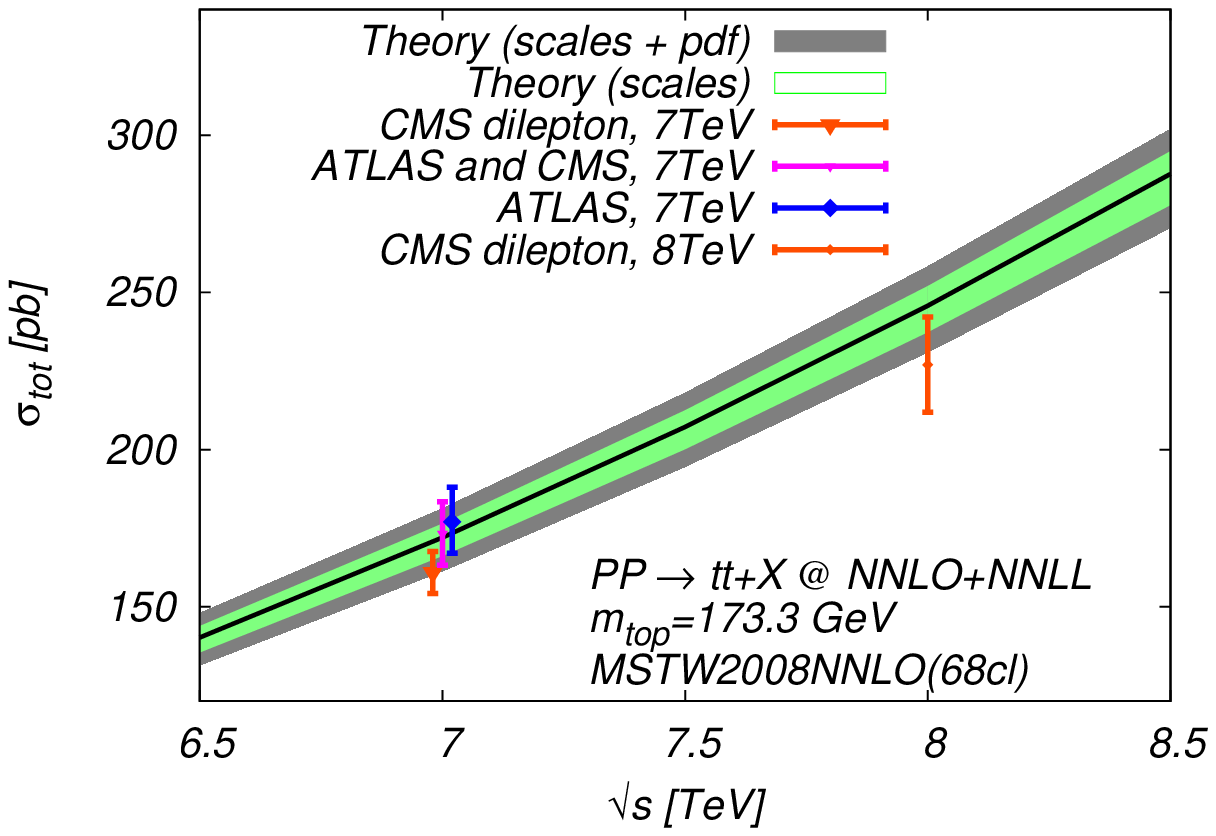}%
  \caption{Theoretical prediction for the LHC as a function of
the collider c.m. energy, compared to available measurement
from ATLAS and/or CMS at 7 and 8 TeV. }
\label{fig:CFM}
  \end{center}
\end{figure}

We conclude this section with the already mentioned NNLO top quark
cross section results, matched to a NNLL threshold resummed
calculation for this observable \cite{Czakon:2009zw}. For the resummed part an added
complication is the accounting for colour, as all four external
particles are coloured. This issue was solved in 
Refs.~\cite{Kidonakis:1997gm,Contopanagos:1997nh,Bonciani:1998vc},
we shall not go into the technical aspects of this. The 
result of the very impressive, and important calculation
\cite{Czakon:2013goa}  is shown in Fig.~\ref{fig:CFM}

\section{Conclusions}
\label{sec:conclusions}

In these lectures I have discussed many aspects of
QCD, from the conceptual to the practical, that are relevant for
understanding and appreciating its role in the physics of particle
colliders. These aspects are often quite technical in nature, and no
doubt I have done poor justice to them in the limited space available.
Nevertheless, some attention must be given to these, in
order to assess the strengths and weaknesses of theoretical results.
This is of crucial importance when confronting these results
with data. With the focus of theory and measurement turning towards 
precision, having paid this attention should be all the more valuable.

Nevertheless I hope that readers are not  blinded by the
technicalities, but are able to sharpen their intuition regarding
how QCD behaves a bit further. Both technical understanding and
intuition will be fruitful in the
theory-experiment collaborations, joint workshops etc in which they
may find themselves at times, and upon which much
of the success of the LHC research programme depends.

\section*{Acknowledgements}

I would like to thank the organizers of the CERN Summer School for the
excellent environment they created at the school, 
and for their forebearance towards these notes. I thank the students
of the school for their interest and lively participation in the
lectures and question sessions.

\appendix

\section{Conventions and useful formulae}
\label{sec:conv-usef-form}

\subsection*{Units}
\label{sec:units}

We use here $\hbar = 1$ and $c=1$. Energy can be converted
to inverse distance and vice versa by the relation
$1  = 197.3 \mathrm{MeV}\, \mathrm{fm}$, with 1 fm (``fermi''
or ``femtometer'') equal to $10^{-15}$m,
and $c = 2.998 \cdot 10^8\; \mathrm{m/s}$.\\

Cross sections are expressed in nanobarns (nb), picobarns (pb) etc,
where $1 \, \mathrm{b} = 10^{-24} cm^2$.

The metric tensor we use in this course is 
\begin{equation}
  \label{eq:85}
  \eta^{\mu\nu} = \mathrm{diag}(-1,1,1,1)
\end{equation}

\subsection*{Cross sections and decay rates}
\label{sec:cross-sections-decay}

The cross section for the scattering of two incoming massless
particles with momenta $k_1$ and $k_2$ to $n$ particles with
momenta $\{p_i\}$ is given by
\begin{equation}
\label{eq:A:121}
  \sigma = \frac{1}{2s}\overline{\sum_{spins}} \int |{\cal M}|^2\, dPS(n)
\end{equation}
where the bar indicates initial spin averaging and 
\begin{equation}
\label{eq:23}
  dPS(n) = \prod_{j=1}^{n}
  \frac{d^3p_j}{(2\pi)^3\,2E_j}\times 
(2\pi)^4 \delta\left(k_1+k_2 - \sum_i^{n}p_i \right)
\end{equation}
If there are $j$ identical particles among the $n$, there is an extra
factor $1/j!$.
For the case $n=2$, in the center of momentum frame, and with 
both outgoing particles having equal mass $m$
\begin{equation}
\label{eq:A:129}
    dPS(2) = \frac{1}{16\pi } \sqrt{1-\frac{4m^2}{s}}\, d\cos\theta\,,
\end{equation}
where $\theta$ is the polar angle of one of the outgoing particles 
with respect to the collision axis.  

The width for the decay of a particle with mass $Q$
and 4-momentum $k$ to $n$ particles with
4-momenta $\{p_i\}$ reads
\begin{equation}
  \label{eq:A:1}
  \Gamma = \frac{1}{2Q}\overline{\sum}_{\mathrm{spins},(\mathrm{colours}..)} \int |{\cal M}|^2\, dPS(n)\,.
\end{equation}
If there are $j$ identical particles among the $n$, there is an extra
factor $1/j!$.
For the case $n=2$, in the center of momentum frame
\begin{equation}
  \label{eq:A:44}
    dPS(2) = \frac{1}{16\pi Q^2 } \sqrt{\lambda(Q^2,m_1^2,m_2^2)} d\cos\theta\,,
\end{equation}
where $\theta$ is the polar angle of one of the outgoing particles 
with respect to some arbitrary axis,
and $\lambda(x,y,z) = x^2+y^2+z^2-2xy-2xz-2yz$.

\subsection*{Dirac algebra}
\label{sec:dirac-algebra}

Dirac equation in x-space:
\begin{equation}
  \label{eq:A:20}
  (\slash\partial + m)\psi(x) = 0
\end{equation}
Dirac equation in momentum space for $u$ and $v$ spinors:
\begin{equation}
  \label{eq:A:21}
  (i\slash{p}+m)\,u(p,s) = 0, \qquad 
  (i\slash{p}-m)\,v(p,s) = 0
\end{equation}
\begin{equation}
  \label{eq:A:3}
  \{ \gamma^\mu,\gamma^\nu \} = 2\eta^{\mu\nu}
\end{equation}
where on the right hand side the 4 by 4 unit matrix in spinor space is implied.
An often-used identity based on this is 
\begin{equation}
  \label{eq:A:48}
  \slash{p}\slash{p} = p^2 \,.
\end{equation}
Other useful relations:
\begin{equation}
  \label{eq:A:4}
      \gamma^{0} = -i\,\left( \begin{array}{cc}
       1 & 0 \\
       0 & -1 
     \end{array} \right),\qquad 
    \vec{\gamma} = -i\,\left( \begin{array}{cc}
       0 & \vec{\sigma} \\
      -\vec{\sigma}  & 0 
     \end{array} \right)\,,
\end{equation}
\begin{equation}
\label{eq:A:29}
      (\gamma^\mu)^\dagger = \gamma^0\gamma^\mu\gamma^0\,,
\end{equation}
\begin{equation}
  \label{eq:A:6}
  \gamma_5 = i\gamma_0\gamma_1\gamma_2\gamma_3\,,
\end{equation}
\begin{equation}
  \label{eq:A:7}
      \gamma_5 = \left( \begin{array}{cc}
       0 & 1 \\
       1 & 0 
    \end{array}\right)\,,
\end{equation}
\begin{equation}
  \label{eq:A:16}
  \left\{  \gamma_5,\gamma^\mu \right\} = 0\,,
\end{equation}
\begin{equation}
  \label{eq:A:8}
      \gamma^\mu\slash{a}\gamma_\mu = -2 \slash{a},\quad
    \gamma^\mu\slash{a}\slash{b}\gamma_\mu = 4 a\cdot b,\quad
    \gamma^\mu\slash{a}\slash{b}\slash{c}\gamma_\mu = -2 \slash{c}\slash{b}\slash{a}\,,
\end{equation}
\begin{eqnarray}
  \label{eq:A:9}
 &\ &     \mathrm{Tr}(\gamma^\mu) =  \mathrm{Tr}(\gamma_5) = 0\,,\quad
      \mathrm{Tr}(\gamma^\mu\gamma^\nu\gamma^\rho) = 0\,, \nonumber \\
 &\ & \mathrm{Tr}(\gamma_5 \gamma_\mu ) = \mathrm{Tr}(\gamma_5
 \gamma_\mu \gamma_\nu) = \mathrm{Tr}(\gamma_5 \gamma_\mu \gamma_\nu \gamma_\rho) = 0\,,
\end{eqnarray}
\begin{equation}
  \label{eq:A:10}
          \mathrm{Tr}(\gamma^\mu\gamma^\nu) = 4\,\eta^{\mu\nu},\quad
        \mathrm{Tr}(\gamma^\mu\gamma^\nu\gamma^\rho\gamma^\sigma) = 
4\,(\eta^{\mu\nu}\eta^{\rho\sigma}-\eta^{\mu\rho}\eta^{\nu\sigma}+
 \eta^{\mu\sigma}\eta^{\nu\rho})\,,
\end{equation}
\begin{equation}
  \label{eq:A:22}
  \mathrm{Tr}(\gamma_\mu\gamma_\nu\gamma_\rho\gamma_\sigma\gamma_5)
 = -4 i \epsilon_{\mu\nu\rho\sigma}\,,\quad
\epsilon_{0123} = +1\,.
\end{equation}

Conjugate spinor:
\begin{equation}
  \label{eq:A:15}
  \bar\psi = i\psi^\dagger \gamma^0\,.
\end{equation}
With this definition the term
\begin{equation}
  \bar\psi \psi
\end{equation}
is hermitean, since $\gamma^0$ is anti-hermitean.
The chiral (left- and righthanded) projections of a fermion are defined by
\begin{eqnarray}
  \label{eq:A:14}
  \psi_L &=& \left(\frac{1+\gamma_5}{2} \right)\psi\,, \nonumber \\
  \psi_R &=& \left(\frac{1-\gamma_5}{2} \right)\psi\,.
\end{eqnarray}
A parity transform on a spinor is defined by
\begin{equation}
  \label{eq:A:5}
  P: \psi(x) \rightarrow \gamma_0 \psi(\tilde{x})\,,
\end{equation}
with $\tilde{x} = (x^0,-\vec{x})$.
Complex-conjugation of general spinor-trace:
\begin{equation}
\label{eq:A:18}
  (\bar{u}(p)\gamma_{\mu_1}\cdots \gamma_{\mu_n} u(p'))^*
 = (-)^n \;(\bar{u}(p')\gamma_{\mu_n}\cdots \gamma_{\mu_1} u(p))\,,
\end{equation}
\begin{equation}
\label{eq:A:23}
  (\bar{u}(p)\gamma_{\mu_1}\cdots \gamma_{\mu_n}\gamma_5 u(p'))^*
 = -(-)^n \;(\bar{u}(p')\gamma_5\gamma_{\mu_n}\cdots \gamma_{\mu_1} u(p))\,.
\end{equation}

\subsection*{Unitary groups and their Lie algebras}
\label{sec:unitary-groups-are}

The U(1) Lie algebra has only 1 generator $t$, which we
choose hermitean.
Acting on a d-dimensional vector $t$ may be represented
as the d-dimensional unit matrix.
 The group elements are then
\begin{equation}
  \label{eq:A:19}
  \exp\left(i\alpha\, t  \right)\,.
\end{equation}
The SU(2) Lie algebra has 3 generators $t_i,\, i=1,2,3$.
If we choose the $t_i$ hermitean, then in the fundamental
representation $t^{(F)}_i = \sigma_i/2$, with $\sigma_i$ the
Pauli matrices
\begin{eqnarray}
  \label{eq:A:su2mat}
  \begin{array}{ccc}
\sigma_1 = \left(
\begin{array}{cc}
  0 & 1\\
  1 & 0 
  \end{array}
  \right)
&
\sigma_2 = \left(
\begin{array}{cc}
  0 & -i\\
  i & 0 
  \end{array}
  \right)

&
\sigma_3 = \left(
\begin{array}{cc}
  1 & 0\\
  0 & -1 
  \end{array}
  \right)\,.
  \end{array}
\end{eqnarray}
The group elements are 
\begin{equation}
\label{eq:A:34}
 U =  \exp\left(i\xi^i\, t_i  \right)\,.
\end{equation}
Note that for the fundamental representation
\begin{equation}
  \label{eq:A:11}
\sigma_2 U \sigma_2 = U^* = (U^\dagger)^T\,.
\end{equation}
Note that the SU(2) Lie-algebra is isomorphic to the SO(3) Lie-algebra,
whose generators $t_i = -i S_i$ read, in the fundamental
representation:
\begin{eqnarray}
  \label{eq:A:so3mat}
  \begin{array}{ccc}
S_1 = \left(
\begin{array}{ccc}
  0 & 0 & 0\\
  0 & 0 & 1\\ 
  0 & -1 & 0
  \end{array}
  \right)
&
S_2 = \left(
\begin{array}{ccc}
  0 & 0 & -1\\
  0 & 0 & 0\\ 
  1 & 0 & 0
  \end{array}
  \right)

&
S_3 = \left(
\begin{array}{ccc}
  0 & 1 & 0\\
  -1 & 0 & 0\\ 
  0 & 0 & 0
  \end{array}
  \right)
  \end{array}\,.
\end{eqnarray}

The SU(3) Lie algebra has 8 generators $t_i,\, i=1,\ldots 8$,
which are not needed explicitly.
Lie algebra generators in general obey the commutation relations
\begin{equation}
  \label{eq:A:32}
  \left[t_i,t_j  \right] = i\, f_{ijk} \, t_k\,,
\end{equation}
with the $f_{ijk}$ the structure constants for the given group.
The $kj$ matrix element of the generator $t^{(A)}_i$ in the adjoint
representation is defined as
\begin{equation}
  \label{eq:A:33}
  \left[t^{(A)}_i  \right]_{kj} = i\, f_{ijk}\,.
\end{equation}

Note that we can always choose anti-hermitean generators $t'$ by
multiplying the hermitean versions $t$ by $i$. In that case
the group elements for SU(2) e.g. are
\begin{equation}
\label{eq:A:42}
  \exp\left(\xi^i\, t'_i  \right)\,.
\end{equation}

Representations of SU(2), etc. groups are often indicated
by $\underline{2}, \ldots$ indicating the size 
of the matrices of that representation. Trivial or singlet
representation are then indicated with $\underline{1}$.

Some other group theory factors (the so-called Casimir factors):
\begin{eqnarray}
\label{eq:A:55}
  \mathrm{SU(2)}:\quad\quad C_A &=& 2,\quad C_F = \frac{3}{4}\,, \nonumber\\
  \mathrm{SU(3)}:\quad\quad C_A &=& 3,\quad C_F = \frac{4}{3}\,. \nonumber
\end{eqnarray}

For the fundamental representation of $SU(N)$ we have
\begin{equation}
  \label{eq:A:30}
  \mathrm{Tr}\left[t^{(F)}_i t^{(F)}_j \right] = \frac{1}{2}\delta_{ij}\,.
\end{equation}

\subsection*{Standard Model quantities}
\label{sec:stand-model-quant}

The amount of electric charge (in units of $e$) $Q$, the hypercharge
$Y$ and the third component of weak isospin $t_3$ are related by
\begin{equation}
\label{eq:A:12}
  Y = 2(Q-t_3)\,.
\end{equation}
In the Standard Model the generator of the U(1) of hypercharge
is conventionally written as 
\begin{equation}
  \label{eq:A:40}
  \frac{1}{2} Y\,,
\end{equation}
and is then represented on d-dimensional vectors as $1/2$ times
the hypercharge eigenvalue times the unit matrix.

The gauge couplings associated with the $\mathrm{SU}_{I_W}(2)$ 
and $\mathrm{U}_Y(1)$ gauge groups are traditionally denoted
$g$ and $g'$ respectively. In terms of these couplings the
unit of electric charge is
\begin{equation}
  \label{eq:A:41}
  e = \frac{g\, g'}{\sqrt{g^2+g'^2}} = g\, \sin\theta_W = g' \, \cos\theta_W\,,
\end{equation}
with $\sin^2\theta_W \simeq 0.226$.

The gauge fields are mixed as follows:
\begin{eqnarray}
  \label{eq:A:38}
  B_\mu & = & \cos\theta_W A_\mu - \sin\theta_W Z_\mu, \quad
  W^3_\mu = \cos\theta_W Z_\mu + \sin\theta_W A_\mu, \\
 W^1_\mu - i W^2_\mu & = & \sqrt{2} W_\mu^+, \quad
 W^1_\mu + i W^2_\mu  =  \sqrt{2} W_\mu^-\,.
\end{eqnarray}

The Fermi constant is defined by
\begin{equation}
  \label{eq:A:50}
  G_F = \frac{g^2}{4\sqrt{2}m_W^2}=\frac{1}{\sqrt{2}v^2} \simeq 1.2 \times 10^{-5} \, \mathrm{GeV}^2\,.
\end{equation}

Vector boson masses in GeV:
\begin{equation}
  \label{eq:A:45}
  m_Z = \frac{gv}{2\cos\theta_W} = 91.1876 \pm 0.0021,\quad m_W = \frac{gv}{2}=
  80.385\pm 0.015\,.
\end{equation}
The photon and gluon are massless.

Heavy quark masses in GeV:
\begin{equation}
  \label{eq:A:46}
  m_c = 1.5,\quad m_b = 5, \quad m_t = 175\,.
\end{equation}
$u,d,s$ are massless.

Lepton masses in GeV:
\begin{equation}
  \label{eq:A:47}
  m_\tau = 1.7 ,\quad m_\mu = 0.105 ,\quad m_e = 0.0005 \,.
\end{equation}

QCD scale in GeV:
\begin{equation}
  \label{eq:A:13}
  \Lambda_{QCD} \simeq 0.2 \,.
\end{equation}

\subsection*{Assorted Feynman rules}
\label{sec:assort-feynm-rules}

All momenta in vertex Feynman rules are incoming here.

\begin{itemize}
\item Fermion propagator
 \begin{picture}(200,40)
\ArrowLine(50,0)(150,0)
\Text(100,10)[]{$\longrightarrow$ p}
\Text(45,0)[r]{$\alpha$}
\Text(155,0)[l]{$\beta$}
  \end{picture}

  \begin{equation}
    \label{eq:A:36}
\frac{1}{i(2\pi)^4}{(-i\slash{p} + m)_{\beta \alpha}\over p^2+m^2-i\epsilon}
  \end{equation}

\item Electron-electron photon vertex
  \begin{displaymath}
   \begin{array}{cc}
\qquad\qquad\qquad &  
\begin{picture}(110,105)
\Photon(50,100)(50,40){5}{5}
\ArrowLine(0,0)(50,40)
\ArrowLine(50,40)(100,0)
\Text(60,70)[l]{$p_3$}
\Text(20,25)[lb]{$p_1$}
\Text(80,25)[rb]{$p_2$}
\Text(-5,5)[b]{$\alpha$}
\Text(105,5)[b]{$\beta$}
\Text(50,105)[b]{$\mu$}
  \end{picture}
    \end{array}
  \end{displaymath}
  \begin{equation}
    \label{eq:A:35}
i(2\pi)^4 \delta(p_1+p_2+p_3) (-i e)\, \gamma^\mu_{\beta \alpha}
  \end{equation}
where the three momentum vectors (not drawn) are pointing to the vertex.

\item Outgoing fermion: $\qquad$
  \begin{picture}(70,20)(0,5)
 \Photon(0,20)(10,10){2}{2}
 \ArrowLine(0,0)(10,10)
 \ArrowLine(10,10)(70,10)
 \LongArrow(30,20)(40,20)
 \Text(45,20)[l]{$p$}
  \end{picture} $\;\;$ $\overline{u}(p,s)$
$\qquad$ Row spinor

\item Outgoing anti-fermion: $\qquad$
  \begin{picture}(70,20)(0,5)
 \Photon(0,20)(10,10){2}{2}
 \ArrowLine(10,10)(0,0)
 \ArrowLine(70,10)(10,10)
 \LongArrow(30,20)(40,20)
 \Text(45,20)[l]{$p$}
  \end{picture} $\;\;$ $v(p,s)$
$\qquad$ Column spinor

\item Incoming fermion: $\qquad$ $u(p,s)\;\;$
  \begin{picture}(70,20)(0,5)
 \ArrowLine(0,10)(60,10)
 \Photon(60,10)(70,20){2}{2}
 \ArrowLine(60,10)(70,0)
 \LongArrow(30,20)(40,20)
 \Text(45,20)[l]{$p$}
  \end{picture}  
$\qquad$ Column spinor

\item Incoming antifermion: $\qquad$ $\overline{v}(p,s)\;\;$
  \begin{picture}(70,20)(0,5)
 \ArrowLine(60,10)(0,10)
 \Photon(60,10)(70,20){2}{2}
 \ArrowLine(70,0)(60,10)
 \LongArrow(30,20)(40,20)
 \Text(45,20)[l]{$p$}
  \end{picture}  
$\qquad$ Row spinor

\item Outgoing vector boson: $\qquad$
  \begin{picture}(70,20)(0,5)
 \ArrowLine(0,20)(10,10)
 \ArrowLine(10,10)(0,0)
 \Photon(10,10)(70,10){2}{7}
 \LongArrow(30,20)(40,20)
 \Text(45,20)[l]{$k$}
  \end{picture} $\quad$ $\epsilon^*_\mu(k,\lambda)$

\item Incoming vector boson:$\quad$ $\epsilon_\mu(k,\lambda)$ $\;$
  \begin{picture}(70,20)(0,5)
 \Photon(0,10)(60,10){2}{7}
 \ArrowLine(60,10)(70,20)
 \ArrowLine(70,0)(60,10)
 \LongArrow(30,20)(40,20)
 \Text(45,20)[l]{$k$}
  \end{picture}  

\end{itemize}

In Feynman diagrams, always start where the charge vector
top of the fermion lines {\it ends} (i.e. start with a 
row spinor), and work your way back against the charge flow.\\[0.2ex]

Integrate over internal momenta: $\int d^4 k_i$\\[0.2ex]

Completeness relations for spin sums over polarization spinors and
polarization vectors, associated with external particles
\begin{equation}
    \label{eq:A:sumu}
    \sum_s u_\alpha(p,s) \overline{u}_\beta(p,s)    
           = (-i\slash{p}+m)_{\alpha\beta}\,,
  \end{equation}
  \begin{equation}
    \label{eq:A:sumv}
      \sum_s v_\alpha(p,s) \overline{v}_\beta(p,s)    
           = (-i\slash{p}-m)_{\alpha\beta}\,,
  \end{equation}
with $m$ the fermion mass, $\alpha,\beta$ are spinor indices.

For photons the sum over the two physical polarizations gives
  \begin{equation}
    \label{eq:A:sumeps}
    \sum_\lambda \epsilon^\mu(k,\lambda) \epsilon^{*\,\nu}(k,\lambda)
 = \eta^{\mu\nu} - \frac{k^\mu \bar{k}^\nu +k^\nu \bar{k}^\mu }{k\cdot
   \bar{k}}\,,
  \end{equation}
where the sum is over the physical spin only, and we define
\begin{equation}
  \label{eq:A:37}
  k^\mu = (k^0,\vec{k}),\qquad \bar{k}^\mu = (-k^0,\vec{k})\,.
\end{equation}
The photon propagator is ($\kappa$ is the gauge parameter here)
\begin{equation}
\label{eq:A:31}
  \Delta_{\mu\nu}(k) = \frac{1}{i(2\pi)^4}\frac{1}{k^2-i\epsilon}
\left(\eta_{\mu\nu}-\left(1-\frac{1}{\kappa^2}\right)\frac{k_\mu k_\nu}{k^2}\right)\,.
\end{equation}
The scalar field propagator is simply
\begin{equation}
\label{eq:A:17}
  \Delta(p) = \frac{1}{i(2\pi)^4}\frac{1}{p^2+m^2-i\epsilon}\,.
\end{equation}
For massive vector bosons the sum over the three physical polarizations 
gives 
\begin{equation}
  \label{eq:A:43}
      \sum_\lambda \epsilon^\mu(k,\lambda) \epsilon^{*\,\nu}(k,\lambda)
 = \eta^{\mu\nu} + \frac{k^\mu k^\nu }{M^2}\,.
\end{equation}

\subsection*{Loop integrals}
\label{sec:loop-integrals}

\begin{equation}
\label{eq:A:24}
  \frac{1}{A_1 A_2} = 
\int_0^1 dx_1\int_0^1 dx_2 \frac{\delta(1-x_1-x_2)}{\left[x_1 A_1 + x_2 A_2 \right]^2}\,.
\end{equation}
%% \begin{equation}
%% \label{eq:A:25}
%%   \frac{1}{\left[q^2+m^2-i\epsilon\right]}  \frac{1}{\left[(k-q)^2+m^2-i\epsilon\right]}
%% = \int_0^1 dx
%% \frac{1}{\left[(q-xk)^2 + (x-x^2)k^2+m^2 -i\epsilon\right]^2}
%% \end{equation}
The result for the integral
\begin{equation}
\label{eq:A:28}
  I(n,\alpha) = \int d^nq \frac{1}{(q^2 + m^2 -i\epsilon )^\alpha}
\end{equation}
is
\begin{equation}
\label{eq:A:27}
  I(n,\alpha) =  i\pi^{n/2} \frac{\Gamma(\alpha-(n/2))}{\Gamma(\alpha)}
  \left(m^2 \right)^{(n/2)-\alpha}\,.
\end{equation}
The Euler gamma function $\Gamma(z)$ has the following relevant properties:
\begin{eqnarray}
  \label{eq:A:26}
  \Gamma(z+1) &=& z\Gamma(z),\quad \Gamma(1) = 1\,,  \\
 \ln \Gamma(1 + z) &\simeq & -z \gamma_E + {\cal O}(z^2),\quad
 \gamma_E = 0.577\ldots
\end{eqnarray}

% \bibliographystyle{utphys}
% \bibliography{laenen_qcd}

\providecommand{\href}[2]{#2}\begingroup\raggedright\endgroup

\end{document}